\documentclass[11pt]{article}
\pdfoutput=1

\usepackage{cancel,slashed}
\usepackage{bbm}
\usepackage{bm}
\usepackage{amsmath,amsfonts,amssymb,mathtools,nicefrac,nccmath,cases}
\usepackage{graphicx}
\usepackage{cite}
\usepackage{physics}
\usepackage{skak}

\usepackage{dsfont}
\usepackage{latexsym}
\usepackage{color}
\usepackage[hyperfootnotes=false,linktocpage]{hyperref}
\usepackage{transparent}
\usepackage[hang,flushmargin]{footmisc}
\usepackage{blkarray}
\usepackage{multirow}
\usepackage{empheq}
\usepackage{comment}
\usepackage{wasysym}
\usepackage{tikzsymbols}

\usepackage{caption}
\usepackage{subcaption}
\usepackage[normalem]{ulem}

\newcommand{\bmat}{\left(\begin{array}}
\newcommand{\emat}{\end{array}\right)}

\def\bN{\mathbbm{N}}
\def\Z{\mathbbm{Z}}
\def\R{\mathbbm{R}}
\def\C{\mathbbm{C}}

\def\P{\mathbbm{P}}

\def\a {\alpha}

\def\1{{\bf 1}}
\def\2{{\bf 2}}
\def\3{{\bf 3}}
\def\4{{\bf 4}}
\def\6{{\bf 6}}

\def\half{\frac{1}{2}}

\def\targ#1#2{\genfrac{[}{]}{0pt}{}{#1}{#2}}
\def\targ2#1#2{\genfrac{}{}{0pt}{}{#1}{#2}}
\def\half{{\textstyle\frac{1}{2}}}

\definecolor{mygr}{rgb}{0,0.6,0}
\definecolor{mygrey}{rgb}{0,0.1,0.2}
\definecolor{myblue}{rgb}{0,0.5,0.9}
\definecolor{myblue2}{rgb}{0,0.5,0.5}
\definecolor{myblue3}{rgb}{0,0.7,0.9}
\definecolor{myblue4}{rgb}{0,0.6,0.6}
\definecolor{myorange}{rgb}{1,0.5,0}
\definecolor{mypurple}{rgb}{0.6,0,1}
\definecolor{mygolden}{rgb}{1,0.8,0.2}
\definecolor{mycyan}{rgb}{0,1,1}
\definecolor{mymagenta}{rgb}{1,0,1}
\definecolor{mykiwi}{rgb}{0.8,1,0.5}
\definecolor{mybrown}{cmyk}{0.14, 0.42, 0.56, 0.2}
\definecolor{myturq}{cmyk}{0.99, 0, 0.2, 0.4}
\definecolor{myaubergine2}{cmyk}{0.4, 0.5, 0, 0.1}
\definecolor{myaubergine}{cmyk}{0.6,0.85,0,0}
\definecolor{CycleGreen}{cmyk}{0.52,0,1,0}
\definecolor{CycleBrown}{cmyk}{0, 0.4, 0.9, 0.2}

\DeclareFontFamily{U}{rcjhbltx}{}
\DeclareFontShape{U}{rcjhbltx}{m}{n}{<->rcjhbltx}{}
\DeclareSymbolFont{hebrewletters}{U}{rcjhbltx}{m}{n}

\DeclareMathSymbol{\lamed}{\mathord}{hebrewletters}{108}
\DeclareMathSymbol{\mem}{\mathord}{hebrewletters}{109}
\DeclareMathSymbol{\ayin}{\mathord}{hebrewletters}{96}
\DeclareMathSymbol{\tsadi}{\mathord}{hebrewletters}{118}
\DeclareMathSymbol{\qof}{\mathord}{hebrewletters}{113}
\DeclareMathSymbol{\resh}{\mathord}{hebrewletters}{114}
\DeclareMathSymbol{\pe}{\mathord}{hebrewletters}{112}
\DeclareMathSymbol{\pesofit}{\mathord}{hebrewletters}{80}
\DeclareMathSymbol{\samekh}{\mathord}{hebrewletters}{115}
\DeclareMathSymbol{\tav}{\mathord}{hebrewletters}{116}
\DeclareMathSymbol{\vav}{\mathord}{hebrewletters}{119}
\DeclareMathSymbol{\het}{\mathord}{hebrewletters}{120}
\DeclareMathSymbol{\yod}{\mathord}{hebrewletters}{121}
\DeclareMathSymbol{\zayin}{\mathord}{hebrewletters}{122}
\DeclareMathSymbol{\alephdot}{\mathord}{hebrewletters}{128}
\DeclareMathSymbol{\tsadisofit}{\mathord}{hebrewletters}{90}
\DeclareMathSymbol{\shin}{\mathord}{hebrewletters}{152}

\newtheorem{conjecture}{Conjecture}

\def\CN {{\cal N}}
\def\CM {{\cal M}}
\def\ii           {{\rm i}}

\def\d {{\rm d}}
\def\be{\begin{equation}}
\def\ee{\end{equation}}
\def\bea{\begin{eqnarray}}
\def\eea{\end{eqnarray}}
\def\bes{\begin{subequations}}
\def\ees{\end{subequations}}

\def\oh{\frac{1}{2}}

\def\re{\mbox{Re}\, }
\def\im{\mbox{Im}\, }

\usepackage{multicol}
\usepackage{float}
\usepackage{caption}
\def\p {{\partial}}
\def\pbar {{\bar\partial}}

\newcommand{\cK}{\mathcal{K}}
\newcommand{\cM}{\mathcal{M}}
\newcommand{\cN}{\mathcal{N}}

\newcommand{\cT}{\mathcal{T}}

\newenvironment{eqn}{\begin{equation}\begin{aligned}}{\end{aligned}\end{equation}\noindent}
\newenvironment{eqn*}{\begin{equation*}\begin{aligned}}{\end{aligned}\end{equation*}\noindent}

\makeatletter
\newsavebox\myboxA
\newsavebox\myboxB
\newlength\mylenA

\newcommand*\xoverline[2][0.75]{%
\sbox{\myboxA}{$\m@th#2$}%
\setbox\myboxB\null
\ht\myboxB=\ht\myboxA%
\dp\myboxB=\dp\myboxA%
\wd\myboxB=#1\wd\myboxA
\sbox\myboxB{$\m@th\overline{\copy\myboxB}$}
\setlength\mylenA{\the\wd\myboxA}
\addtolength\mylenA{-\the\wd\myboxB}%
\ifdim\wd\myboxB<\wd\myboxA%
   \rlap{\hskip 0.5\mylenA\usebox\myboxB}{\usebox\myboxA}%
\else
    \hskip -0.5\mylenA\rlap{\usebox\myboxA}{\hskip 0.5\mylenA\usebox\myboxB}%
\fi}
\makeatother

\begin{document}
\pagestyle{plain}

\makeatletter
\@addtoreset{equation}{section}
\makeatother
\renewcommand{\theequation}{\thesection.\arabic{equation}}

\pagestyle{empty}
\rightline{IFT-UAM/CSIC-22-13}
\vspace{0.5cm}
\begin{center}
\Huge{{4d strings at strong coupling} 
\\[10mm]}
\normalsize{Fernando Marchesano\,$^{\symrook}$  and Max Wiesner\,$^{\symknight}$\\[12mm]}
\small{
${}^{\symrook}$ Instituto de F\'{\i}sica Te\'orica UAM-CSIC, c/ Nicol\'as Cabrera 13-15, 28049 Madrid, Spain \\[2mm] 
${}^{\symknight}$ Center of Mathematical Sciences and Applications, Harvard University, \\ 20 Garden Street, Cambridge, MA 02138, USA
\\[10mm]} 
\small{\bf Abstract} \\[5mm]
\end{center}
\begin{center}
\begin{minipage}[h]{15.0cm} 

Weakly coupled regions of 4d EFTs coupled to gravity are particularly suitable to describe the backreaction of BPS fundamental axionic strings, dubbed EFT strings, in a local patch of spacetime around their core. We study the extension of these local solutions to global ones, which implies probing regions of strong coupling and provides an estimate of the EFT string tension therein. We conjecture that for the EFT string charge generators such a global extension is always possible and yields a sub-Planckian tension. We substantiate this claim by analysing global solutions of 4d strings made up from NS5-branes wrapping Calabi--Yau threefold divisors in either type IIA or heterotic string theory. We argue that in this case the global, non-perturbative data of the  backreaction can be simply encoded in terms of a GLSM describing the compactification, as we demonstrate in explicit examples.

\end{minipage}
\end{center}
\newpage
\setcounter{page}{1}
\pagestyle{plain}
\renewcommand{\thefootnote}{\arabic{footnote}}
\setcounter{footnote}{0}


\tableofcontents

\section{Introduction}
\label{s:intro}

An interesting approach to characterise certain aspects of Effective Field Theories (EFTs) is to study solutions to their field equations that describe objects beyond the perturbative spectrum. This has been recently used in the context of the Swampland Programme \cite{Vafa:2005ui} (see \cite{Brennan:2017rbf,Palti:2019pca,vanBeest:2021lhn,Grana:2021zvf} for reviews) to translate proposals for Swampland criteria into more familiar physics. A neat example is 4d $\CN=1$ EFTs coupled to gravity, where the physics of backreacted $\half$BPS strings and membranes connects the Weak Gravity Conjecture (WGC) \cite{Arkani-Hamed:2006emk} to other Swampland Conjectures in a very direct manner \cite{Lanza:2020qmt}. Particularly remarkable is the connection between the Swampland Distance Conjecture (SDC) and the local 4d backreaction of fundamental axionic strings, dubbed EFT strings in \cite{Lanza:2021udy}. It was proposed in \cite{Lanza:2020qmt,Lanza:2021udy} that any infinite distance limit of a 4d EFT coupled to gravity can be realised as the local backreaction of an EFT string, and then pointed out how this allows one to derive the SDC from the WGC applied to strings. In general, it was found that the physics of local EFT string solutions characterises  the EFT itself along infinite distance limits. It is in this sense that properties of objects with non-trivial backreaction serve to probe the EFT and uncover its structure and limitations, an idea that has been recently applied in different contexts \cite{Klaewer:2016kiy,Dolan:2017vmn,Hebecker:2017wsu,Buratti:2018xjt,Draper:2019zbb,Draper:2019utz,Bonnefoy:2019nzv,Gendler:2020dfp,Buratti:2021fiv,Alim:2021vhs,Cribiori:2022cho}.

Backreacting solutions have been mostly used to probe weakly-coupled regions of the EFT, which is where we have better control of the solution and where Swampland conjectures are usually tested. In the case of EFT strings, their solution has in fact only been described in a local patch of spacetime around the string core, which by construction probes a slice of a weakly-coupled EFT regime. Extending the solution to larger regions of spacetime necessarily implies covering regions of strong coupling, where the axionic shift symmetry that defines the string charge is heavily broken. While the description of the solution is much more involved in these regions, as long as it corresponds to a physical object one should be entitled to apply the same philosophy as in the weak-coupling regimes, and translate the properties of these objects into EFT data. Given that this strategy has already proven to be quite fruitful in connecting different Swampland conjectures at weak coupling, a complete EFT string solution may give us a window to test Swampland criteria at strong-coupling regimes as well. 

In this paper we initiate the study of 4d $\half$BPS EFT string solutions beyond weak coupling, and find a series of results that lead to Conjecture \ref{conj:SESC}. The question addressed by this conjecture is whether a local EFT string solution can be extended or not to the whole of spacetime, meaning a finite-energy, regular solution over an infinite plane transverse to the string worldsheet. This problem is non-trivial for two reasons: First, because the naive extension of a local EFT string solution diverges at large distances from  the string core -- a divergence that may or may not be regulated by strong coupling effects. Second, because the deficit angle induced by the string solution increases as we move away from the string core, and so there is the risk that at some finite distance it exceeds $2\pi$. At this point the gravitational backreaction of the string overcloses the transverse space and it does not make sense to continue the solution towards spatial infinity. What our results suggest is that global EFT string solutions with deficit angle below $2\pi$ exist, but that this is only guaranteed for the EFT strings with the smallest charges, dubbed elementary EFT strings. Typically their deficit angle at infinity will be a fraction of $2\pi$, so when considering higher charges there is only a finite number of them not overclosing the transverse space. In other words, there is only a finite number of EFT strings whose global tension is sub-Planckian.

Our approach to study global EFT string solutions is highly influenced by F-theory \cite{Vafa:1996xn}, and the way in which it regularises the backreaction of 7-branes. The influence from F-theory is two-fold: First, we consider the construction of solutions by patching basic building blocks, as in \cite{Bergshoeff:2006jj}. Adapting the results of \cite{Bergshoeff:2006jj} to our setup highlights the interplay between the duality/monodromy group $\Gamma$ of the EFT moduli space and the tension of the global BPS string solution. It  also signals the necessity of additional strings -- dubbed regulator strings -- to form a finite-energy, global string solution. In analogy with F-theory global solutions, these additional strings carry charges that do not commute with those of EFT strings, which reflects that the non-Abelian properties of the EFT monodromy group $\Gamma$  are crucial to extend globally BPS string solutions. Second, we encode the 4d string backreaction in terms of a field theory probing the solution, in analogy to the D3-brane gauge theory in F-theory setups \cite{Banks:1996nj}. In our case the auxiliary theory is a 2d Gauge Linear Sigma Model (GLSM) that probes the backreaction of NS5-branes wrapping divisors of a Calabi--Yau threefold, as can be seen by applying the results of \cite{Quigley:2011pv,Blaszczyk:2011ib} to our setting. It turns out that in terms of GLSM data the EFT string backreaction simplifies tremendously, allowing us to describe it beyond weak coupling and to compute its total energy. 

Thanks to this approach we are able to analyse global EFT string solutions in explicit examples, and to draw several lessons from them. First, we notice that the global EFT string solution is essentially characterised by the local solution that it extends, and that more complicated solutions are built from superpositions of elementary string solutions. This indicates that EFT string charges and their conjugates under the monodromy group $\Gamma$ are the ones that determine the spectrum of fundamental 4d strings, also beyond weak coupling. This has consequences for the structure of $\Gamma$, that we formulate in Conjecture \ref{conj:abelian}, as well as an interpretation in terms of the cobordism Conjecture \cite{McNamara:2019rup} applied to this class of theories. Second, we find that elementary strings form a continuous family of global string solutions which together cover the whole of the moduli space. This suggests that global 4d string solutions can be used to probe strong coupling regions of the theory. In particular, they can be used to estimate the EFT string tension in such regimes. Finally, we consider global solutions corresponding to $\half$BPS strings that nevertheless lie outside the cone of EFT string charges. Even if at long distances they appear to be fundamental objects, we find that when approaching the string core they must necessarily be composite. The full microscopic description of these objects however remains mysterious, and further analysis seems to be needed to unveil their nature. 

The paper is organised as follows. In section \ref{s:nuclear} we discuss local 4d EFT string solutions at weak coupling, and how they can be  extended to global solutions probing strong coupling regions of the EFT. We do so by first discussing a toy example based on F-theory, from which we extract general rules for our setting. We arrive to Conjecture \ref{conj:SESC}, which encapsulates the main lessons obtained in this work, and to Conjecture \ref{conj:abelian}, which can be thought of as a consequence of it. In section \ref{s:4d2d} we deploy our techniques to construct global 4d BPS string solutions based on NS5-branes, by exploiting their description in terms of a GLSM, and we analyse the case of the quintic. Two further Calabi--Yau examples are analysed in section \ref{s:examples}, where we show how the results from the two previous sections serve to specify the spectrum of global 4d string solutions. This not only applies to the 4d EFT strings, but also to BPS non-EFT strings, whose solution is non-perturbative even locally. We draw our conclusions in section \ref{s:conclu}.

Several technical details are relegated to the appendices. Appendix \ref{ap:N=2} discusses how to adapt strings in $\CN=2$ vector multiplet moduli spaces to the $\CN=1$ language used in the main text.  Appendix \ref{ap:conju} works out the conjugacy classes and the relation to the global string tension for the toy example of section \ref{sec:toy}. Appendix \ref{ap:GLSM} provides some useful background on GLSMs. Appendix \ref{apsec:quintic} computes the area of the quintic K\"ahler moduli space. Appendix \ref{app:monodromies} works out the different relations among the monodromy generators that appear in the  examples of section \ref{s:examples}.


\section{String solutions at strong coupling}
\label{s:nuclear}

Weakly coupled regions of 4d EFTs are particularly suited to describe local patches of back-reacted fundamental string solutions. As stressed in \cite{Lanza:2020qmt,Lanza:2021udy}, a certain kind of local solutions, dubbed EFT strings, display a set of remarkable properties that allows them  to probe the physics of such weakly coupled regions. In this section we review the main features of local EFT string solutions, and discuss their extension to strong coupling regimes, or equivalently to global solutions in 4d. Based on our analysis of section \ref{s:4d2d} and the examples of section \ref{s:examples}, we propose that such a global extension is always possible, with the precise statement summarised in Conjecture \ref{conj:SESC}.

\subsection{4d EFT strings}

Let us consider a 4d ${\cal N} =2$ or ${\cal N}=1$ EFT with a cut-off $\Lambda$. Out of the fields of the EFT, we consider a subset of chiral multiplets whose scalar component $\{t^i\}$ can be treated as moduli,\footnote{For ${\cal N} =2$ theories the $\{t^i\}$ can be identified with scalars within the vector multiplets, see Appendix \ref{ap:N=2}. For most ${\cal N} =1$ theories all chiral multiplets are expected to enter the superpotential \cite{Palti:2020qlc}, and therefore the F-term scalar potential $V$. However, if we are in an EFT regime such that both the Hubble scale $H\sim \sqrt{V}/M_{\rm P}$ and the mass scales corresponding to $\{t^i\}$ are negligible compared to $\Lambda$, we may ignore the presence of such a potential in our analysis, at least when describing varying field configurations for the $\{t^i\}$ at scales close to $1/\Lambda$. Under this assumption, we may also treat the space of constant-field configurations in $\{t^i\}$ as a moduli space $\CM$.} and such that they have a periodic real coordinate: $t^i \sim t^i + 1$, $\forall i$. The relevant piece of the effective action is then 
\be
S=M^2_{\rm P}\int \left(\frac{1}{2}R*1-K_{i\bar j}\,\d t^i \wedge*\d\bar t^{\bar j} \right)\, ,
\label{effaction}
\ee 
where $K_{i\bar j} = \p_i \pbar_{\bar j} K$ and $K$ is the K\"ahler potential. From here we may analyse string-like solutions of our EFT, in the same spirit as in \cite{Greene:1989ya}, see also \cite{Green:1993zr}. Following \cite{Lanza:2020qmt,Lanza:2021udy}, we split the 4d coordinates into $(t,x,z,\bar z)$ with $z\in \mathbb{C}$, and impose 2d Poincar\'e invariance on $(t,x)$. That is,  we allow the varying fields $t^i$ to depend only on $z,\bar z$ and choose a metric Ansatz of the form 
\be\label{metric}
M_{\rm P}^2 \d s^2= -\d t^2+\d x^2+ e^{2D}\d z\d\bar z\, ,
\ee
where $D$ also only depends on the coordinates $(z,\bar z)$. The equations of motion read
\be
K_{\bar i j}\p\pbar t^j+K_{\bar i j k}\p t^j \wedge {\bar\p}t^{k}=0\, ,
\label{eomstring}
\ee
and so a simple class of $\oh$BPS solutions correspond to holomorphic $\pbar t^i = 0, \forall i$ and to anti-holomorphic $\p t^i =0, \forall i$ profiles. In the following we will focus on the holomorphic ones $t^i (z)$, that in the above setup can be thought of as maps from $\C$ to the field space $\CM$ corresponding to the $\{t^i\}$. An anti-holomorphic version of a holomorphic string-like solution describes a string with opposite charges and preserving the opposite half of supersymmetry. Finally, from Einstein's equations it  follows that 
\be
e^{2D}=|f(z)|^2 e^{-K}\, ,
\label{stringwarp0}
\ee
with $f(z)$ a holomorphic non-vanishing function \cite{Greene:1989ya}. 

In \cite{Lanza:2020qmt,Lanza:2021udy} a particular set of string solutions were analysed, that correspond to fundamental axionic strings, dubbed EFT strings. Here fundamental means that these are objects that cannot be resolved within the EFT, and whose backreaction does not overclose the transverse space. In other words they satisfy
\be\label{EFTregime}
 \Lambda^{2} <  T < 2\pi M^{2} _{\rm P}\, ,
\ee 
where $T$ is the string tension. Axionic means that the profile for $\{t^i\}$ near the string core probes a region of $\CM$ where $K$ displays perturbative axionic shift symmetries, such that all non-perturbative effects breaking them are exponentially suppressed. This set of axionic symmetries includes the one along the quantised charges of the string, which corresponds to a monodromy of the form 
\be\label{tmon}
t^i\rightarrow t^i +e^i\, , \qquad \qquad e^i \in \mathbbm{Z}\, ,
\ee
around the string core. Finally, the string tension $T$ depends on the point of $\CM$, and the inequality \eqref{EFTregime} must be satisfied along the whole region probed by the EFT string solution. 

With these definitions, solutions describing an EFT string located at $z=0$ are given by \cite{Lanza:2021udy}
\be\label{tsol}
t^i=t_0^i+\frac{1}{2\pi\ii}e^i\log \left(\frac{z}{z_0}\right)\, ,
\ee
with $t_0^i, z_0$  constant.
Writing $t^i = a^i+ \ii  s^i$ and  $z=re^{\ii\theta}$ we can split this solution as
\begin{subequations}
\label{solsplit}
\bea\label{imt}
s^i & = & s_{0}^i-\frac1{2\pi}e^i\log \left(\frac{r}{r_0}\right)\, ,\\
\label{axionmon}
a^i &= &\frac{\theta}{2\pi}\,e^i+\text{const}\, ,
\eea
\end{subequations}
where $a^i$ are interpreted as axions and $s^i$ as their saxionic partners. Such a solution should in fact be seen as a local one, in the sense of a map from the disc $D(r_0) \equiv \{ |z| < r_0\} \subset \C$ to $\CM$. In a local patch, by appropriately redefining $z$ we may set the holomorphic function in \eqref{stringwarp0} to a constant, such that
\be\label{stringwarp}
e^{2D}=e^{K_0-K}\,  , \qquad \text{with} \quad K_0\equiv K(t_0^i)\,.
\ee

Not all solutions of this form correspond to EFT strings. For this to be the case, the disc image $\bm{t}(D) \subset \CM$ must lie within a region where the EFT string definition is satisfied. That is, let us assume that at the point $\bm{t}_0 \equiv \bm{t}(z_0) \in \CM$ the theory displays  a perturbative axionic shift symmetry along the direction \eqref{axionmon}, only broken by non-perturbative corrections of the form $e^{2\pi \ii m_i t^i}, m^i \in \Z$ charged under the monodromy \eqref{tmon}, all of them sufficiently suppressed. For the EFT string solution to be valid, this statement must still hold when we take $r \to 0$ in \eqref{imt}, and probe a whole region of saxionic values of the theory. As discussed in \cite{Lanza:2021udy}, in specific setups this criterion selects a particular cone of axionic string charges, dubbed ${\cal C}^{\text{\tiny EFT}}_{\rm S}$ therein. The generators of this cone are identified as {\it elementary EFT strings}.

The elements of ${\cal C}^{\text{\tiny EFT}}_{\rm S}$ that are not elementary admit one or several decompositions of the form $\bm{e} = \sum_a \bm{e}_a$, with $\bm{e}_a \in {\cal C}^{\text{\tiny EFT}}_{\rm S}$. In this case one can also build the  multi-string solution:
\be
t^i=t_0^i+\frac1{2\pi\ii} \sum_a e_a^i \log\left(\frac{z-z_a}{z_0-z_a}\right)\, ,
\label{multisol}
\ee
where $e_a^i$ are the charges of the string located at $z_a$ in the $z$-plane, so that in the limit $z_a \to 0$ one recovers \eqref{tsol}.  If $s^i_0$ is large, then $s^i$ will remain large in the domain $\bigcap_{a}D_a$, where $D_a=\{|z-z_a|\leq |z_0-z_a|\}$, which is non-vanishing if the strings are sufficiently close to each other.

In these perturbative regions of the theory it is relatively simple to compute the string tension, since one may resort to a dual multiplet description of the string action. Indeed, using the axionic shift symmetries we can dualise their axion fields $a^i$ into two-form potentials ${\cal B}_{2\, i}$, to which the strings couple electrically. Similarly, using that the K\"ahler potential only depends on the chiral fields $t^i$ through their saxionic component $s^i$, we define the corresponding dual saxions 
\be\label{dualfields}
\ell_i=-\frac12 \frac{\p K}{\p s^i}\, ,
\ee
and using supersymmetry we find that a string of charge ${\bf e} = \{ e^i\}$ has the following tension \cite{Lanza:2019xxg}
\be\label{thetension}
T_{\bf e}\equiv M^2_{\rm P}\,e^i\ell_i\, .
\ee
A similar result is obtained if one computes the linear energy density of the EFT string solution on a disc of radius $r$ \cite{Lanza:2021udy}
\be
\label{Eback}
\begin{aligned}
{\cal E}_{\rm back}(r)&=M_{\rm P}^2\,e^i [\ell_i(r) - \ell_i(0)] = M_{\rm P}^2\int_{D(r)} \bm{t}^* (J_{\CM}) = M_{\rm P}^2\int_{\bm{t}(D(r))} J_{\CM}   \, ,
\end{aligned}
\ee
where $\ell_i(r)$ is the dual saxion value  that corresponds to $s^i(r)$ in \eqref{imt}, and $J_{\CM} = \ii K_{i\bar\jmath}\,\d t^i\wedge \d \bar t^{\bar\jmath}$ is the K\"ahler form in $\CM$, which we integrate over the image of $D(r)$ under \eqref{tsol}. Using the additional property that for EFT string flows $e^i \ell_i (0) =0$, one finds that an EFT string tension can be computed as the area of an appropriate disc image $\bm{t}(D)$ in $\CM$. This is highly reminiscent, but not equivalent, to the picture considered in \cite{Greene:1989ya,Green:1993zr}, in which the tension of a cosmic string was computed in terms of its profile $\bm{t}(z)$, as the area of $\bm{t}(\C) \subset \CM$. 

To connect with the prescription of \cite{Greene:1989ya,Green:1993zr} one should first extend the EFT string solution  to the whole of the complex plane, which is challenging for several reasons. First, if we fix the values $t_0^i$, $r_0$ and naively extend the solution \eqref{tsol} for $r > r_0$ we will hit a pole for the linear energy density \eqref{Eback} at $r_{\rm strong} \simeq r_0\, e^{2\pi s_0^i/e^i}$, which is where we approach a boundary of the saxionic cone. Before that happens, though, our assumption that non-perturbative effects $e^{2\pi \ii m_i t^i}$ are negligible is no longer valid, and we enter a strongly-coupled region in which the axionic shift symmetry is substantially broken by them. As a result the K\"ahler potential should be non-trivially modified by these terms, as well as the holomorphic profile \eqref{tsol}, that now will also include an arbitrary polynomial in $z$. In principle, all these new effects could regularise the naive EFT string solution for arbitrarily large radii, and therefore yield a finite linear energy density ${\cal E}_{\rm back}(r)$ for any $r>r_0$. However, they can never modify the EFT string solution in such a way that an EFT vacuum is recovered asymptotically, due to the non-trivial deficit angle induced by the string backreaction. 

In order to relate the EFT string solution with a vacuum of the EFT, one may follow \cite{Lanza:2021udy} and consider a closed EFT string on a loop of radius $L > 1/\Lambda$, on top of a constant-field configuration or vacuum specified by $\bar{\bm{t}} \in \CM$. Near the string core the backreaction will be as in \eqref{tsol}, or rather as a coarse-grained version that will stop the saxionic flow at a minimal radius $R_\Lambda = 1/\Lambda$. At distances larger than $L$ the string backreaction will start to die off as the one of a codimension-three object, and so we will quickly connect with the vacuum solution at $\bar{\bm{t}}$. So essentially this configuration yields a map from a disc $D(L) \subset \C$ to $\CM$, which is given by \eqref{tsol} with $\bm{t}_0 = \bar{\bm{t}}$ and $z_0 = M_{\rm P} L$, coarse-grained to an accuracy $r_\Lambda = M_{\rm P}/\Lambda$. The corresponding saxionic string flow starts at $\bar{\bm{s}}$ and ends at ${\bm{s}}_\Lambda\equiv \bm{s}(r_\Lambda) = \bar{\bm{s}} + \frac{\bf{e}}{2\pi} \log(\Lambda L)$. Microscopically, the total energy of the string loop is given by $2\pi L T_{\bf e}(\bar{\bm{s}})$, with $\bar{\bm{s}} = \Im \bar{\bm{t}}$. Macroscopically, the corresponding energy density splits as
\be
T_{\bf e}(\bar{\bm{s}}) \simeq  \cT_{\bf e}\left(\Lambda,\bar{\bm{s}}\right)  +  {\cal E}_{\rm back}\left(\Lambda,L,\bar{\bm{s}}\right)\, ,
\label{splitEloop}
\ee
where the lhs is independent of the EFT cut-off $\Lambda$, but the terms in the rhs are not. For larger values of $\Lambda$ the string backreaction is more accurate and it contains more energy, while the effective tension $ \cT_{\bf e}\left(\Lambda,\bar{\bm{s}}\right) = T_{\bf e}(\bm{s}_\Lambda)$ decreases. In the formal limit $\Lambda \to \infty$ we have  $ \cT_{\bf e}\left(\Lambda,\bar{\bm{s}}\right) \to 0$, and all the energy is stored in the backreaction, while for $\Lambda \leq 1/L$ the opposite occurs. 

While \eqref{splitEloop} has a clear interpretation for field space points $\bar{\bm{t}} \in \CM$ with approximate axionic shift symmetries, its meaning becomes less clear when these are broken. Indeed, in that case we no longer have a simple axion-independent expression of the form \eqref{thetension}, and in fact the dual saxions $\ell_i$ are not well defined. In such regimes, we may instead use the above loop configuration to estimate the string tension in terms of the linear energy density of the backreaction:
\be
{\cal E}_{\rm back}\left(L,\bar{\bm{t}}\right) \equiv {\cal E}_{\rm back}\left(\Lambda \to \infty,L,\bar{\bm{t}}\right)\, . 
\label{1stdefT}
\ee
Here ${\cal E}_{\rm back}\left(L,\bar{\bm{t}}\right)$ is the linear energy density of a string solution on a region $D(L) \subset \C$, such that the EFT string core lies at $z=0$ -- around where we should recover the solution \eqref{tsol} -- and we impose the boundary condition $\bm{t}(z=M_{\rm P} L) = \bar{\bm{t}}$. With this definition one in principle has a well-defined prescription to estimate the EFT string tension $T_{\bf e}(\bar{\bm{t}})$ away from the weak-coupling regime, and in particular to see if this tension blows up or stays finite. As we will argue, ${\cal E}_{\rm back}\left(L,\bar{\bm{t}}\right)$ is independent of $L$ and finite for all possible values of $\bar{\bm{t}}$. Its interpretation in terms of the EFT string tension $T_{\bf e}(\bar{\bm{t}})$ is however subtle, because at strong coupling regimes \eqref{1stdefT} also contains the contribution from the backreaction of additional strings that act as regulators.

\subsection{Strong coupling and regulator strings}
\label{ss:regulator}
 
 The simplest approach to extend 4d EFT string solutions to strong coupling is to consider holomorphic maps of the form $\bm{t}: \C \to \CM$, such that near the origin they reduce to \eqref{tsol}. One may then restrict the corresponding string solution to a region $D \subset \C$ containing the origin, obtaining the map $\bm{t}|_{D}: D  \to \CM$ discussed above, with $ \bar{\bm{t}} \simeq \bm{t}_0 \equiv \bm{t}(z_0)$ for a given $z_0 \in \partial D$. Fixing $z_0$ and scanning over the family of maps $\bm{t}$ will change the value of $\bar{\bm{t}}$, exploring strongly-coupled regions. In particular, given a map $\bm{t}(z)$ one can consider the family $\bm{t}_\lambda (z) \equiv \bm{t}(\lambda z)$, $ \lambda \in \C^*$. As we change $\lambda$, ${\bm{t}}_0$ will cover the holomorphic curve $\im \bm{t}(\C) \subset \CM$, where $\im$ denotes the image of $\C$ in $\CM$ under the map $\bm{t}$.  Because $\im t_\lambda(D) \subset \im \bm{t}(\C)$, all these solutions will be of finite energy if $\im \bm{t}(\C)$ has finite area. In practice, this finite-energy condition  implies that we can perform the one-point-compactification extension $\bm{t}: \P^1 \to \CM$, which is the kind of holomorphic maps that we will consider in the following.

 \subsubsection{A toy example}
\label{sec:toy}
 
Describing a 4d BPS string in terms of a holomorphic map $\bm{t}: \P^1 \to \CM$ corresponds to the approach taken in \cite{Greene:1989ya}, where the particular choice of moduli space was taken to be
\be
\CM = SL(2, \Z) \backslash SL(2, \R)/U(1)\, ,
\label{MSL}
\ee
with K\"ahler potential $K = - \log (\im \tau )$, $\tau \in \CM$. In the following we  use \eqref{MSL} as a toy example, in order to illustrate the general features that appear in the more involved examples of the next sections. Particularly interesting for our purposes are the results of \cite{Bergshoeff:2006jj}, where two elementary string-like solutions are identified in terms of which all other BPS solutions can be constructed. These two building blocks are described by a one-to-one holomorphic map $\tau(z): \C \to \CM$ and the holomorphic function $f(z): \C \to \C$ that describes the warp factor via \eqref{stringwarp0}:
 \begin{align}\label{1A}
  1A: \quad \tau=j^{-1}\left[\frac{(z-z_\rho)(z_i-z_{i\infty})}{(z-z_{i\infty})(z_i-z_\rho)}\right]\,, \quad f=\eta^2 (z-z_{i\infty})^{-\frac{1}{12}} (z-z_i)^{\frac{1}{4}}(z-z_\rho)^{-\frac{1}{6}}\, .
\end{align}
\begin{align}\label{1B}
1B: \quad \tau=j^{-1}\left[\frac{(z-z_\rho)(z_i-z_{i\infty})}{(z-z_{i\infty})(z_i-z_\rho)}\right]\,, \quad f=\eta^2 (z-z_{i\infty})^{-\frac{1}{12}} (z-z_i)^{-\frac{1}{4}}(z-z_\rho)^{\frac{1}{3}}\, .
\end{align}
Here $\eta(\tau(z))$ is the Dedekind $\eta$-function, and $j^{-1}: \C \to \CM$ is the inverse of the modular $j$-function: $j(\rho) = 0$, with $\rho \equiv e^{\frac{2\pi i}{3}}$, $j(i) =1$ and $j(i\infty) = \infty$. Each of these three points in $\CM$ correspond to several conjugacy classes of the modular group $\Gamma = SL(2, \Z)$, generated by the elements  $T$ and $S$. As discussed in appendix \ref{ap:conju}, $\rho \in \CM$ corresponds to the classes $\{ [TS], [ST^{-1}], [-TS], [-ST^{-1}]\}$, $i \in \CM$ to the classes $\{[S], [-S]\}$ and $i\infty \in \CM$ to $\{[T^n]\}$. In the 4d string solutions above, one is choosing a conjugacy class and then a representative for $\rho$, $i$ and $i\infty$ which we will dub as $M_\rho, M_i, M_{i\infty} \in SL(2, \Z)$, respectively. These three elements specify the monodromy of the backreacted string solution around the three preimages $z_\rho$, $z_i$, $z_{i\infty} \in \C$. They can then be interpreted as the locations of three 4d string cores, which respectively realise each of these monodromies. 

Not all monodromy choices are allowed or are in-equivalent for a solution of the form
  $\bm{t}: \P^1 \to \CM$. Similarly to  vortex configurations, we have that:
\begin{itemize}
    \item[-] Because $\P^1$ is compact, the total monodromy of the configuration must cancel. 
    \item[-] Choices related by a global conjugation $M_\a \mapsto h M_\a h^{-1}$ $\forall \a$, with $h \in \Gamma$ a fixed element of the modular group, are equivalent. 
\end{itemize}

Due to this second property, one may gauge-fix one of the monodromies of the configuration. In the above solutions we have that $[M_{i\infty}] = [T]$, which corresponds to the minimal infinite-order monodromy of this toy example, so we may fix $M_{i\infty}=T$. One can then consider different choices of $M_i$ and $M_\rho$ such that the total monodromy cancels, see appendix \ref{ap:conju}. The above solutions \eqref{1A} and \eqref{1B} can be interpreted as those two choices for which $\tau: \P \to \CM$ is a one-to-one map and, as a result, $\im \tau(\P) = \CM$ has minimal area, as compared to a solution which covers $\CM$ multiple times. From here one can observe an interesting relation between the area of $\CM$ and the different monodromy orders of the solution.

\begin{figure}[htb]
    \centering
     \begin{subfigure}[b]{0.465\textwidth}
         \centering
         \includegraphics[width=\textwidth]{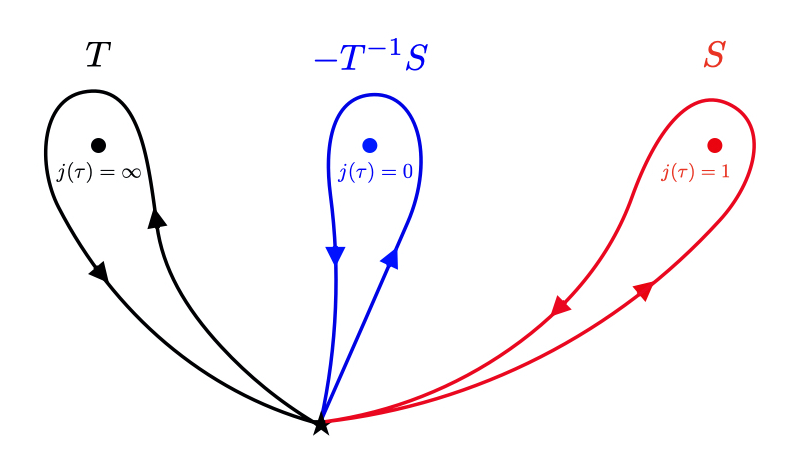}
         \caption{1A}
         \label{fig:1A}
     \end{subfigure}
     \hspace{0.03\textwidth}
     \begin{subfigure}[b]{0.465\textwidth}
         \centering
         \includegraphics[width=\textwidth]{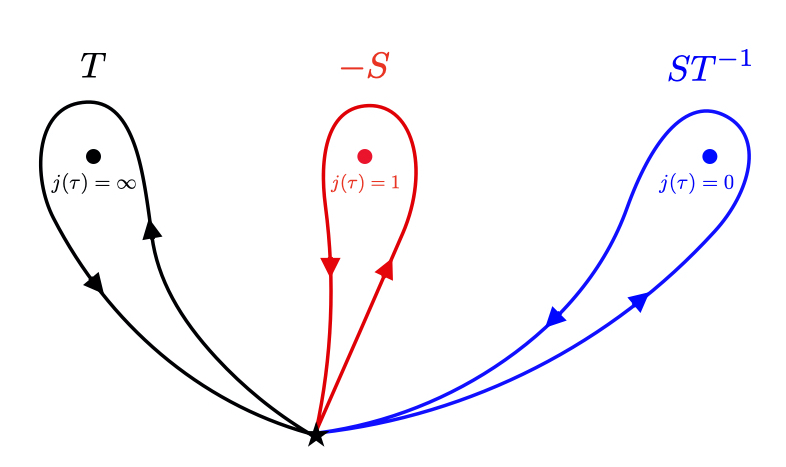}
         \caption{1B}
         \label{fig:1B}
     \end{subfigure}
     \caption{These figures illustrate the 1A (left) and 1B (right) building blocks by showing the relative positions of the three strings associated to the points $\tau=i\infty, i, \rho$ in moduli space. In addition the respective monodromy action when encircling the strings counterclockwise is shown.}
         \label{fig1}
\end{figure}

Indeed, let us consider the configuration $1A$. If we choose $ \re (z_{i\infty}) < \re(z_{\rho}) < \re(z_i)$ the total monodromy, computed by a counter-clockwise contour, is given by $M_{\rm total} = M_{i\infty} M_{\rho} M_i$, as illustrated in figure~\ref{fig:1A}. The specific monodromies of the solution correspond to  $M_{\rho} = - T^{-1}S$, $M_i = S$, $M_{i\infty} = T$ which are of order $N=6$, $4$ and $\infty$. For each monodromy $M_\a$, its order $N_\a$ is related to the tension/deficit angle localised at its string core, which is given by \cite{Bergshoeff:2006jj}
\be
|\delta_\a| \equiv \frac{|T_\alpha|}{M_{\rm P}^2} = \left[\det \left(\log M_\a\right)\right]^{1/D} = 2\pi \frac{p_\a}{N_\a} \, ,
\label{deltorder}
\ee
with $D=2$ the dimension of the representation and $p_\a \in \bN$. Notice that \eqref{deltorder} is only well-defined modulo $2\pi$ because larger, super-Planckian tensions $T > 2\pi M_{\rm P}^2$ mean that we do not have a conical string solution that can extend to spatial infinity. The sign of $\delta_\alpha$ is such that the total tension of the configuration  vanishes, in the sense that 
\be
M_{\rm P}^2 \int_{\bm{t}(\C)} J_{\CM} +  \sum_\a \delta_\a = 0\, ,
\label{tadpole}
\ee
which following \cite{Bergshoeff:2006jj} can be interpreted as a BPS equation. To motivate this, one considers a configuration over $\C$ in which all localised string tensions are non-negative, except one which is placed at infinity and is considered non-physical. The positive localised tensions and the backreaction energy stored in $ M_{\rm P}^2 \int_{\bm{t}(\C)} J_{\CM}$ add up, accounting for the total tension/deficit angle induced by the configuration at infinity, which must be below $2\pi$. At the same time, the monodromies of each localised string are combined into a total monodromy. BPSness and sub-Planckian tension imply that these two quantities are directly related to each other, and that they can be simultaneously cancelled by a string at infinity whose local deficit angle obeys \eqref{deltorder}. Notice that this implies that the curve area $ M_{\rm P}^2 \int_{\bm{t}(\C)} J_{\CM}$ is tightly constrained by the monodromy orders within the modular group. 

In the case at hand $ M_{\rm P}^2 \int_{\bm{t}(\C)} J_{\CM}  = M_{\rm P}^2 \int_\CM J_{\CM} = \frac {\pi}{6}$, and \eqref{tadpole} is satisfied for both building blocks, where $p_\a =1$. In the case $1A$ one  takes $z_i \to \infty$, and so \eqref{tadpole} reads $\frac{\pi}{6} + \frac{\pi}{3} - \frac{\pi}{2} = 0$. In the solution $1B$, illustrated in figure~\ref{fig:1B}, if one chooses $\re(z_{i\infty}) < \re(z_{i}) < \re(z_\rho)$ the monodromies correspond to $M_{\rho} = - ST^{-1}$, $M_i = -S$, $M_{i\infty} = T$, and have order $N=3$, $4$ and $\infty$, respectively. By taking $z_\rho \to \infty$ to host a negative tension,  \eqref{tadpole} is satisfied as $\frac{\pi}{6} + \frac{\pi}{2} - \frac{2\pi}{3} = 0$. 

Both building blocks can be considered as BPS extensions of a local EFT string solution. Indeed, according to our previous discussion, we should identify the infinite-order-monodromy string located at $z_{i\infty}$ with an EFT string core. Following our conventions let us then set $z_{i\infty} = 0$. In addition, for the 
 solution $1A$ we must take the limit $z_i \to \infty$. It then reduces to
 \begin{align}\label{1As}
  1A: \quad \tau=j^{-1}\left(\frac{z-z_\rho}{z}\right)\,, \quad f=\eta^2 z^{-\frac{1}{12}}(z-z_\rho)^{-\frac{1}{6}}\, ,
\end{align}
and it is completely specified by the value of $z_\rho \in \C^*$. Similarly the solution $1B$ becomes
 \begin{align}\label{1Bs}
1B: \quad \tau=j^{-1}\left(\frac{z_i}{z}\right)\,, \quad f=\eta^2 z^{-\frac{1}{12}} (z-z_i)^{-\frac{1}{4}}\, ,
\end{align}
and only depends on the parameter $z_i$.  As said, the total linear energy density of each configuration is given by the sum of the positive tensions in \eqref{tadpole}: $T_{1A} = \frac{\pi}{2} M_{\rm P}^2$ and $T_{1B} = \frac{2\pi}{3} M_{\rm P}^2$. 
 
Care should however be taken when interpreting these building blocks as valid BPS string solutions. Indeed, one may give a higher dimensional interpretation for each string profile by relating it to an elliptic fibration over $\C$ specified by $\tau(z)$ \cite{Greene:1989ya}. We then recover a complex surface $S$ in which we can globally define the (2,0)-form $\Omega = f(z) dz \wedge dz'$, where $z'$ is the fibre coordinate. If $S$ is a patch of a K3 surface, as one would expect from a BPS string configuration, then $\Omega$ should not have any zeros or poles. However whenever we have a finite-order monodromy with a localised positive tension a pole is developed in its location. As outlined in \cite{Bergshoeff:2006jj} the way out is to combine the   blocks $1A$ and $1B$ to build more general solutions, in which only string cores with vanishing or negative tension survive, and the latter can be  sent to spatial infinity. 
 
A simple construction of this sort is obtained by gluing a block $1A$ and  $1B$  to obtain a solution with warp factor
\begin{eqn}
    2AB:\quad f&=\eta^2(z-z^A_{i\infty})^{-\frac{1}{12}} (z-z^A_i)^{\frac{1}{4}}(z-z^A_\rho)^{-\frac{1}{6}}(z-z^B_{i\infty})^{-\frac{1}{12}} (z-z^B_i)^{-\frac{1}{4}}(z-z^B_\rho)^{\frac{1}{3}}\,.
\end{eqn} 
The superscript $A(B)$ indicate that the strings located at $z^{A(B)}$ originate from the $1A$($1B$) building block. To get rid of the positive-tension cores, we set $z_i^A=z_i^B$ such that the tension of the $z_i$-strings cancel each other, and $z_\rho^A=z_\rho^B$. We can then send $z_\rho^A=z_\rho^B\rightarrow \infty$ to obtain
\begin{align}\label{2AB}
    2AB:\quad j(\tau)= (z-z_{i\infty}^A)^{-1}(z-z_{i\infty}^B)^{-1}\,,\quad f=\eta^2 (z-z^A_{i\infty})^{-\frac{1}{12}} (z-z^B_{i\infty})^{-\frac{1}{12}}\,.
\end{align}
In this way we get rid of all finite-tension string cores at finite $z$ and are left with two strings with monodromy of infinite order. Accordingly, the tension of this solution is entirely contained in the backreaction and given by $T_{2AB} = \frac{\pi}{3} M_{\rm P}^2$. Notice that the resulting map $\mathbf{\tau}: \C \rightarrow \cM$ is a double cover of the fundamental domain $\cM$, or equivalently a one-to-one map from $\C$ to two glued copies of $\CM$. Since we chose two distinct building blocks to build the $2AB$ solution, the monodromy $M_{i\infty}^{(A)}$ around the string at $z_{i\infty}^A$ is the $S$-conjugate of the monodromy $M_{i\infty}^{(B)}$ around the string at $z_{i\infty}^B$, i.e. $M_{i\infty}^{(B)}=S M_{i\infty}^{(A)} S^{-1}$. Let us identify the string core at $z_{i\infty}^A$ as the EFT string with $\tau(z^A_{i\infty})=i\infty$. From the perspective of this string the value of $\tau$ at $z_{i\infty}^B$ is given by $\tau(z^B_{i\infty})=0$ which reflects the difference of the monodromies around the two strings cores. 

\begin{figure}[htb]
    \centering
     \begin{subfigure}[b]{0.465\textwidth}
         \centering
         \includegraphics[width=\textwidth]{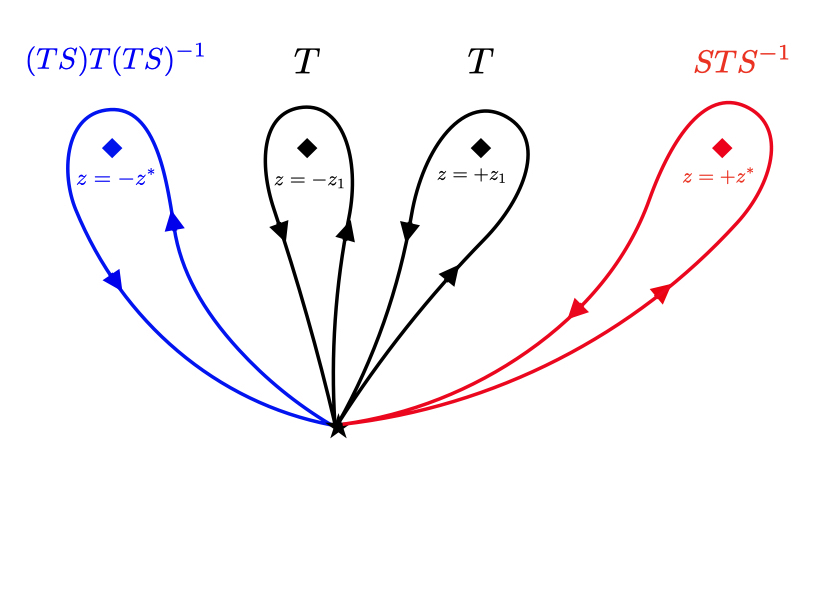}
         \label{fig:2times2AB1}
     \end{subfigure}
     \hspace{0.03\textwidth}
     \begin{subfigure}[b]{0.465\textwidth}
         \centering
         \includegraphics[width=\textwidth]{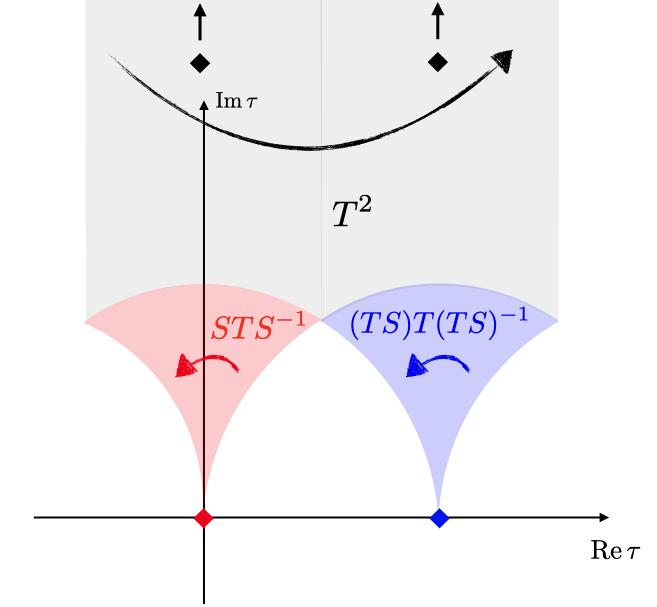}
         \label{fig:2times2AB2}
     \end{subfigure}
    \caption{The left figure shows the location of the string cores in the setup corresponding to gluing two $2AB$ solutions together. The two EFT string cores with monodromy $T$ are located at $z=\pm z_1$ whereas the two additional cores are located at $z=\pm z^*$. Notice that the monodromy around the latter differs by conjugation with $T$. The right figure shows the corresponding image in the upper half plane $\text{Im}\,\tau>0$ with the two EFT string cores corresponding to the cusp at $\tau=i\infty$ and the other two corresponding to the cusps at $\tau=0$ and $\tau=1$. }
    \label{fig:2times2AB}
\end{figure}

It is instructive to use the $2AB$ solution to generate solutions of higher EFT string charge, that correspond to multi-string solutions locally of the form \eqref{multisol}. Take for instance two $2AB$ solutions corresponding to two EFT strings with monodromy $T$ located at $z=\pm z_1$ for $|z_1|<|z_0|$. For $|z|\ll |z_0|$ the profile for $\tau$ is then given by 
\begin{align}
   \tau(z)=\tau_0+\frac{1}{2\pi i}\left[\log \left(\frac{z-z_1}{z_0}\right) + \log \left(\frac{z+z_1}{z_0}\right)\right]\,.
\end{align}
Then there are two further string cores located at $\pm z^*$ with $|z^*|>|z_1|$ such that the full profile for $\tau$ satisfies 
\begin{align}
    2\times 2AB:\quad  j(\tau)=\left(z^2-z_1^2\right)\left(z^2-(z^*)^2\right)\,.
\end{align}
Thus in total there are four string cores, as illustrated in figure \ref{fig:2times2AB}. From our previous discussion we know that the monodromies around the string cores at $\pm z^*$ need to be conjugates of $T$. Whereas the monodromy around the  core at $+z^*$ is given by $M_{+z*}=STS^{-1}$, the monodromy around $-z^*$ is additionally conjugated by $T$, i.e. $M_{-z^*} = (TS) \,T\, (TS)^{-1}$. The combined monodromy around all cores as shown in figure \ref{fig:2times2AB} is then 
\begin{align}
    M_{2\times 2AB}= \left[(TS)\,T\,(TS)^{-1}\right] T^2 \left[STS^{-1}\right] = (TSTS^{-1})^2\,, 
\end{align}
which we recognise as the square of the monodromy at infinity for the single 2AB solution, corresponding, as expected, to twice of the tension. The resulting curve in the moduli space is thus a four-fold cover of the $j$-line, as indicated in figure~\ref{fig:2times2AB}.

Instead of discussing additional solutions built from $1A$ and $1B$ (cf. \cite{Bergshoeff:2006jj}), let us  see what these results imply for an EFT string on a closed loop of radius $L$. As discussed above, we can estimate such a string tension by computing \eqref{1stdefT}, namely the linear energy density stored in a region $D(L) \subset \C$. For this we need to specify the map $\tau: \C \to \CM$, and then restrict it to the said region. In the case of the building blocks of this toy example, specifying the profile for $\tau$ can be done by relating the remaining free parameter in \eqref{1As} or \eqref{1Bs} with the vacuum expectation value $\bar{\tau}$. For this note that our previous boundary condition translates into $\tau(z=\hat{L}) = \bar{\tau}$, with $\hat{L} \equiv M_{\rm P}L$, which respectively imply
\be
 1A: \quad z_\rho = \hat{L} \left( 1 - j(\bar{\tau})\right) \,, \qquad 1B: \quad z_i = \hat{L}\, j(\bar{\tau}) \, . 
\ee
One therefore observes that, as we proceed towards strong coupling, the location of the finite-order-monodromy string approaches the location of the EFT string. Indeed, let us for instance take the $1B$ case. As $\bar{\tau} \to \rho$, $j(\bar{\tau}) \to 0$ and so eventually the string locus $z_i$ enters the region $D(L)$. Therefore, as we describe our closed-loop EFT string in a vacuum that corresponds to a strongly coupled region specified by $\bar{\tau}$, a second string core will be nucleated in its vicinity, also wrapping a closed loop, in some sort of bound state. When we enter a region in  moduli space such that $|j(\bar{\tau})| \leq (\Lambda L)^{-1}$,\footnote{This is a naive estimate in which we ignore warping effects to compute distances in the transverse space $\C$. At strong coupling such warping corrections may become significant, but the said  effect will still occur.} the EFT will be unable to resolve such a bound state, and in this sense the EFT string will cease to exist by itself. Notice that the total monodromy associated to the bound state is different from the initial EFT string monodromy, but because both strings form closed, trivial loops this does not violate charge conservation.  Finally, the linear energy density of this bound state will be larger than that of the EFT string at weak coupling, but it will be bounded from above by either $T_{1A}$ or $T_{1B}$, which are below $2\pi M_{\rm P}^2$. In this sense, this bound state has a regulating effect on the EFT string tension, as compared to the naive extension of the solution \eqref{tsol}. We will therefore dub this new string core that nucleates at strong coupling as {\it regulator string}. As we will argue in the following, the appearance of regulator strings is not a particularity of this toy example, but a general feature of 4d BPS strings at strong coupling.

Indeed, as mentioned before it is not clear that one can interpret the solutions $1A$ and $1B$ as sensible 4d BPS string solutions. Nevertheless, regulator strings also appear when we consider the $2AB$ configuration discussed around \eqref{2AB}. Indeed, let us consider a $1A$ EFT string wrapped on a loop of radius $L$, and for this set  $z^A_{i \infty}=0$ at the origin of $D(L)\subset \mathbbm{C}$. In this case, the vev $\bar \tau$ is in one-to-one correspondence with the free parameter $z_{i\infty}^B$ where the $S$-conjugate EFT string is located, as
\be
2AB: \quad z_{i\infty}^B = \hat{L} - \left[\hat{L} j(\bar{\tau}) \right]^{-1}\, .
\ee
It is now this $S$-conjugate string that acts as a regulator. As we send $j(\bar \tau)\rightarrow \hat{L}^{-2}$ this regulator string will eventually nucleate within $D(L)$. If we translate this into the physics of the vacuum, it signals a transition to a different phase of the theory in which the EFT string gets replaced by a bound state of itself with its $S$-conjugate, and so there is a jump in the BPS string charge.

This behaviour at strong coupling is reminiscent of the strong coupling regime of $\cN=2$ gauge theories. Consider for instance the Coulomb branch of $\cN=2$ $SU(2)$ super Yang--Mills theory parametrised by a complex scalar $u\in \mathbbm{P}^1$, associated to the breaking $SU(2)\rightarrow U(1)$. For $u\to \infty$ a weak coupling regime is reached in which the relevant BPS particle is the $W$-boson with electric and magnetic charge $(2,0)$. We can think of this BPS particle as the analogue of the EFT string in the $2AB$ string setup and of $u$ as the analogue of $\bar \tau$. As we change the Coulomb branch parameter $u\rightarrow \Lambda^2$, with $\Lambda$ the dynamically generated scale of the $SU(2)$ gauge theory, the magnetic monopole with charge $(1,0)$ becomes light. In the strong coupling phase $u<\Lambda^2$ the $W$-boson itself is not part of the BPS spectrum anymore. Instead the BPS spectrum consists of the monopole and the dyon with charge $(2,1)$, i.e. the bound state of $W$-boson and monopole. Compared to the $2AB$ string solutions, we can regard the monopole as the analogue of the regulator string at $z_{i\infty}^B$ and the dyon as analogue to the bound state of the EFT string and its $S$-conjugate that is formed once the latter nucleates.

\subsubsection{The general picture}  
\label{sss:general}

Let us now consider some other moduli space $\CM$ which is a K\"ahler manifold of complex dimension one, with monodromy group $\Gamma$. The monodromy group $\Gamma$ is a representation of the fundamental group of $\CM$ with all singular and orbifold points removed, which implements the duality group of the theory. Here, and in the following, we assume that $\CM$ is an exact moduli space, i.e. any non-perturabtive corrections that become important beyond weak coupling do not generate a non-perturbative superpotential but only correct the K\"ahler potential. String solutions similar to those in the toy example consist of holomorphic maps $\bm{t}: \P \to \CM$. Therefore, for such string solutions to have sub-Planckian tension, the area of $\CM$ must be smaller than $2\pi$ in Planck units. In general $\CM$ will have a set of singular points, and each of them can be associated with one or several conjugacy classes of  $\Gamma$. A global string solution not only means to provide the map $\bm{t}$, but also to specify a monodromy action at each singular point, within the corresponding conjugacy class. The preimages of such points under the map $\bm{t}$ will correspond to specific locations in $\C$, which we identify with the different string cores of the solution, each of them implementing the chosen local monodromy. 

First, to describe an EFT string solution there should be at least an infinite-order conjugacy class, whose singular point corresponds to a weakly-coupled region of the theory (for instance a large volume or large complex-structure limit), and is located at infinite distance in moduli space. We identify the preimage of this point under $\bm{t}$ with the EFT string core of our solution, which we place at $z=0 \in \C$. We consider the conjugacy class that corresponds to the minimal EFT string charge, and fix a particular representative within this class that corresponds to \eqref{tmon}. Second, to satisfy the condition \eqref{tadpole} we need a finite-order conjugacy class that plays the role of a non-physical string with a localised negative tension. Following the above scheme, we place the preimage of the corresponding singular point at $z=\infty$, which we can always do by means of a M\"obius transformation on $\C$. Third, having a map $\bm{t}: \P \to \CM$ implies that the monodromies involved in the string solution can be combined into the identity, as in figure \ref{fig1}. That is:
\begin{equation}
    M_{\rm total} = M_{\text{\tiny EFT}} \left[\prod_\a M_{\a}\right] M_{z=\infty} = {\rm Id}\, ,
    \label{totalM}
\end{equation}
where $\a$ labels additional string cores, for which we assume a particular arrangement. Because a finite-order monodromy cannot be the inverse of an infinite-order one, there should be at least a third singular point in $\CM$, beyond the two already accounted for, whose preimage in $\C^*$ will translate into the location of an additional string core. This third string, which we assume to implement an infinite-order monodromy,\footnote{If this or other monodromy points are of finite order then one could patch several copies of $\CM$, so that all strings with localised tension can be either cancelled among them or sent to spatial infinity.} is a regulator string, and specifying its position in $\C$ fixes the residual $PSL(2,\C)$ freedom of the map $\bm{t}: \P \to \CM$. If instead we consider the EFT string on a loop of radius $L$, the $PSL(2,\C)$ freedom can be fixed by determining the value of $\bm{t}$ at $z =  \hat{L}$, which we identify with the vev $\bar{\bm{t}}$. The positions of all regulator strings depend on $\bar{\bm{t}}$, so for certain regions of $\CM$ they will enter the region $D(L)$, and our finite-energy loop configuration will describe a bound state of an EFT string with one or more regulator strings.

An interesting byproduct of the above scheme is that it relates the area of $\CM$ with the order of the conjugacy classes of the monodromy group $\Gamma$. Indeed, let us consider a BPS configuration such that \eqref{totalM} is satisfied, so that \eqref{tadpole} must also be true. If moreover we assume that all monodromies are of infinite order except $M_{z=\infty}$, which is of order $N_{\infty}$ and corresponds to a negative tension $\delta_{z=\infty} < 0$, then 
\begin{equation}
    \text{Area} (\CM) = \frac{2\pi}{M_{\rm P}^{2}} \frac{p_\infty}{N_{\infty}}\, ,
    \label{areaM}
\end{equation}
for some $p_\infty \in \bN$.  If further finite order monodromies are present, the term $M_{\rm P}^{-2}\sum_\a |\delta_\a|$ should be subtracted from the rhs of \eqref{areaM}. In any case we find that, if the theory admits BPS string configurations over $\P^1$, the area of the moduli space is quantised in terms of the orders of the conjugacy classes of the monodromy group, and it is always finite.

If $\CM$ is a K\"ahler manifold of complex dimension $d$, then we expect its singularities to appear along divisors of complex dimension $d-1$. A clear example is the vector multiplet moduli space of type II Calabi-Yau compactifications, that is known to be quasi-projective \cite{MR1368632}. In this case the singular divisors form the so-called discriminant locus $\Delta$, and each component hosts at least one conjugacy class of the monodromy group $\Gamma \subset {\rm Sp}(2d+2)$. The different properties of these spaces and their singular structure have been recently analysed from the viewpoint of paths of infinite distance in \cite{Grimm:2018ohb,Grimm:2018cpv,Corvilain:2018lgw}, to which we refer for more details.\footnote{Our notation follows \cite{Aspinwall:2001zq}, which differs from \cite{Grimm:2018ohb,Grimm:2018cpv,Corvilain:2018lgw}, in the sense that for us the discriminant locus $\Delta$ only contains divisors in the interior of $\CM$, at which the underlying SCFT becomes singular. Besides this discriminant locus there are further divisors at the boundary of $\CM$, associated to additional conjugacy classes of $\Gamma$. In particular outside of our $\Delta$ there are the divisors at infinite distance, whose monodromies correspond to EFT string charges.} In the following sections we will consider them as well, but from a perspective more suited to describe string solutions.

In this setup, a 4d BPS string solution over $\C$ of finite energy will correspond to a holomorphic map $\P \to \CM$. It will specify a two-cycle $\Sigma \subset \CM$ that will intersect several of the singular divisors, and the preimage of those intersections in $\C$ will reflect as the different string core loci of the solution, each of them with a monodromy and a local deficit angle $\delta_\alpha$. To describe an EFT string solution we moreover need a non-compact two-cycle $\Sigma$, whose asymptotic behaviour is captured by its intersection with some infinite-order divisors of $\CM$ at infinite distance in moduli space. We interpret such intersection numbers as the EFT string charges in \eqref{tsol}, and the preimage of the intersection point under the map $\bm{t} : \P \to \Sigma \subset \CM$ as the EFT string core, which we place at $z=0 \in \C$. Following the one-modulus scheme we assume a single intersection  with a divisor of finite-order monodromy, corresponding to a string core of negative tension,  whose preimage under the map $\bm{t} : \P \to \Sigma \subset \CM$ we set at $z = \infty$. Finally, in a finite-energy solution the choice of monodromies should be such that \eqref{totalM} is satisfied, and as above this implies that {\it i)} the area of $\Sigma$ in $\CM$ is finite and given by the rhs of \eqref{areaM} and {\it ii)} at least an additional point of intersection with a third divisor exists. The preimage of this point in $\C^*$ corresponds to the location of one of the regulator strings of the 4d solution, and its location fixes the residual $PSL(2,\C)$ freedom of the map $\bm{t} : \P \to \Sigma$. Again, a different way to fix the freedom in the map $\bm{t}$ is to restrict it to the region $D(L) \subset \C$ that appears in the EFT string loop configuration, and specify the value $\bm{t} (\hat{L})$. Alternatively, one may look at the map $\bm{t} : \C \to \Sigma$ in the vicinity of the EFT string core at $z=0$, which necessarily will be of the form \eqref{tsol} or \eqref{multisol} for some choice of $\bm{t}_0 \equiv \bm{t}(z_0)$. There, fixing the $PSL(2,\C)$-freedom of $\bm{t}$ amounts to specifying the value of $z_0 \in \C^*$, and so all regulator string positions in $\C$ depend on this local complex parameter. 

Notice that this notion of 4d EFT string solution is more restrictive than the one considered in \cite{Lanza:2021udy}. In there, the string profile was only defined in a weakly-coupled asymptotic region of $\CM$, such that $f(z) = $ constant and \eqref{tsol} are good approximations. In other words, the string profile was only defined in a local patch as $\bm{t}|_D : D \to \CM$, with the image of the disc $D \subset \C$ contained in a weakly-coupled region of $\CM$, and containing an infinite-distance point to which the EFT string core is mapped. Clearly, if we have a global 4d EFT string solution $\bm{t} : \P \to \Sigma \subset \CM$ of the sort described above, it is trivial to construct a local one by simply restricting the map $\bm{t}$ to a disc  $D(r_0) \equiv \{ |z| < r_0\} \subset \C$. Then, if all the regulator strings lie outside of this disc, more precisely if $r_0 \ll |z^{\rm reg}_\alpha|, \forall \alpha$, we should recover a local solution of the form \eqref{tsol}, or more generally a multi-EFT-string solution \eqref{multisol}.

Going in the opposite direction is not that obvious. A priori a local EFT string solution $\bm{t}|_D : D(r_0) \to \CM$ of the form \eqref{tsol} may not allow for an extension to the whole complex plane, and in particular exhibit the features described above. First, it should be that $\bm{t}(D)$ can be extended to a globally well-defined complex curve $\Sigma_{\bm{e}} \subset \CM$, for a given EFT string charge $\bm{e}$ and all possible choices of $\bm{t}_0$. Second, for solutions of sub-Planckian area the singular divisors that $\Sigma_{\bm{e}}$ intersects should allow for a choice of monodromies such that \eqref{totalM} is satisfied. Third, in this set of intersections only a single finite-order-monodromy divisor should be involved,  corresponding to a negative-tension string in 4d, that determines the area of $\Sigma_{\bm e}$ as in \eqref{areaM}.  Based on our results in the following sections, we conjecture that global EFT string solutions with these features are always possible, a statement that we package in the following proposal:

\vspace{.5em}
\begin{conjecture}[{\bf Strong EFT String Completeness}]\
\label{conj:SESC}

\noindent
In any 4d $\CN=1$ or $\CN =2$ EFT compatible with weakly coupled gravity, any string charge in ${\cal C}^{\text{\tiny EFT}}_{\rm S}$ is represented by a  family of  BPS string solutions over $\C$ of finite tension, uniquely extending all local EFT string solutions. For elementary EFT strings the tension is sub-Planckian. 

\end{conjecture}
\vspace{1em}

 A weaker form of this conjecture was introduced in \cite{Lanza:2021udy}, stating that for any charge in ${\cal C}^{\text{\tiny EFT}}_{\rm S}$ a local EFT string solution exists. The above proposal essentially claims that there is no obstruction to extend such a local EFT string solutions to a global one of finite energy, at the expense of introducing regulator strings in the global solution. One may perform the extension for each EFT string charge $\bm{e} \in {\cal C}^{\text{\tiny EFT}}_{\rm S}$ and for each local boundary condition $\bm{t}_0 \equiv \bm{t}(z_0)$ in \eqref{tsol}. Geometrically, the charge $\bm{e}$ describes the intersection vector of a two-cycle $\Sigma \subset \CM$ with the cone of divisors at infinite field distance that correspond to ${\cal C}^{\text{\tiny EFT}}_{\rm S}$. Let us denote by the class $[ \Sigma_{\bm{e}} ]$ the set of holomorphic two-cycles with such an intersection. By changing $\bm{t}_0$ we move within the infinite class of representatives in $[\Sigma_{\bm{e}}]$, describing an infinite family of global BPS string solutions. For non-elementary charges we can decompose $\bm{e} = \sum_a \bm{e}_a$, with $\bm{e}_a \in {\cal C}^{\text{\tiny EFT}}_{\rm S}$ and build a multi-string solution of the form \eqref{multisol} that is a superposition of elementary solutions, enhancing the family of solutions associated to a charge ${\bm e}$. The definition of local EFT string solution in Conjecture \ref{conj:SESC} assumes that all mutually-commuting EFT string monodromies have their core positions $z_a$ close to each other as compared to the position of a regulator string, so that the local solution \eqref{multisol} is a good approximation in the domain $\cap_a D_a$. In that case, our findings suggest that all the parameters of the global solution are contained in this local patch, and in this sense the extension to $\C$ is unique.\footnote{Notice that in general one can also have EFT string charges that do not commute with each other, like in the $2AB$ solution of the toy example. There, non-commuting set of charges correspond to different weak coupling limits, and so to different cones ${\cal C}^{\text{\tiny EFT}}_{\rm S}$. Then, as in the toy example, the non-commuting charges should act as regulators of each other, and the local EFT string solution refers to a set of mutually commuting ones.}
 
 The uniqueness of the extension of the local EFT string solution is directly related to the statement in \cite{Seiberg:2010qd,Banks:2010zn} that in 4d EFTs compatible with weakly coupled gravity the K\"ahler form $J_\CM$ must be exact, and therefore all complex curves are non-compact. Indeed, if a compact complex curve $C$ existed in the bulk of $\CM$, then we would be able to combine it with a two-cycle $\Sigma$ that extends a local EFT string solution, and build a different extension of the same local solution. One could also think in the opposite direction, and try to build a compact two-cycle from two different global extensions  $\Sigma_{\bm{e}}$ and $\Sigma'_{\bm{e}}$ of the same local solution. Finally, let us consider again the decomposition  $\bm{e} = \sum_a \bm{e}_a$ of a non-elementary EFT string charge in a set of commuting charges. Since these are mutually BPS objects exerting no force on each other, it must hold that $T_{\bm{e}} = \sum_a T_{\bm{e}_a}$, and so elementary EFT strings are the lightest ones. The above conjecture then essentially states that their tensions are of the form $T_{\bm{e}} = 2\pi\frac{p}{N} M_{\rm P}^2$, with $p < N$.

Notice that because $T_{\bm{e}}$ cannot be made parametrically small compared to the Planck mass, there will be a limited number of EFT string charges such that $T_{\bm{e}} < 2\pi M_{\rm P}^2$. Whenever the global tension $T_{\bm{e}}$ is super-Planckian, the physical intuition that leads to \eqref{tadpole} and that relates $T_{\bm{e}}$ to the monodromy at spatial infinity fails. However, Conjecture \ref{conj:SESC} proposes that one should still be able to build formal solutions over $\C$ from patching elementary string solutions as in the toy example, including the interplay between monodromies and the total area. This exercise is not only academic, since it describes the physics of EFT strings as they approach regions of strong coupling. Indeed, that the global tension $T_{\bm{e}}$ is super-Planckian does not mean that the corresponding EFT string automatically forms a black hole when it nucleates. Recall that the tension of a  string in a loop of length $L$ in a vacuum $\bar{\bm{t}}$ can be estimated as ${\cal E}_{\rm back}\left(L,\bar{\bm{t}}\right)$, cf.\eqref{1stdefT}. Since this tension is nothing but the area of the region $D(L)$ under the map $\bm{t}_{\bm{e}} : \C \to \Sigma_{\bm{e}}$ and this only depends on the boundary conditions at $\partial D(L)$,  $T_{\bm{e}}(\bar{\bm{t}}) \equiv {\cal E}_{\rm back}\left(L,\bar{\bm{t}}\right)$ turns out to be independent of $L$. Because the area of $\bm{t}_{\bm{e}}  (D(L))$ is always smaller than the area of $\Sigma_{\bm{e}}$, we have that $T_{\bm{e}}(\bar{\bm{t}}) \leq T_{\bm{e}}, \forall \bar{\bm{t}}$, and so the tension of the global string solution $T_{\bm{e}}$ has to be understood as an upper bound for a nucleated string tension in a particular vacuum. As long as $T_{\bm{e}}(\bar{\bm{t}}) < 2\pi M_{\rm P}^2$, the extended solution will describe the EFT string tension beyond weak coupling. More precisely, as pointed out before in the interior of the moduli space this quantity will generically measure the tension of a bound state of strings, including the EFT string and several regulator strings.

The above conjecture is substantiated by the results of the following sections, where we perform a general analysis of 4d strings in type IIA and heterotic Calabi--Yau compactifications, and more precisely of those strings built from wrapping NS5-branes on internal divisors. In this  setup, we observe further interesting features that are not captured by Conjecture \ref{conj:SESC}, but that could also be part of the general picture. First, even if the moduli space $\CM$ is in general quite complicated, we manage to formulate the discussion of global string solutions in terms of a quasi-toric space $\CM_\tau$ with a much simpler metric than $\CM$. This auxiliary space is in one-to-one correspondence with $\CM$, and contains the information of the action of the monodromy group $\Gamma$ and its associated divisors, including the discriminant locus $\Delta$. Thanks to this auxiliary space, one is able to describe 4d string solutions as
\begin{equation}
    \C \stackrel{\bm{\tau}}{\to} \CM_\tau \stackrel{\mathfrak{M}}{\to} \CM\, ,
    \label{split}
\end{equation}
where $\tau$ is a holomorphic map, and $\mathfrak{M}$ is a one-to-one map that captures the effect of quantum corrections, dubbed {\em mirror map}. The complex curves $\bm{\tau}(\C) = \Xi_{\bm{e}} \subset \CM_\tau$ contain all the topological information of the string solutions, in particular their intersections with singular divisors. As a result, one can characterise the EFT string charges and the regulator strings of a global string solution in terms of $\Xi_{\bm{e}} = \mathfrak{M}^{-1} (\Sigma_{\bm{e}})$. 

Thanks to this simplified picture, one is able to construct explicit solutions and test their properties analytically. We examine such properties for a set of explicit examples, finding not only that the content of Conjecture \ref{conj:SESC} is verified, but also further interesting observations. For instance, we find that for any elementary EFT string charge $\bm{e}$, the family of two-cycles within $[ \Sigma_{\bm{e}} ]$ forms a complex-dimension one foliation of $\CM$ over the corresponding divisor at infinity. This in particular implies that an elementary EFT string solution can reach any point in moduli space $\CM$, by appropriately choosing the local weak-coupling conditions $z_0$, $\bm{t}_0$ in \eqref{tsol} and then extending the solution towards strong coupling regions. It would be interesting if this was a general property of EFT string solutions, as it would mean that one has a physical object by means of which one can probe the bulk of the moduli space of an EFT. 

Some interesting consequences can be derived from Conjecture \ref{conj:SESC}. For instance, notice that \eqref{totalM} needs to be satisfied for each elementary EFT string which, using the Distant Axionic String Conjecture \cite{Lanza:2020qmt,Lanza:2021udy} can be associated with a generator of the infinite-distance divisors of $\CM$. So each of these generators translates into a constraint on the elements of the monodromy group $\Gamma$. In \eqref{totalM}, each element of the product corresponds to a specific singular divisor of $\CM$, or equivalently to a conjugacy class of $\Gamma$, which in turn corresponds to a unique element of the abelianisation of the monodromy group $\Gamma^{\rm ab} \coloneqq \Gamma/[\Gamma,\Gamma]$. So upon abelianisation \eqref{totalM} reads
\begin{equation}
\left[m_{\text{\tiny EFT}}m_{z=\infty}\right]^{-1} =  \prod_\a m_{\a} \, ,
    \label{totalMab}
\end{equation}
where $m_\a \in \Gamma^{\rm ab}$ represents the abelianised element $M_\a$, etc. Note that because $M_{z=\infty}$ is of finite order, so must be $m_{z=\infty}$. If elementary string solutions intersect all the different components $\Delta_\a$ of the discriminant locus, all the corresponding  elements $m_\a$ will be involved in the constraints \eqref{totalMab}. Let us then assume that the set $\{m_{\text{\tiny EFT}}, m_{z=\infty}\}$ and finite-order elements $\{m_\beta\}$  generate a finite-index subgroup  of $\Gamma^{\rm ab}$, as it happens in the examples of section \ref{s:examples}.\footnote{In fact, it holds the stronger condition that $\{M_{\text{\tiny EFT}}, M_{z=\infty}, M_{\beta}\}$  generate a finite-index subgroup $\Gamma_\text{\tiny EFT}$ of $\Gamma$, with $M_\beta$ of finite order.} We are then led to:  

\vspace{.5em}
\begin{conjecture}\
\label{conj:abelian}

\noindent
Let $\CM$ be the moduli space of a 4d supersymmetric EFT compatible with weakly coupled gravity, and $\Gamma$ its monodromy group. Then its abelianisation is of the form $\Gamma^{\rm ab} = \Z^m \times  \Gamma^{\rm ab}_{\rm fin}$, where $m \leq N = \#$ generators of the cone of  divisors at infinity, and $\Gamma^{\rm ab}_{\rm fin}$ is a finite group. 

\end{conjecture}
\vspace{1em} 

This applies in particular to the vector multiplet moduli space  of type II Calabi--Yau compactifications. If in there we interpret $\Gamma$ as a realisation of the fundamental group $\pi_1(\CM)$, we have that $H_1(\CM, \Z) \simeq \Z^m$. This is in agreement with the statement that $\pi_1(\overline{\CM}) =0$, where $\overline{\CM}$ is a finite cover of the moduli space compactified  by the addition of the divisors at infinity. As pointed out in \cite{Dierigl:2020lai}, for our toy example one can interpret $\pi_1(\overline{\CM}) =0$ in terms of the Cobordism conjecture \cite{McNamara:2019rup}, in the sense that adding the set of EFT strings to the theory trivialises the first homotopy group of the moduli space, as first conjectured in \cite{Ooguri:2006in}. In fact, this reasoning not only allows to recover the naive set of elementary EFT string charges generating ${\cal C}^{\text{\tiny EFT}}_{\rm S}$ in the toy example, but also the full spectrum of conjugate EFT charges and the  structure of non-Abelian braid statistics associated to them \cite{Dierigl:2020lai}. It would be very interesting to see if these results can be generalised to arbitrary monodromy groups resulting from string  compactifications to 4d.

Notice that what the main new assumption behind Conjecture \ref{conj:abelian} boils down to, is that the `endpoints' of the EFT string solution $m_{\text{\tiny EFT}}$ and $m_{z=\infty}$ together with some torsional elements generate a finite-index subgroup of $\Gamma^{\rm ab}$. Physically, one could interpret this as that the string charges corresponding to the discriminant locus $\Delta$ are not fundamental 4d string charges of the EFT. The motivation for this is the remaining components of $\Delta$ correspond to finite-distance conifold-like singularities of the moduli space \cite{Melnikov:2019tpl}, that as in \cite{Strominger:1995cz} should be resolved by integrating in a finite number of states into the EFT. The fundamental string charges instead correspond to divisors at infinite distance in $\CM$ and to finite-order monodromies.


\section{4d string backreaction from a 2d field theory perspective}
\label{s:4d2d}

In this section we develop our strategy to study 4d EFT string solutions beyond weak coupling regions. Our setup will be type IIA and heterotic Calabi--Yau compactifications, and 4d strings  built from wrapping NS5-branes on internal four-cycles. Following \cite{Quigley:2011pv,Blaszczyk:2011ib}, we  encode the NS5-brane backreaction in terms of the GLSM describing such background, and in particular in a non-trivial profile for the complex FI-terms of the GLSM. With this profile one can describe the set of regulator strings that are involved in a global EFT string solution, as we show for the quintic in this section and for two further examples in section \ref{s:examples}.

\subsection{The analogy with 7-branes in F-theory}
\label{ss:FtheoryD3}

Accounting for the backreaction of strings in 4d EFTs is quite reminiscent of the description of D7-brane backreaction in ten-dimensional type IIB string theory, as it is clear from the toy example considered in section \ref{sec:toy}. Both objects, EFT strings in 4d and 7-branes in 10d, are complex co-dimension one objects and thus give rise to a logarithmic profile for the complex scalar to which they couple magnetically. In type IIB it is well-known how to describe the D7-brane backreaction away from their location, by keeping track of the type IIB axio-dilaton in terms of a line bundle ${\cal L}$ that describes the variation of the complex structure of an elliptic curve over their transverse space.\footnote{For a Weierstrass model described by the hypersurface $y^2=x^3+fx+g$ in $\mathbb{P}_{2,3,1}$ the projective coordinates $(x,y)$ of $\mathbb{P}_{2,3,1}$ are then sections of $\mathcal{L}^2$ and $\mathcal{L}^3$, respectively.} In F-theory, strong coupling effects including $D(-1)$-instantons, are then geometrised by considering compactification manifolds that are elliptic fibrations over some K\"ahler base $B_n$. The Einstein equations then relate the first Chern class of the line bundle ${\cal L}$ to the first Chern class of $B_n$, as $ c_1({\cal L}) = c_1 (B_n)$.

An equivalent way to describe the backreaction of the 7-branes on the type IIB axio-dilaton $\tau$ is by realising that $\tau$ can be thought of as the complexified gauge coupling of the $\cN=4$ super Yang--Mills theory with gauge group $SU(2)$, realised as the low-energy worldvolume theory of a stack of two D3-branes in type IIB string theory \cite{Banks:1996nj}. In the presence of O7-planes one can  split the stack of two D3-branes into a single D3-brane plus its orientifold image D3$'$. The orientifold breaks the supersymmetry on the D3-brane to $\cN=2$, and separating the D3-brane and its image from the O7-locus additionally breaks the gauge group as $SU(2)\rightarrow U(1)$. From the worldvolume perspective this separation can be interpreted as giving a vev to an adjoint scalar. The vev of this scalar, i.e. the Coulomb branch parameter, can then be interpreted as  $z \in \C$, with the orientifold  located at the origin.

The backreaction of a system of $n_{D7}$ parallel D7-branes can thus be described by the Coulomb branch of an $\cN=2$ $SU(2)$ gauge theory with $n_{D7}$ flavours. More precisely, the local axio-dilaton $\tau(z)$ can be identified with the gauge coupling of the D3-brane gauge theory, with Coulomb branch parameter $z$ probing the directions transverse to the 7-branes. Identifying $\tau(z)$ (and its magnetic dual $\tau_D(z)$) with the periods of a $T^2$ as in Seiberg--Witten theory, one can again describe this system as a torus fibration over the space transverse to the 7-branes. In particular, taking into account the effect of the D(-1) instantons, one observes that the O7-plane splits into two $[p,q]$ 7-branes. The resulting profile for the axio-dilaton can then be identified with a solution built from the building blocks discussed in section \ref{sec:toy}, with the $[p,q]$ 7-branes taking over the role of the regulator strings.

Whereas in the case of string theory 7-branes all this is well-established, our goal in this section is to find a similar description of 4d string backreaction, in terms of a lower-dimensional gauge theory. In the following, we want to provide such a description for the case that the 4d EFT arises from type IIA or heterotic string theory compactifications on Calabi--Yau threefolds. More precisely, we will analyse the backreaction of NS5-branes from the perspective of a 2d field theory probing the directions transverse to them.

\subsection{EFT strings as GLSM anomalies}
\label{ss:EFTGLSM}

To that end, let us consider those 4d $\cN=2$ and $\cN=1$ EFTs obtained by respectively compactifying type IIA  or heterotic string theory on a Calabi--Yau threefold $Y_3$. In this case a subsector of the cone of EFT string charges arises from NS5-branes wrapping divisors corresponding to the generators of the K\"ahler cone $\cK$ of $Y_3$ \cite{Lanza:2021udy}
\begin{align}\label{Kahlercone}
    \cK = \text{span}\langle D_1, \dots, D_{h^{1,1}}\rangle \,. 
\end{align}
Denoting the complexified K\"ahler moduli of $Y_3$ by $t^i=b^i +is^i$, $i=1,\dots, h^{1,1}$ a string wrapping the four-cycle 
\begin{align}
    D_{\mathbf{e}} = \sum e^i D_i\,,
\end{align}
will classically induce a profile
\begin{align}\label{talog}
    t^i = t^i_0 +\frac{e^i}{2\pi i }\log\left(\frac{z}{z_0}\right)\,,
\end{align}
where we placed the NS5-brane at $z=0$ in the transverse space $\C$. We would now like to interpret this backreaction from the perspective of a field theory probing the transverse direction. For this, note that in the large-volume limit the complexified K\"ahler moduli $t^i$ can be identified with the complex Fayet-Iliopoulos parameters $\tau^i$ of a gauged linear sigma model with gauge group $U(1)^{h^{1,1}}$ which flows to the non-linear sigma model with target space $\R^{1,3} \times Y_3$. See appendix \ref{ap:GLSM} for some background on GLSMs.

Our proposal is to describe the backreaction of the NS5-branes through this 2d field theory probing the space transverse to the NS5-branes. To draw the analogy to the case of D7-branes, 
we identify the GLSM realised on a 2d worldsheet probing the NS5-brane background as the analogue of the 4d worldvolume theory of the D3-brane probing the D7-brane background. In this context the complex FI-parameters can be viewed as the analogues of the gauge coupling $\tau$ for D3-brane gauge theory. The crucial question is now how to translate the effect of the flavours, i.e. the D7-branes, and the Coulomb branch parameter of the D3-brane gauge theory into quantities of the GLSM.\footnote{Let us stress that this is just an analogy between these two setups, but that the physical systems described in this way are \text{not} the same. In particular, the two setups are not related via duality and we do not claim that the GLSM arises e.g. from the D3-brane upon compactification to 2d.} 

The GLSM associated to the compactification of type IIA/heterotic string on $Y_3$ contains,\footnote{We use $(0,2)$ language to account for both  heterotic and type IIA string  compactifications at the same time.} as reviewed in appendix \ref{ap:GLSM}, $h^{1,1}$ gauge field strength multiplets $\Upsilon_i$, $i=1,\dots, h^{1,1}$, together with neutral and charged bosonic chiral multiplets, $S_i$, $i=1,\dots, h^{1,1}$ and $\Phi^\alpha$, $\alpha=1,\dots, n$, respectively. The charged fields carry charge $Q_\alpha^i$ under the $i$-th $U(1)$ gauge factor. In addition, we have $d$ Fermi-multiplets $\Gamma^a$, $a=1,\dots, d$, cf. \eqref{ap:Gamma}, with $U(1)_i$ charge $Q_a^i$. Notice that for the GLSM to preserve $(2,2)$ supersymmetry we need $d=n$ and the same charge spectrum for the $\Gamma^a$ as for the $\Phi^\alpha$. For each gauge factor, we can now introduce an FI-term in the action 
\begin{align}\label{FIGLSM}
    S_\text{FI} = \frac{1}{8\pi i}\int d^2z d\theta^+ \Upsilon_i \tau^i + \text{h.c.}\,,
\end{align}
where $\tau^i = \theta^i + i\varrho^i$. For $\varrho^i\rightarrow \infty$ the GLSM then flows to a Non-Linear Sigma Model (NLSM) with target space $Y_3$.  In this limit, we can identify the FI parameters of the GLSM with the complexified K\"ahler moduli of $Y_3$, i.e. $t^i \simeq \tau^i$. However, away from the large volume limit the map between the complex FI-parameter space $\cM_{\tau}$ and quantum K\"ahler moduli space $\cM_{\rm qK}$
\begin{eqn}\label{mirrormap}
    \mathfrak{M}: \quad \cM_{\tau} \;&\rightarrow \;\cM_{\rm qK}
\end{eqn}
receives corrections from worldsheet instantons
\begin{align}
      t^i = \mathfrak{M}(\tau^i) = 
      \frac{1}{2\pi i} \log q_i + \mathcal{O}(q_i)\,, \qquad q_i=e^{2\pi i \tau^i}\,. 
\end{align}
Compared to $\cM_{\rm qK}$, the complex manifold $\cM_{\tau}$ has a much simpler structure. The FI-parameter space spanned by the coordinates $q_i$ is a subset of the algebraic torus $T_{h^{1,1}}=(\mathbbm{C}^*)^{h^{1,,1}}$ that can be obtained by taking $T_{h^{1,1}}$ and removing the discriminant locus $\Delta$ where the GLSM becomes singular, i.e. $T_{h^{1,1}}\backslash \,\Delta$. While $\cM_{\tau}$ has this simple toric structure, the same is not true for $\cM_{\rm qK}$ due to worldsheet instanton corrections.
Many of the details of $\cM_{\rm qK}$ are encoded in the one-to-one map $\mathfrak{M}$, which can be computed at the exact level by calculating the sphere partition function of the GLSM, using localisation techniques \cite{Jockers:2012dk,Gomis:2012wy}. In case the CY $Y_3$ has a mirror, the map \eqref{mirrormap} can equivalently be determined using mirror symmetry. Indeed, under mirror symmetry the K\"ahler moduli $t^i$ map to flat coordinates on the complex-structure moduli space of the mirror $\tilde Y_3$, whereas $q_i$ map to parameters describing its complex-structure deformations. Calculating the periods of $\tilde Y_3$ one then obtains the map $\mathfrak{M}: \tau^i \mapsto t^i$. Since it can be determined via mirror symmetry, in the following we will refer to \eqref{mirrormap} as \textit{mirror map}.\footnote{Let us stress that \eqref{mirrormap} is still a map from the FI-parameter space to the quantum K\"ahler moduli space $\cM_{\rm qK}$ of type IIA on $Y_3$. In particular, it is \textit{not} a map between the type IIB complex structure moduli space and $\cM_{\rm qK}$.}

As it stands, the GLSM is associated to type IIA/heterotic string theory in a spacetime given by $Y_3$ times four-dimensional Minkowski. The presence of NS5-branes wrapping divisors in $Y_3$ leading to strings, however, induces a profile for the K\"ahler moduli $Y_3$ which varies along a complex plane within  four-dimensional Minkowski space. In favourable cases such as heterotic standard embedding, the K\"ahler deformation space is an exact moduli space since WS instantons only correct the K\"ahler potential but do not induce a non-perturbative superpotential. In the following we will mostly focus on these cases and hence do not have to concern ourselves with the effect of non-perturbative superpotentials on the extension of local solutions of the form \eqref{talog}. In order to describe the variation of the K\"ahler moduli with the GLSM we should first also account for the complex plane transverse to the string. We thus effectively want a GLSM description associated to a compactification on the four-fold
\begin{align}\label{X4factor}
    X_4 = Y_3 \times \C\,. 
\end{align}
We can incorporate the additional factor $\C$ by first considering its one-point compactification into a $\P^1$. Since 
\begin{align}\label{P1}
    \P^1 = \frac{\C^2\backslash \{0\}}{\C^*}\, ,
\end{align}
we can include this extra factor in the GLSM description by adding two extra chiral bosonic fields $(\Phi^0, \tilde \Phi^0)$ (accounting for the $\mathbbm{C}^2$ factor) together with two extra chiral Fermi fields $(\Gamma^0, \tilde\Gamma^0)$, as required by supersymmetry. In addition, we add one extra gauge group $U(1)_0$ with field strength $\Upsilon_0$ together with one neutral chiral multiplet $S_0$ to realise the $\C^*$ quotient in \eqref{P1}. The additional matter is charged only under $U(1)_0$ with charges $Q_0^0=\tilde Q_0^0=1$ for both the bosonic and fermionic multiplets, in order to achieve the factorised form of $X_4$ in \eqref{X4factor} prior to the addition of NS5-branes. We also include an additional FI-term to the action 
\begin{align}
    S_\text{FI}^{(0)} = \frac{1}{4}\int d^2z d\theta^+ \Upsilon_0 (\theta^0+i\varrho^0)+ \text{h.c.}
\end{align}
As the other matter fields are not charged under $U(1)_0$, the D-term constraint for $D_0$  reads
\begin{align}
    \varrho^0 = | \phi^0|^2 + |\tilde \phi^0|^2\,,
\end{align}
where $\phi^0, \tilde \phi^0$ are the leading bosonic components of $\Phi^0, \tilde \Phi^0$ in Wess-Zumino gauge. The resulting space is thus indeed a $\P^1$ with $\varrho^0$ describing its radius. Since we want the transverse space to be non-compact we choose $\varrho^0\rightarrow \infty$ and solve the D-term constraint by sending $\tilde \phi^0\rightarrow \infty$. We can thus freely tune the vev of $\phi^0$ which we can interpret as the coordinate $z \in \C$ on the space transverse to the NS5-branes. In the following, we trade $\phi^0$ for $z$ to make the relation to the space-time interpretation more clear.

Coming back to the analogy with D3-branes in type IIB, we see that the vev of the scalar in the multiplet $\phi^0\equiv z$ takes over the role of the vev of the adjoint scalar in the D3-brane gauge theory, i.e. the Coulomb branch parameter. To make the dictionary complete we need to understand how the FI-parameters depend on the scalar $z$ in the presence of NS5-branes, giving rise to a profile as in \eqref{talog} in the classical regime. To see this, we can make use of the relation observed in \cite{Quigley:2011pv,Blaszczyk:2011ib}  between NS5-branes in heterotic string compactifications and FI-terms, applied to the present setup. 

As is well-known from higher dimensional cases, space-time filling NS5-branes in heterotic string theory wrapping complex co-dimension two cycles in a compact manifold can be interpreted as point-like gauge instantons of the heterotic gauge theory. In the GLSM description one can equally interpret them as gauge theory instantons. As a consequence, the 2d theory is not gauge invariant anymore but suffers from an anomaly given by the term (cf. \eqref{ap:anomaly})
\begin{align}
       \delta S = \frac{\mathcal{A}^{IJ}}{16 \pi}\left(\int d^2z d\theta^+ \Lambda_I \Upsilon_J + {\rm h.c.}\right)\, , \qquad I,J = 0, 1, \dots, h^{1,1}\, .
\end{align}
That is, the presence of the NS5-brane introduces a non-trivial anomaly coefficient $\mathcal{A}^{IJ}$ that needs to be cancelled by other means. As shown in \cite{Quigley:2011pv} this anomaly can be cured by introducing a logarithmic dependence of the FI-terms on the chiral field $\Phi^\alpha$.

Let us see how this works for our case of interest: we want to consider NS5-branes that are not space-time filling but are extended along two of the non-compact directions and wrap a divisor of the internal CY manifold $Y_3$. However, we are interested in capturing the backreaction of the NS5-brane along the two non-compact directions transverse to it. In the GLSM description of this example, we already took these directions into account by introducing the extra matter fields $(\Phi^0, \tilde \Phi^0)$ and the additional gauge group $U(1)_0$. As discussed, in the limit $\varrho^0\rightarrow \infty$ we recover the geometry \eqref{X4factor}, where the second factor is parametrised by $z = \phi^0$. We can now consider a basis of divisors on this product manifold given by 
\begin{align}
    [D'_0] = [Y_3]\,,\qquad [D'_i] = [\C \times D_i]\,,\quad i=1,\dots, h^{1,1}(Y_3)\,,
\end{align}
where $D_i$ are the K\"ahler cone divisors of $Y_3$ defined in \eqref{Kahlercone} and $[D_0']$ is just the class of $Y_3$ inside $X_4$. The NS5-branes of interest wrap four-cycles which are intersections of these
\begin{align}\label{divisorGLSM}
    \mathcal{S}_{e^i}(z^0) = e^i \left(D'_i . D'_0(z^0)\right)\, \subset X_4\,. 
\end{align}
Here the representatives in $[D_0']$ are labelled by the coordinate $z^0\in \C$ corresponding to the location of the NS5-brane in the non-compact factor $\C$ of $X_4$. From the perspective of the GLSM associated to $X_4$ we can interpret this NS5-brane as a gauge theory instanton shifting the anomaly polynomial $\mathcal{A}^{IJ}$ as 
\begin{align}
    \mathcal{A}^{0i} \rightarrow \mathcal{A}^{0i} - e^i \,,\qquad  \mathcal{A}^{i0} \rightarrow \mathcal{A}^{i0} - e^i\,,
\end{align}
with all other coefficients $\mathcal{A}^{00}, \mathcal{A}^{ij}$ unchanged. Notice that this shift preserves the symmetry of the anomaly coefficient $\mathcal{A}^{IJ}$.  In the space-time interpretation, the shift in the anomaly corresponds to an additional contribution to the Bianchi identity, of the form
\begin{align}\label{Bianchi}
    dH_3 = \frac{\alpha'}{4} \left({\rm Tr}(R\wedge R) - {\rm tr}(F\wedge F)\right) + e^i \delta^{(2)} (D_0'(z_0)) \wedge D_i' \,.
\end{align}
The anomaly introduced through the presence of the point-like instanton now needs to be cancelled which is equivalent to solving the Bianchi identity in the presence of the last term in \eqref{Bianchi}. In other words, cancelling the anomaly amounts to solving the Bianchi identity with an NS5-brane localised in $\C$, which by supersymmetry implies taking into account its backreaction. 

Following \cite{Quigley:2011pv} one may cancel the GLSM anomaly by introducing a logarithmic dependence of the FI-parameters on the fields $\Phi^i$. More precisely, consider the modified FI-term 
\begin{align}\label{FIlogarithmic}
    S_\text{FI}' =  \frac{1}{4}\int d^2 z d\theta^+ \Upsilon_i\left(\tau^i_0 + \frac{e^i}{2\pi i} \log \Phi^0\right) + \text{h.c.}
\end{align}
The variation of this term is given by
\begin{align}
    \delta S_\text{FI}' = -\frac{e^i Q_0^0}{8 \pi} \int d^2z d\theta^+  \Lambda_0 \Upsilon_i+ \text{h.c.}\,
\end{align}
which partially cancels the $\mathcal{A}^{0i}$ component of the anomaly coefficient. However, since $\mathcal{A}^{IJ}$ is manifestly symmetric, we need an additional ingredient to cancel the anomaly completely. To that end \cite{Quigley:2011pv} noticed that one can add the following term to the GLSM action
\begin{align}\label{ST}
    S_{T} = -\frac{1}{4\pi}\int d^2 y d\theta^+ T^{IJ} V_{I+} V_{J-}\,,
\end{align}
with $T^{IJ}$ an anti-symmetric tensor. As shown in \cite{Quigley:2011pv} the variation of this term yields 
\begin{align}
     \delta S_T = -\frac{T^{IJ}}{8 \pi} \int d^2z d\theta^+  \Lambda_I \Upsilon_J+ \text{h.c.}
\end{align}
The anomaly is thus cancelled in case we have 
\begin{align}
    T^{IJ} = \sum_{\alpha =0}^n e^{[I}_\alpha Q^{J]}_\alpha \,,\qquad \sum_{\alpha=0}^n e^{(I}_\alpha Q^{J)}_\alpha - \frac{1}{2} \mathcal{A}^{IJ}=0\,,
\end{align}
 which in our setup only has contributions for $(I,J)=(i,0)$. To sum up, the anomaly due to the presence of the NS5-brane is then cancelled if we allow for logarithmic FI-terms as in \eqref{FIlogarithmic} 
\begin{align}\label{FIlog}
    \tau^i = \tau^i_0 +\frac{e^i}{2\pi i} \log \Phi^0 \rightarrow  \tau^i_0 +\frac{e^i}{2\pi i} \log\left(\frac{z}{z_0}\right)\,,
\end{align}
and add a term of the form \eqref{ST} with $T^{0i}=e^i$ and $T^{i0}=-e^i$. Notice that in the last step in \eqref{FIlog} we identified the leading scalar piece of $\Phi^0$ with the coordinate $z$ on $\C$ transverse to the NS5-brane. In the presence of the NS5-branes, the FI-terms thus have a similar profile as the K\"ahler moduli \eqref{talog} at leading order. This should not come as a surprise since for $\tau^a\rightarrow i \infty$ the FI-parameters of the GLSM reduce in the IR to the K\"ahler moduli of the CY. However, away from the large volume limit this identification receives corrections encoded in the mirror map. Since the logarithmic behaviour of the FI-terms exactly cancels the anomaly, we can take the profile \eqref{FIlog} to be valid also far away from $z=0$. It is then the mirror map that encodes the corrections to the $t^i$ profile \eqref{talog} in such a region.

As we vary $z \in \C$ we cover a 2-cycle $\Xi$ in the FI-parameter space of the GLSM which covers regions where the GLSM is not well-approximated by a NLSM with target space $Y_3$. Since the GLSM also allows us to study these phases, it gives us an ideal way to test the strong coupling regimes of the solutions associated to NS5-strings in heterotic/type IIA compactifications to 4d.  

\subsubsection*{The dictionary between 7-branes and NS5-branes}

Let us briefly digress and further comment on the analogy between our NS5-brane setup and the gauge theory description of D7-brane backreaction in 10d type IIB string theory. In the latter, one describes the profile of the type IIB axio-dilaton in terms of the gauge coupling function $\tau$ of the $\cN=2$ SYM theory on a D3-brane worldvolume. The axio-dilaton profile can then be calculated by computing the periods of the auxiliary torus in SW theory as a function of the Coulomb-branch parameter, which is identified with the coordinate on the space transverse to the D7-branes. In our setup, something similar happens. Here, we identify the coordinate on the transverse space with the expectation value of the matter field $\phi^0$, i.e. a parameter on the Higgs branch. As a consequence of the presence of NS5-branes, we find that the complexified FI-terms of the 2d gauge theory depend on this parameter, providing us with a map 
\begin{eqn}
    \bm{\tau}: \,\quad &\C \rightarrow \mathcal{M}_{\tau}\, ,
\end{eqn}
where $\mathcal{M}_\tau$ is the FI-parameter space and the map $\bm{\tau}$ is given by \eqref{FIlog}. As mentioned before, the translation to the K\"ahler moduli space $\CM$ goes via the mirror map $\mathfrak{M}$,  which can be determined by calculating the periods of the mirror CY $\tilde Y_3$. We thus effectively calculate the periods of an auxiliary manifold $\tilde Y_3$ that shares its moduli space with the GLSM, as a function of the Higgs branch parameter $z$. The analogy to the D3-brane gauge theory description of 7-branes in type IIB string theory can now be summarised in the following dictionary:
\begin{center}
\begin{tabular}{c c c}
 \textbf{7-branes in 10d} && \textbf{NS5-strings in 4d} \\\hline
type IIB axio-dilaton & $\longleftrightarrow$ & K\"ahler moduli $t^i$ of $Y_3$ \\
Coulomb branch parameter &  $\longleftrightarrow$ & Higgs branch parameter $\phi^0$ \\
Periods of torus &  $\longleftrightarrow$ & Periods of mirror $\tilde Y_3$
\end{tabular}
\end{center}
\subsection{Beyond weak coupling}
\label{ss:BWC}

As we just discussed, representing the backreaction of NS5-strings in terms of a GLSM allows us to also describe the corresponding 4d string solutions beyond the weak-coupling limit. In such a regime the GLSM does not flow to a NLSM with target space $Y_3$ anymore, but instead to some possibly non-geometric theory such as Landau--Ginzburg theories. Still, in the quantum theory one can transition smoothly between these different phases, although there still exist complex co-dimension one singularities in the complex FI-parameter space $\mathcal{M}_\tau$. 

In GLSMs associated to NLSMs with target space a Calabi--Yau manifold, these singularities either correspond to so-called $\sigma$-vacua or to mixed $\sigma$-Higgs vacua. These vacua differ from the vacua of the theory described by the NLSM in the following way: in the limit $\tau^a \rightarrow i \infty$, we have the vacuum condition 
\begin{align}
    \sum_{\alpha=1}^n Q^\alpha_i |\phi^\alpha|^2 = \text{Im}\,\tau^i\,,
\end{align}
which can be solved by giving vevs to (a subset of) the matter fields $\phi^\alpha$. Given the bosonic potential of the GLSM \eqref{app:Ubos} these vevs induce a mass for the neutral chiral $S_i$ multiplets introduced above \eqref{FIGLSM}. Let us denote the scalar component of these multiplets by  $\sigma_i$. Integrating out these massive fields one obtains the NLSM. For the $\sigma$-vacua the opposite happens, and the $\sigma_i$ fields acquire a large vev inducing a mass for the fields $\phi^\alpha$. 

For simplicity let us focus in the following on $(2,2)$ GLSMs corresponding to Calabi--Yau threefold compactifications of type IIA or heterotic string theory with standard embedding. In this case, integrating out the massive fields $\phi^\alpha$ one recovers the effective one-loop D-term condition \cite{Morrison:1994fr}
\begin{align}\label{vacuacondition}
\prod_{\alpha | Q^i_{\alpha}>0} \left(\sum_{j} Q^j_\alpha \sigma_j\right)^{Q^i_\alpha} = q_i  \prod_{\alpha|Q^i_{\alpha}<0} \left(\sum_j Q_\alpha^j \sigma_j\right)^{-Q_\alpha^i}\,,
\end{align}
where we introduced the exponentiated FI-terms $q_i = e^{2\pi i \tau^i}$. Notice that we do not sum over the index $i$, such that \eqref{vacuacondition} gives $h^{1,1}$ conditions. We thus have $h^{1,1}$ equations to determine the vevs of the $h^{1,1}$ $\sigma_i$-fields at the $\sigma$-vacua. However, in case the GLSM is associated to a CY compactification, the above equation is homogeneous in the $\sigma$-fields and the system of equations \eqref{vacuacondition} is over-determined. Therefore, solving \eqref{vacuacondition} necessarily leads to a constraint for the parameters $q_i$ and solutions to \eqref{vacuacondition} generically only exist in co-dimension one in the complex FI-parameter space $\cM_\tau$. A point $\bm{q}\in \cM_\tau$ for which \eqref{vacuacondition} has a solution thus corresponds to a flat $\sigma$-direction and hence to a singularity of the GLSM. The complex co-dimension one locus of singularities solving \eqref{vacuacondition} is typically referred to as the principal component of the discriminant divisor. The full discriminant divisor $\Delta$ comprises further GLSM singular loci of complex co-dimension one, like  mixed Higgs-$\sigma$-vacua. For instance, singular loci appear in the vicinity of each of the boundaries of the classical K\"ahler cone of $Y_3$. Taking these and further singular loci into account one obtains an intricate singularity structure of the FI-parameter space.  Under the mirror map $\mathfrak{M}$, these singular divisors of $\CM_\tau$ translate into the singular divisors of the actual moduli space $\CM$, that host the  conjugacy classes of the monodromy group $\Gamma$ as described in section \ref{sss:general}. By abuse of language, we will denote by $\Delta$ both {\it i)} the full discriminant divisor of $\CM_\tau$ where the GLSM becomes singular and {\it ii)} its image under $\mathfrak{M}$, which is the set of singular divisors at finite distance in the bulk of $\CM$ \cite{Melnikov:2019tpl}. 

Given the logarithmic profile for the FI parameters \eqref{FIlog} in $\CM_\tau$, we expect the solution to hit these singularities for some value of $z$. Since they are associated with monodromies in moduli space, from the spacetime perspective we interpret them as the regulator strings of section \ref{ss:regulator}, i.e. string cores that appear when the NS5-brane solution reaches strong coupling regimes. As in our general picture, these systems with multiple string cores correspond to holomorphic maps from the complex plane transverse to the string into the moduli space $\CM$. Using the splitting \eqref{split} one can instead characterise them in terms of a map into the complex FI-parameter space:
\begin{eqn}
    \bm{q}: \quad &\P^1 \rightarrow \mathcal{M}_\tau \,, \\
    &z \mapsto \bm{q}(z) = \exp(2\pi i \bm{\tau}(z))\,. 
    \label{FIexp}
\end{eqn}
Describing a 4d string solution via this map is much simpler than using the map $\bm{t}(z) = \mathfrak{M}(\bm{\tau}(z))$, due to the simple solution \eqref{FIlog}. Notice that such a simple profile essentially says that in terms of the FI-parameter space there is a single source associated to the NS5-brane core, and that regulator  strings do not act as point-like sources. As one can see in explicit examples like the quintic, this assumption is justified whenever regulator strings correspond to unipotent monodromies at finite distance in moduli space, that is to components of $\Delta$. In that case they correspond to conifold-like singularities that are smoothed out by integrating in a finite numbers of degrees of freedom \cite{Strominger:1995cz}, which should not affect the string solution.  The precise mapping is therefore determined by the asymptotic charge ${\bm{e}}$ of the NS5-string, and is given by
\begin{align}\label{qofz}
    q_i(z) = \left(\frac{z}{z_0}\right)^{e^i}\bar q_i \,.
\end{align}
Here, we used \eqref{FIlog} and defined $\bar q_i= \exp(2\pi i \tau^i_0)$. The equation \eqref{qofz} then gives rise to a two-cycle in the FI-parameter space
\begin{align}
    \Xi_{\bm{e}} = \left\{q_i - \left(\frac{z}{z_0}\right)^{e^i}\bar q_i=0 \right\} \subset \mathcal{M}_\tau \,,
\end{align}
where we do not sum over the index $i$. Under the mirror map \eqref{mirrormap} this two-cycle is mapped to a two-cycle $\Sigma_{\bm{e}} \equiv \mathfrak{M} (\Xi_{\bm{e}}) \subset \cM_{\rm qK}$ in the quantum K\"ahler moduli space. Let us denote by the class $[\Xi_{\bm{e}}]$ the set of two-cycles which have in common their asymptotic behaviour in terms of the string charge $\bm{e}$, and $[\Sigma_{\bm{e}}]$ their image under the mirror map. A representative of the class is determined by $z_0$ and $\bar q_i$. The tension of the global string solution in Planck units is then given by the area  of the two-cycle $\Sigma_{\bm{e}}$ as calculated from the induced metric
\begin{align}\label{areatension}
  \frac{T_{\bm{e}}}{M_{\rm P}^2} = M_{\rm P}^2 \int_{\Sigma_{\bm{e}}} g_{t \bar t} dt\wedge d\bar t\,.
\end{align}
One can also compute this area by pulling back $J_\CM$ into $\CM_\tau$ and integrating it over $\Xi_{\bm{e}}$. There one can see that the area of a two-cycle in the class $[\Xi_{\bm{e}}]$  (or similarly in $[\Sigma_{\bm{e}}]$) is independent of the representative. Indeed, varying $\bar{q}_i$ in \eqref{qofz} will sweep a three-chain $\Pi_3 \subset \CM_\tau$ to which we can apply Stokes' theorem. Then one can use K\"ahlerity and that  $\mathfrak{M}^*(J_{\CM})$ vanishes in $\partial \Pi_3$ except on the two  representatives of $[\Xi_{\bm{e}}]$. As a result, the tension of the string solution does not depend on the position of the various strings. There is therefore a correspondence between the string charges and the classes of  genus zero holomorphic two-cycles.

It is instructive to construct a basis 
\begin{align}
\{\Xi_1, \Xi_2, \dots , \Xi_ {h^{1,1}}\}
\end{align} 
of two-cycles in $\cM_\tau$ by considering the elementary strings, obtained from NS5-branes with charge 
\begin{align}
    {\bm{e}}_i = \left(0, \dots, 0,\underbrace{1}_{i\text{-th}},0, \dots,0\right)\, .
\end{align}
To each of this basis element we then have a two-cycle $\Sigma_i \equiv \bm{t}(\Xi_{{\bm{e}}_i})$ in $\cM_{\rm qK}$. The charge ${\bm{e}}$ of a string then translates into the degree of $\Sigma_{\bm{e}}$ with respect to the basis of primitive 2-cycles $\Sigma_i$. 

The degree of these curves is in fact constrained. To see this, let us consider the finite cover $\tilde{\cM}_{\rm qK}$ of $\cM_{\rm qK}$ for which the monodromy group is neat (cf. page 41 of \cite{Cecotti:2020rjq}), i.e. all monodromies in $\tilde \cM_{\rm qK}$ are unipotent. This means that we remove all finite order monodromies of $\cM_{\rm qK}$ when moving to the cover $\tilde \cM_{\rm qK}$. We can further consider the covers of the curves $\Sigma_{i}$ defined via the projection
\begin{eqn}
    \pi: \qquad \tilde{\cM}_{\rm qK} &\to \cM_{\rm qK}\,,\\
    \pi(\tilde \Sigma_{i})&=\Sigma_{i}\,.  
\end{eqn}
Our general discussion in section \ref{sss:general} implies that the area of the curves  $\tilde{\Sigma}_i \subset \tilde \cM$ as measured by \eqref{areatension}  is given by $2\pi$ times an integer. Physically, the multi-cover $\tilde \Sigma_{i}$ can be achieved  by considering a setup with multiple NS5-strings and their corresponding regulators. The condition that the two-cycles $\Sigma_i$ have area $2\pi$ or below constrains the  two-cycle degree in $\cM_{\rm qK}$. Above some maximal degree, we will have a global  string solution with super-Planckian tension, which overcloses the transverse space.

We can also consider strings $S_{\bm{e}}$ with charge $\bm{e}$  wrapped on a loop of radius $L$. In this case, the backreaction again yields a map 
\begin{align}
    \bm{q}|_D : \quad D(L) \rightarrow \mathcal{M}_{\tau}\,.
\end{align}
 Here, $D(L) \subset \C$ is a bounded region transverse to the loop such that $\hat{L} \equiv M_{\rm P} L \in \partial D$, along which we get a non-trivial profile for our fields. The map $\bm{q}|_D $ embeds $D(L)$  in a two-cycle $\Xi_{\bm{e}} \subset \CM_\tau$, and via the mirror map \eqref{mirrormap} into a curve in the class $[\Sigma_{\bm{e}}]$. The correct representative is chosen such that 
\begin{align}
 \left(\mathfrak{M} \circ \bm{\tau} \right)(z=\hat{L}) \in \Sigma_{\bm{e}} \,,
\end{align}
which given the map \eqref{qofz} uniquely determines a representative of $[\Sigma_{\bm{e}}]$. The embedding of $\text{Im}(\mathfrak{M} \circ \bm{\tau})$ in the class $[\Sigma_{\bm{e}}]$ is thus well-defined. The tension of a string with charge $\bm{e}$ at the point $(\mathfrak{M} \circ \bm{\tau})(z=\hat{L})$ is determined by the area of $\im (\mathfrak{M} \circ \bm{\tau})[D(L)] \subset \Sigma_{\bm{e}}$:
\begin{align}
    \frac{T_{\bm{e}}(\bar{\bm{t}})}{M_{\rm P}^4} \simeq \text{Area}\,\left(\im (\mathfrak{M} \circ \bm{\tau})[D(L)] \right) \leq \text{Area}\,\left(\Sigma_{\bm{e}}\right) =  \frac{T_{\bm{e}}}{M_{\rm P}^4} \,.
\end{align}

\subsection{The quintic}
\label{ss:quintic}

To illustrate the above picture, let us consider type IIA compactified on the quintic. In this case there is a unique EFT string constructed from NS5-branes, arising from wrapping them on the single divisor class $D$ of the quintic.  The corresponding K\"ahler modulus is related to the single FI parameter $\tau$ of the GLSMs with gauge group $U(1)$ and charges 
\begin{align}\label{Qquintic}
    \underline{Q} = (-5,1,1,1,1,1)\,. 
\end{align}
As in the general case, to describe the backreaction of the NS5-brane along the transverse complex plane $\C$, we need to introduce an additional $U(1)$ gauge factor and two extra matter fields $\Phi^0, \tilde \Phi^0$. Identifying the scalar in the multiplet $\Phi^0$ with the coordinate $z \in \C$, the backreaction of the NS5-brane leads to a profile for $\tau$ given by 
\begin{align}
    \tau= \tau_0 + \frac{1}{2\pi i} \log \frac{z}{z_0} \,. 
\end{align}
The singular locus of the quintic is determined by \eqref{vacuacondition} which using \eqref{Qquintic} reads 
\begin{align}
    \sigma^5 = 5^5 q \sigma^5\ \implies \  1-5^5 q=0\, , 
\end{align}
with $q\equiv \exp(2\pi i \tau)$. 
The singular locus thus consists of a single point $q=\frac{1}{5^5}$ in the complex FI-parameter space. This point corresponds to the conifold point in the mirror quintic \cite{Berglund:1993ax}. From the viewpoint of the backreaction of the NS5-brane wrapped on $D$ we associate a regulator string to this singularity,  located at 
\begin{align}
    z_{\rm reg}= \frac{e^{-2\pi i \tau_0}z_0}{5^5}\,. 
\end{align}
The monodromies $M_{q=0}, M_{\Delta}$ around the large volume point/divisor $\{q=0\}$ and the conifold locus $\{\Delta=0\}$ generate the monodromy group. We can represent the two monodromies by their action on the central charges $(Z(\mathcal{O}_{Y_3}), Z(\mathcal{O}_D), Z(\mathcal{C}), Z(\mathcal{O}_{\rm pt}))$ of $D(2p)$-branes as described in appendix~\ref{app:monodromies} by 
\begin{align}
    M_{q=0}=\left(
\begin{array}{cccc}
 1 & 1 & 5 & 0 \\
 0 & 1 & 5 & 0 \\
 0 & 0 & 1 & 1 \\
 0 & 0 & 0 & 1 \\
\end{array}
\right) \,,\qquad M_{\Delta} = \left(
\begin{array}{cccc}
 1 & 0 & 0 & 0 \\
 -5 & 1 & 0 & 0 \\
 0 & 0 & 1 & 0 \\
 -1 & 0 & 0 & 1 \\
\end{array}
\right)\,. 
\end{align}
For $z\rightarrow \infty$ the backreacted solution approaches $q=\infty$,  corresponding to the mirror of the Landau--Ginzburg point, which is a point with monodromy
\begin{align}
    M_{q=\infty} =\left(
\begin{array}{cccc}
 1 & -1 & 0 & 0 \\
 5 & -4 & -5 & 5 \\
 0 & 0 & 1 & -1 \\
 1 & -1 & 0 & 1 \\
\end{array}
\right)\,. 
\end{align}
This monodromy is of order five, i.e. $(M_{q=\infty})^5 =1$. From here one can see that the backreaction of an NS5-brane wrapped on $D$ gives a one-to-one map from the transverse space into the quantum K\"ahler moduli space of the quintic. 
Indeed, the image of this map includes the three special points in the quintic moduli space corresponding to the (mirror of) the large complex structure point, the conifold point and the Landau--Ginzburg point. In the string backreaction  these points are interpreted respectively as the NS5-brane core, the regulator string and the negative tension string at spatial infinity, that compensates the deficit angle produced by the backreaction. The analogue of \eqref{totalM} is thus simply given by the identity $ M_{q=0} M_{\Delta} M_{q=\infty}=\text{Id}$.

At the level of $\cM_\tau$ we can determine the relative positions of the EFT and regulator strings, but in order to describe the full backreaction we need to know the details of the holomorphic function $f(z)$ and the map \eqref{mirrormap} to $\cM_{\rm qK}$. By itself, the analysis of the singular loci in $\cM_\tau$ does not tell us anything about the tension localised at the core of the regulator strings, but this information is encoded in the monodromy action. In the case of the conifold point we have a finite distance singularity with unipotent monodromy, as is typically the case for the component of the discriminant locus $\Delta$. As a result, the regulator string should not introduce a local deficit angle. In other words, a `conifold' string is expected to become tensionless as we approach the point  $q=\frac{1}{5^5}$ in moduli space.

In the present example we can easily confirm this expectation by using the mirror map  \eqref{mirrormap} in the vicinity of the conifold point, which can be obtained by  considering the periods of the mirror quintic close to it. Schematically these read \cite{Candelas:1990qd}
\begin{align}\label{Pic} 
\vec \Pi_c =\left(\begin{matrix}e^{2\pi i t_c}+ \mathcal{O}(e^{4\pi i t_c}) \\ a+ \mathcal{O}(e^{2\pi i t_c}) \\ b+ \mathcal{O}(e^{2\pi i t_c}) \\ t_c e^{2\pi i t_c} +\text{const} + \mathcal{O}(e^{2\pi i t_c}) \end{matrix}\right) \,.
\end{align}
Here $a\,,b$ are constants and we introduced the complex scalar field 
\begin{align}
    t_c\equiv \partial_{q_c}^2 \mathcal{F} = \frac{1}{2\pi i} \partial_{q_c}^2 \left(q_c^2 \log q_c + \text{const.} \right)=  \frac{1}{2\pi i} \log q_c + \text{const.}\,,
\end{align}
where $\mathcal{F}$ is the local prepotential in the vicinity of $q_c=0$, and the constant $c$ arises from the constant terms in the period \eqref{Pic}. Notice that $q_c$ is related to $q$ via 
\begin{align}
q_c = q-5^{-5}\,. 
\end{align} 
In the limit $t_c\rightarrow i \infty$, i.e. $q_c\rightarrow 0$, we reach the conifold point and it appears an approximate continuous shift symmetry $t_c\rightarrow t_c + \delta$ with $\delta\in \mathbbm{R}$, as the K\"ahler potential 
\begin{align}
e^{-K}\simeq  |q_c|^2 \log |q_c|^2 = -2 e^{-4\pi \text{Im} t_c} \text{Im}\, t_c + c +\mathcal{O}(e^{2\pi i t_c})\,.
\end{align}
is independent of $\re t_c$ at leading order. Notice that relative to the leading term, the terms that break the shift-symmetry are not exponentially suppressed, but only by $\mathcal{O}(t_c^{-1})$. Thus, the breaking terms already appear at the perturbative level, unlike in the case of EFT strings. Nevertheless using such approximate shift symmetry we can apply the approach in  \cite{Lanza:2020qmt,Lanza:2021udy} and dualise the saxion $\text{Im}\, t_c$ to a linear multiplet $\ell^c$ 
\begin{align}
 \ell^c = -\frac{1}{2} \frac{\partial K}{\partial \text{Im}\,t_c} = \frac{1}{2} \frac{8\pi \text{Im}\,t_c e^{-4\pi \text{Im}\,t_c}+e^{-4\pi \text{Im}\,t_c}}{-2\text{Im}\,t_c e^{-4\pi i\text{Im}\,t_c}+c} \,.
\end{align}
In the limit $t_c\rightarrow i\infty$ the linear multiplet $\ell^c$ thus scales as 
\begin{align}\label{tensionconifold}
\ell^c\to \frac{4\pi }{c} \text{Im}\,t_c\; e^{-4\pi \text{Im}\,t_c} + \mathcal{O}\left(e^{-4\pi \text{Im}\,t_c}\right)\,.  
\end{align} 
At the same time, the charge of this string under the 2-form dual to $\text{Re}\,t_c$ is given by 
\begin{align}
\mathcal{Q}^2 = M_{\rm P}^2 \partial_{\text{Im}\,t_c}^2 K = -\frac{\partial \ell^c}{\partial \text{Im}\,t_c} \simeq \frac{8\pi^2}{c} \text{Im}\,t_c\; e^{-4\pi \text{Im}\,t_c} \rightarrow 0 \, \qquad \text{as}\qquad t_c\rightarrow i \infty \,. 
\end{align} 
Thus, asymptotically, we get a tensionless string that is weakly coupled under the associated two-form gauge symmetry. We can also calculate its charge-to-mass ratio, finding 
\begin{align}
\frac{\mathcal{Q}^2}{T^2} \sim \frac{1}{\frac{2\pi }{c} \text{Im}\,t_c\; e^{-4\pi \text{Im}\,t_c}} \rightarrow \infty \,.
\end{align} 
This behaviour is actually not unexpected since at finite-distance singularities the ellipsoid describing the charge-to-mass ratio of particle states is expected to degenerate \cite{Bastian:2020egp}. For strings this should happen whenever there is a tensionless non-critical string.  

Finally, similarly to EFT strings \cite{Lanza:2021udy} we can now calculate the energy stored in a disk of radius $r$ centred around the conifold string. Using the approximate shift symmetry we can perform the same calculation as in the infinite distance case: 
\begin{eqn}
\mathcal{E}^c_\text{back}  = M_{\rm P}^2 \int_{D(r)} \mathcal{J}_\mathcal{M} = M_{\rm P}^2 \int_{D(r)} \mathrm{d}\ell^c \wedge \text{Re}\, t_c = M_{\rm P}^2 \left[\ell^c(r)-\ell^c(0)\right]= \frac{4\pi}{c} r^2 \log{r^2}\, ,
\end{eqn}
where from \eqref{tensionconifold} we know that $\ell^c(0)=0$. Notice that as expected the string does not have any energy localised at $r=0$  and therefore does not yield to a deficit angle at its position. 

While the regulator string is itself tensionless as the previous analysis revealed, the same is not true for the string at spatial infinity associated to the Landau--Ginzburg point. The monodromy associated to this string is of order 5 and hence we have a deficit angle associated to this string, which in order to satisfy \eqref{tadpole} must be negative:
\begin{align}
    \delta_\infty =\frac{T_\infty}{M_{\rm P}^2} = -  \frac{2\pi p}{5}\, , 
\end{align}
for some $p\in \bN$. This deficit angle compensates the one created by the combined backreaction of the NS5-brane core and the regulator string. Due to the identification of the tension of the global string solution with the area of the associated two-cycle \eqref{areatension}, we find that the volume of the two-cycle $\Sigma_1$ associated to a single NS5-brane charge in the single divisor class of the quintic is given by 
\begin{align}
   M_{\rm P}^2\, \text{Area}(\Sigma_1) = \frac{2\pi p}{5} \,.
\end{align}
The example of the quintic also nicely illustrates the constraint on the degree of the equation \eqref{qofz} describing the two-cycles. Indeed, the asymptotic deficit angle of the minimal string solution is given by $|\delta_\infty|=\frac{2\pi p}{5}$, so by taking several copies one would exceed a tension of $2\pi M_{\rm P}^2$. In particular, by taking a five-fold cover of the quintic K\"ahler moduli one  obtains an area
\begin{align}\label{AreaSigma5}
   \text{Area}(\Sigma_5)= 5\cdot  \text{Area}(\Sigma_1) = \frac{2\pi p}{M_{\rm P}^{2}}\,,
\end{align}
in accordance with previous results in the literature. Indeed, by the results of \cite{MR2344329,MR2191883,Douglas:2006zj}, $p$ corresponds to the Euler characteristic of the first Hodge bundle $\mathcal{F}^3=H^{3,0}$ in the mirror quintic. Applying the results of \cite{Greene:1989ya,Green:1993zr} to the quintic, we show in appendix \ref{apsec:quintic} that one can fix $p=1$.


\section{Examples}
\label{s:examples}

In this section we consider examples of string solutions beyond the simple one-modulus case discussed in the previous section. More precisely, we consider type IIA (or heterotic string with standard embedding) string theory compactified on Calabi--Yau threefolds with $h^{1,1}=2$ and $h^{1,1}=3$, and analyse the backreaction of different NS5-strings. While the examples that we look at are quite simple, the string solutions already exhibit interesting features.

Indeed, one key aspect of examples with a higher-dimensional moduli space $\CM$ is that several regulator strings may be involved in the backreaction of an NS5-brane. Geometrically, this is because the two-cycle $\Sigma$ associated to the string solution typically intersects several times the singular divisors of $\CM$. In particular, when $\Sigma$ intersects the same divisor several times the corresponding regulator strings correspond to conjugate, non-commuting monodromies that are the analogues of mutually non-local 7-branes in F-theory. Thanks to this non-commutative feature one may combine several  infinite-order monodromies into a finite-order one,  as needed for a sub-Planckian BPS string configuration.

\subsection{\texorpdfstring{Calabi--Yau with $h^{1,1}=2$}{Calabi--Yau with h11=2}}
\label{sec:h112}

\begin{figure}[!t]
    \centering
     \begin{subfigure}[b]{0.46\textwidth}
         \centering
         \includegraphics[width=\textwidth]{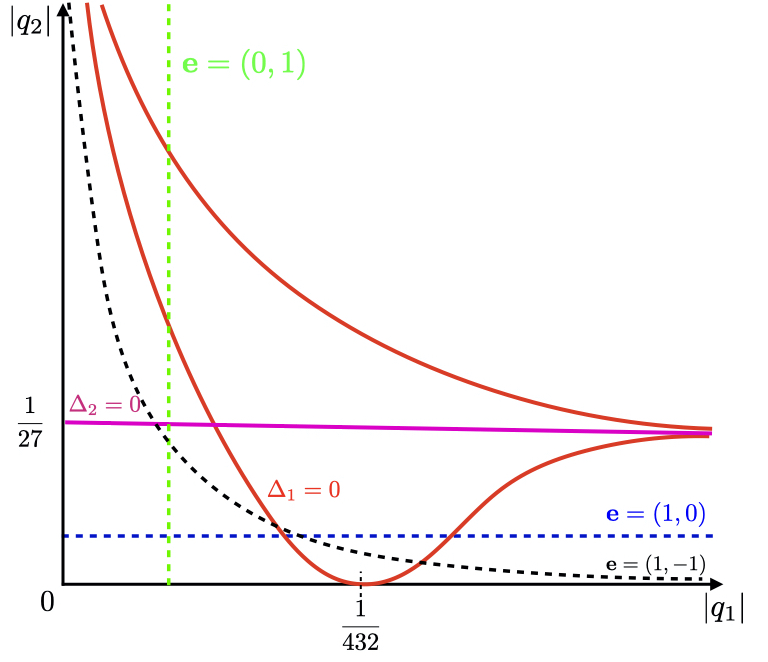}
         \caption{}
         \label{fig:modulispaceq}
     \end{subfigure}
     \hspace{0.03\textwidth}
     \begin{subfigure}[b]{0.48\textwidth}
         \centering
         \includegraphics[width=\textwidth]{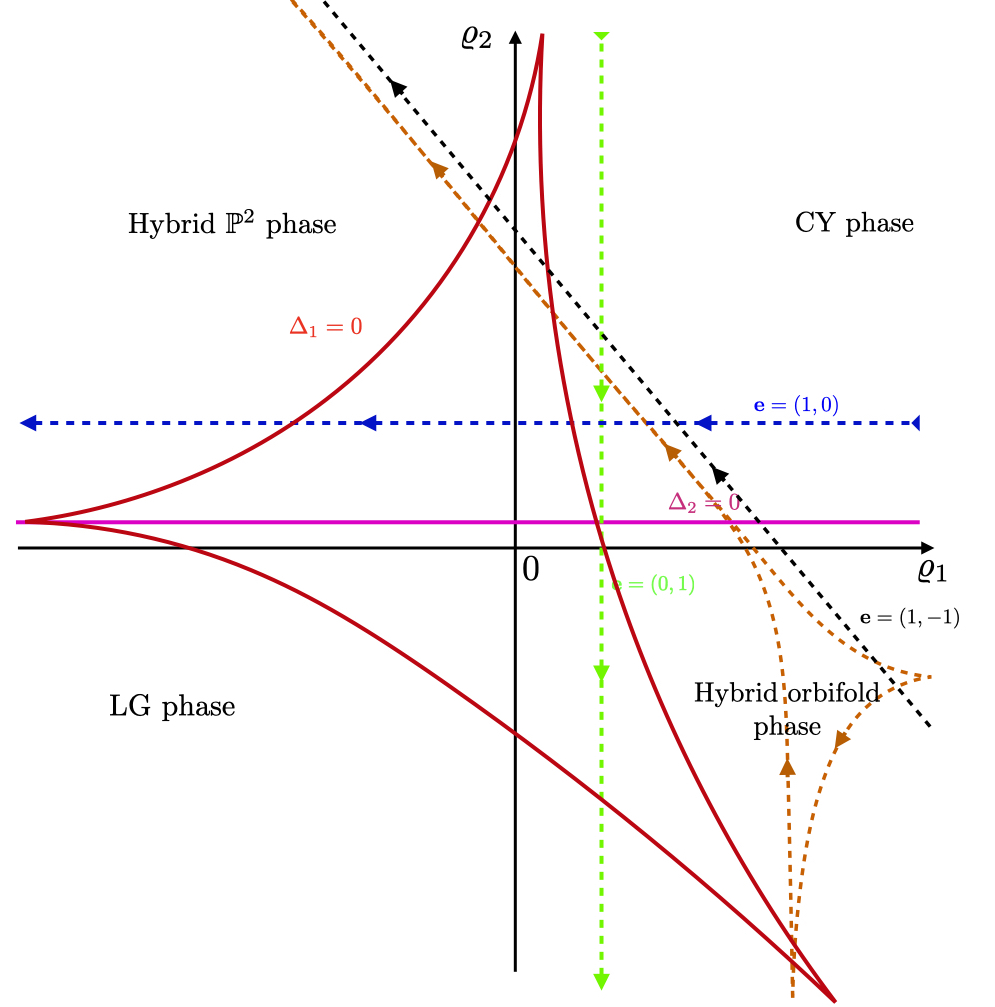}
         \caption{}
         \label{fig:modulispacelogq}
     \end{subfigure}
    \caption{Two sketches of the FI-parameter space. On the left we show the singular divisors $\{\Delta_1=0\}$ (red) and $\{\Delta_2=0\}$ (magenta) for $q_1,q_2\in \mathbb{R}$ in the $(|q_1|,|q_2|)$-plane. The right shows the same in terms of $\varrho^i = - \frac{1}{2\pi} \log |q_i|$. On the right we also indicated the four different weak-coupling phases of the GLSM. The dashed lines correspond to the projection of the solution profiles for EFT strings with charges $\bm{e}=(1,0)$ (blue) and $\bm{e}=(0,1)$ (green). We have also included a non-EFT string with charge $\bm{e}=(1,-1)$, plotting both the naive profile (black) and the actual profile (orange) that reproduces the large-volume NS5-brane monodromy. The arrows indicate the embedding of the space-time $\mathbb{P}^1$ into the two-cycle in the FI-parameter space. Following the arrows one moves from the string core at $z=0$ to $z=\infty$. From the figure it is easy to see that the set of two-cycles associated to an elementary string sweeps the whole FI-parameter space. The same occurs for the actual moduli space $\CM$, where they form a complex-dimension-one foliation over the corresponding divisor at infinity.}
    \label{fig:modulispace}
\end{figure}

Let us consider the Calabi--Yau threefold $Y_3$ obtained as the degree-18 hypersurface in the weighted projective space $\P_{1,1,1,6,9}$ \cite{Candelas:1994hw}, whose EFT strings were analysed in \cite{Lanza:2021udy}. This threefold arises as the target space of a NLSM in the IR limit of a GLSM with gauge group $U(1)^2$ and seven matter fields $(\Phi_\alpha, \Gamma^\alpha)$\,, $\alpha=1,\dots 7$ with charge matrix given by 
\begin{align}
\underline{Q} = \left(\begin{matrix} -6 & 2 & 3&1&0&0&0 \\ 0&0&0&-3&1&1&1\end{matrix}\right)\,. 
\end{align}
From here, it is apparent that $Y_3$ can be described as a smooth Weierstrass model with base $B_2 = \P^2$. Accordingly, we have two divisor classes corresponding to the zero section $E$ of the Weierstrass model and the vertical divisor $D_2$ obtained by pulling back the hyperplane class $h\in H^{2}(\P^2)$ to the fibre, i.e. $D_2=\pi^* h$, where 
\begin{align}
 \pi: Y_3 \rightarrow B_2\,,
\end{align}
is the projection from the fibre to the base of $Y_3$. Since $Y_3$ is a smooth Weierstrass model, the K\"ahler form of $Y_3$ can be expanded as 
\begin{align}
 J_{Y_3} = t^1 \underbrace{(E+3D_2)}_{=:D_1} + t^2 D_2\,,
\end{align}
where we introduced the second K\"ahler cone generator $D_1$ that is associated to the zero section $E$ via a shift by $\pi^*c_1(B_2) = 3D_2$. The quantum K\"ahler moduli space and the associated monodromies of this model have been investigated in detail in e.g. \cite{Aspinwall:2001zq}. We can now consider NS5-branes wrapped on the Nef divisors 
\begin{align}\label{Dgeneralh2}
D = e^1 D_1 + e^2 D_2\,, 
\end{align}
for $e^1, e^2\in \mathbbm{Z}_{\geq 0}$. As discussed in \cite{Lanza:2021udy} in the large volume regime the backreaction of these NS5-branes gives a logarithmic profile of the form \eqref{tsol} for the two K\"ahler moduli $(t^1,t^2)$. Associated to these moduli we have the two complex FI parameters $\tau^i$ of the GLSM for the two $U(1)$ gauge factors. Let us denote their exponentiated form by $q_i = \exp(2\pi i \tau^i)$, as in \eqref{FIexp}. In terms of these variables the backreaction is given by \eqref{qofz}, which for the present case reads 
\begin{align}\label{backh2}
 q_1 = \left(\frac{z}{z_0}\right)^{e^1}\bar{q}_1 \,,\qquad q_2 = \left(\frac{z}{z_0}\right)^{e^2}\bar{q}_2\,,
\end{align}
where as before $z \in \mathbbm{C}$ is the 4d coordinate transverse to the NS5-brane core, located at $z=0$. As in the general case discussed in section \ref{sss:general}, we now have additional strings that regulate the backreaction of the NS5-string. These regulator strings are governed by the singularities in the FI-parameter space.   The principal component of the singular locus can be obtained from \eqref{vacuacondition} which in this case yields two equations: 
\begin{align}\label{vacuaconditionh2}
    \left(\sigma_1 -3\sigma_2\right)=432 q_1\sigma_1 \,,\qquad \sigma_2^3=q_2\left(\sigma_1-3\sigma_2\right)^3\,. 
\end{align}
These equations have a solution provided that
\begin{align}\label{Deltah2}
 \Delta_1 \equiv \left(1-432q_1\right)^3 -27q_2 \cdot (432 q_1)^3 = 0 \,,
\end{align}
which defines the principal component of the discriminant divisor. Besides \eqref{Deltah2} there is a second singularity that can be obtained as the principal component of the discriminant locus for the one-parameter system corresponding to the base $\mathbb{P}^2$. This second singular locus is defined by 
\begin{align}
    \Delta_2 \equiv 1 +27q_2=0 \,. 
\end{align}
The discriminant locus in $\cM_\tau$ can be visualised by its projection to the $(|q_1|,|q_2|)$  plane  as in figure \ref{fig:modulispace}, or equivalently to the  $(\varrho^1,\varrho^2)$-plane with $\varrho^i = - \frac{1}{2\pi} \log |q_i|$. More precisely, the shape of the amoeba corresponding to $\Delta_1=0$ is obtained from \eqref{vacuaconditionh2} by considering $(q_1,q_2)$ as a function of $\zeta =\frac{\sigma_2}{\sigma_1}\in \mathbb{R}$. The part of the divisor $\{\Delta_1=0\}$ that does not correspond to $\zeta\in \mathbb{R}$ is projected into the interior of the amoeba. In figure \ref{fig:modulispaceq}, the string solutions \eqref{backh2} reduce to real lines determined by the projection of the complex graph
\begin{align}
q_2 = \left(\frac{q_1}{\bar q_1}\right)^{e^2/e^1}\bar{q}_2\,. 
\end{align}
to the $(|q_1|,|q_2|)$-plane. 
On the one hand, the solution to the equation $\Delta_2=0$ indicates the presence of  regulator strings at the $e^2$ solutions of the equation 
\begin{align}\label{ze2}
 \left(\frac{z}{z_0}\right)^{e^2} = -\frac{1}{27  \bar q_2}\,.
\end{align}
On the other hand, the equation $\Delta_1=0$ is a polynomial of order three in $q_1$ and therefore its solution will in general yield multiple strings.

For the present case the monodromy group $\Gamma\subset Sp(6,\mathbb{Z})$ is generated by three elements which we denote by
\begin{align}\label{generatorGammah2}
   \Gamma =\langle M_{q_2=0}, M_{\Delta_1}, M_{\Delta_2} \rangle\,,
\end{align}
which, respectively, correspond to the monodromies around the large volume divisors $\{q_2=0\}$ and the two components of the discriminant locus $\{\Delta_i=0\}$. We give an explicit representation for $\Gamma$ in appendix~\ref{app:monodromies}, where one can see that each of these three elements are unipotent monodromies. One might have been tempted to also add $M_{q_1=0}$, i.e. the monodromy around the divisor $\{q_1=0\}$, to the set of generators. However, as we show in appendix~\ref{app:monodromies},  $M_{q_1=0}$ can be written as a combination of $M_{\Delta_2}$ and $M_{q_2=0}$. 
Given the monodromy group, we can now analyse different classes of string solutions, classified by their NS5-brane charges \eqref{Dgeneralh2}:

\subsubsection*{\texorpdfstring{EFT strings  $\bm{e} =(1,0)$}{EFT strings  e=(1,0)}}

Let us first focus on the elementary EFT string charge $\bm{e} =(1,0)$, i.e. we have a string wrapping only the divisor $D_1$. The monodromy $M_\text{\tiny EFT}$ appearing in \eqref{totalM} is thus given by $M_{q_1=0}$. In this case \eqref{backh2} tells us that $q_2$ is just a constant. For $q_2 =\bar{q}_2$ constant, the defining equation of the singular divisor $\{\Delta_1=0\}$ has three solutions, given by 
\begin{eqn}\label{q0pm}
 q_1^0 &= \frac{1}{432(1-27\bar{q}_2)} \left[1+3\bar{q}_2^{1/3} - 9 \bar{q}_2^{2/3} \right]\,,\\
 q_1^+ &= \frac{1}{432(1-27\bar{q}_2)} \left(1-3\bar{q}_2^{1/3} \right)\left(1+\frac{3}{2} (1+i\sqrt{3})\bar{q}_2^{2/3}\right)\,,\\
  q_1^- &=  \frac{1}{432(1-27\bar{q}_2)} \left(1-3\bar{q}_2^{1/3} \right)\left(1+\frac{3}{2} (1-i\sqrt{3})\bar{q}_2^{2/3}\right)\,. 
\end{eqn}
Accordingly, there are three regulator strings in this case, located at 
\begin{align}\label{z0pm}
 z^{0, \pm} = \frac{q_1^{0,\pm}}{ \bar q_1}\,z_0 \,.
\end{align}
Notice that the position of these regulator strings depend on $\bar{q}_1,\bar{q}_2$ and $z_0$.  Let us denote the monodromy around the three strings by $M_{\Delta_1^{0,\pm}}$. These monodromies are in the same conjugacy class $[M_{\Delta_1}]$, associated to the divisor $\{\Delta_1=0\}$. However, as worked out in detail in \cite{Aspinwall:2001zq} the representatives of this class differ  for the three strings at $z^{0,\pm}$ by conjugation with $M_{q_2=0}$. Here  $M_{q_2=0}$ stands for the monodromy associated to the large-volume limit divisor $\{q_2=0\}$, corresponding to the usual $B$-field shift \eqref{tmon} with $(e^1,e^2) = (0,1)$ -- see appendix~\ref{app:monodromies} for an explicit representation. It turns out that for each point in \eqref{z0pm} we can choose 
\begin{align}\label{Mdelta1reps}
    M_{\Delta_1^-}= M_{\Delta_1} \,,\quad M_{\Delta_1^0}= (M_{q_2=0})^{-2} M_{\Delta_1}(M_{q_2=0})^2\,,\quad M_{\Delta_1^+}=(M_{q_2=0})^{-1} M_{\Delta_1}M_{q_2=0}\,. 
\end{align}
To see why this is the case, let us study the separation of the regulator strings as a function of the string solution parameters $(\bar{q}_1, \bar{q}_2)$. For instance, one may ask if and when it is possible that (a subset of) these regulator strings coincide. From \eqref{q0pm} it is clear that this can be only achieved for $\bar{q}_2=0$ where in fact all three strings coincide.\footnote{This could also be achieved by setting $z_0=0, \infty$, but in this case the profile \eqref{backh2} maps the entire $\mathbb{P}^1$ to a single point $q_1=\infty,0$, respectively.} This is already apparent from the form of $\Delta_1$ in \eqref{Deltah2} which at $q_2=0$ reduces to 
\begin{align}
 \Delta_1|_{q_2=0} = (1-432q_1)^3\,,
\end{align}
which has a triple zero at $q_1 =1/432$. Since all three strings correspond to $\{\Delta_1 =0\}$ they are associated to unipotent monodromies and hence do not induce a local deficit angle. Moreover, using the techniques of \cite{Grimm:2018ohb} one can check that each individual string corresponds to a finite distance singularity in moduli space. As we tune $\bar{q}_2$ to smaller values the three strings approach each other and for $\bar{q}_2=0$ eventually coincide at 
\begin{align}\label{z0pmq20}
   z^{0,\pm}|_{\bar{q}_2=0} =\frac{1}{432 \bar{q}_1}z_0\,. 
\end{align}
The monodromy around the resulting string at this point is also of infinite order, but unlike $M_{\Delta_1}$ it must correspond to an infinite-distance monodromy. To see this notice that along the divisor $q_2=0$ the mirror map $\mathfrak{M}$ is given by  \cite{Aspinwall:1999ii}
\begin{align}\label{mirrormapSL2}
    j(t^1) = \frac{1}{1728q_1(1-432 q_1)}\,,
\end{align}
where on the lhs we have the modular $j$-function in the conventions of section \ref{sec:toy}, with argument the complex K\"ahler modulus $t^1$. Accordingly, the divisor $\{q_2=0\}$ is a double cover of the $SL(2, \mathbb{Z})_{t^1}$ fundamental domain, and so is the two-cycle $\Sigma_{(1,0)}$ swept by the string solution in the limit $\bar{q}_2=0$. Via \eqref{mirrormapSL2}, the point $q_1=\frac{1}{432}$ corresponds to the cusp at $(t^1,t^2)=(0,i \infty)$, which lies at infinite distance. Thus, the combined monodromy around the three regulator strings cannot simply be the cube of a representative of $[M_{\Delta_1}]$, but the individual monodromies around the three strings need to be conjugates of each other. Since the three strings coincide precisely on the locus $\bar{q}_2=0$ the monodromies in fact should be conjugated w.r.t. $M_{q_2=0}$ \cite{Aspinwall:2001zq}, as it happens in \eqref{Mdelta1reps}. Using the generators of the monodromy group given in appendix~\ref{app:monodromies}, the combined monodromy around the three strings can be calculated to be
\begin{align}\label{Mreg}
     M_{\rm reg}=M_{\Delta_1^0} M_{\Delta_1^+} M_{\Delta_1^-} =\left(
\begin{array}{cccccc}
 1 & 0 & -3 & 0 & 3 & 0 \\
 -1 & 1 & 3 & 0 & -12 & 0 \\
 0 & 0 & 1 & 0 & -3 & 0 \\
 0 & 0 & -1 & 1 & 1 & 0 \\
 0 & 0 & 0 & 0 & 1 & 0 \\
 0 & 0 & 0 & 0 & -1 & 1 \\
\end{array}
\right)\,,
\end{align}
which satisfies $(\text{Id}-M_{\rm reg})^n = 0$ for $n\geq 4$, and so it indeed corresponds to an infinite-distance monodromy\cite{Grimm:2018ohb}.  

In general, for $\bm{e}=(1,0)$ we can give a simple dictionary between the position of the three regulator strings and the free parameters $(\bar q_1, \bar q_2)$ of the solutions \eqref{backh2}.\footnote{Notice that the parameter $z_0$ can always be absorbed in a redefinition of $\bar q_1$.} The  parameter $\bar{q}_1$ determines the position of the three coincident strings at $\bar{q}_2=0$ via \eqref{z0pmq20}. As we switch on a non-vanishing $\bar{q}_2$ the three regulator strings separate from the point \eqref{z0pmq20} and from each other, as captured by \eqref{z0pm}. This separation is reminiscent of the splitting of the O7-plane locus into 7-branes in type IIB/F-theory. There, Sen's limit is given by  $\bar{\tau}_{\rm IIB}\rightarrow i \infty$, where ${\tau}_{\rm IIB}$ is the type IIB axio-dilaton,  and $\bar \tau_{\rm IIB}$ can be thought of as the analogue of $\frac{1}{2\pi i} \log \bar{q}_2$. Away from this limit, the orientifold locus splits into two components corresponding to a $[0,1]$- and a $[2,-1]$-7-brane. The monodromies around these two branes correspond to different representative of the same conjugacy class, which is why the two branes are considered to be mutually non-local. The relative position of these two 7-branes is determined by $\bar{\tau}_{\rm IIB}$. One may then interpret  the limit $\bar{q}_2\rightarrow 0$ as the analogue of  Sen's limit, and the splitting of the single string into three components  for finite $\bar{q}_2$ as the analogue of the O7-plane splitting in F-theory. Note however that the O7-plane splits into two 7-branes associated to infinite-distance monodromies,  whereas in our setup the component strings are associated to finite-distance divisors. 

\begin{figure}[!t]
    \centering
    \includegraphics[width=0.55\textwidth]{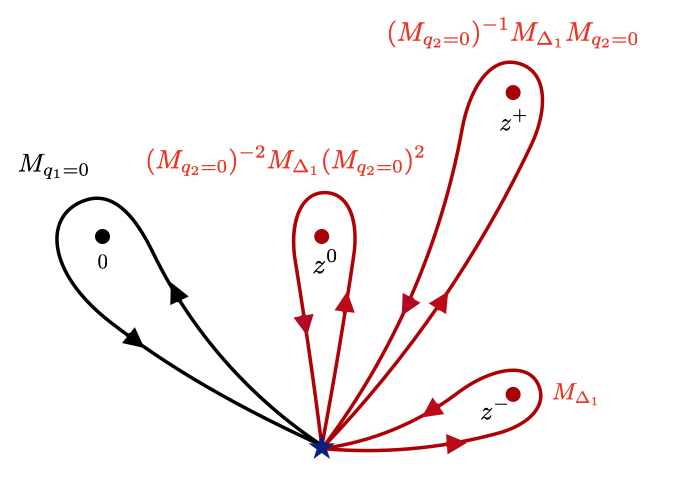}
    \caption{Positions of the regulator strings for the charge $\bm{e}=(1,0)$, for $1/432 \gg |\bar{q}_1|>0$ and $1/27 \gg|\bar{q}_2|>0$ together with the monodromies around them as measured along a loop encircling the strings counterclockwise.}
    \label{fig:stringpositions10}
\end{figure}

For $1/432 \gg |\bar{q}_1|>0$ and $1/27 \gg|\bar{q}_2|>0$ the positions of the EFT string and the three regulator strings is as illustrated in figure \ref{fig:stringpositions10}. The combined monodromy around the regulator strings and the string at $z=0$ corresponds to the inverse of the monodromy $M_{q_1=\infty}$ around the divisor $\{q_1=\infty\}$:
\begin{align}\label{M(1,0)}
    (M_{q_1=\infty})^{-1} = M_{q_1=0} (M_{q_2=0})^{-2} M_{\Delta_1}M_{q_2=0}M_{\Delta_1}M_{q_2=0}M_{\Delta_1}\,,
\end{align}
with $M_{\text{\tiny EFT}}=M_{q_1=0}$ the monodromy around the large volume-divisor $\{q_1=0\}$. The monodromy $M_{q_1=\infty}$ satisfies
\begin{align}
    M_{q_1=\infty}^6=\text{Id}\,. 
\end{align}
In this case the monodromy $M_{z=\infty}$ appearing in \eqref{totalM} is simply given by $M_{z=\infty}=M_{q_1=\infty}$ and we find 
\begin{align}
    M_{\text{\tiny EFT}} \left[\prod_{i\in \{0,\pm\}} M_{\Delta_1^i}\right] M_{z=\infty} = \text{Id}\,,
\end{align}
in accordance with the general expression  \eqref{totalM}.

Finally, let us discuss the tension of this system. As before, we may compute the global tension by  considering the point at $z=\infty$, which in this case corresponds to $q_1=\infty$. At this point we have an orbifold singularity with monodromy $M_{q_1=\infty}$ of finite order given by \eqref{M(1,0)}. Accordingly, the tension of a fully backreacted and regulated NS5-brane on $D_1$ is given by 
\begin{align}\label{Te1}
 \frac{T_{(1,0)}}{M_{\rm P}^2} = \frac{2\pi p_1}{6}\,. 
\end{align}
 The full solution corresponds to a two-cycle $\Sigma_{(1,0)} \subset \CM$. Along this two-cycle only $t^1$ varies and only infinite-order monodromies are involved, so \eqref{Te1} should match the area of $\Sigma_{(1,0)}$. As mentioned above, at $\bar{q}_2=0$ $\Sigma_{(1,0)}$ is a double cover of the $SL(2,\mathbbm{Z})$ $j$-line, corresponding to the fundamental domain of $SL(2,\mathbbm{Z})$ plus its image under the standard $S$-transformation. Each $SL(2,\mathbbm{Z})$ fundamental domain has area $\frac{\pi}{6}$. Therefore the area of the double cover indeed matches \eqref{Te1} for $p_1=1$. 

\subsubsection*{\texorpdfstring{EFT strings  $\bm{e} =(0,1)$}{EFT strings  e =(0,1)}}

\begin{figure}[!t]
    \centering
    \includegraphics[width=0.6\textwidth]{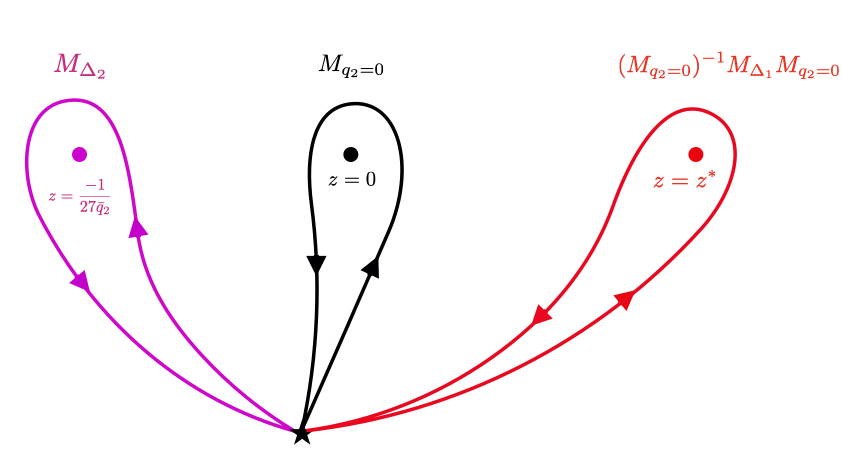}
    \caption{Positions of the regulator strings at $z=\frac{-1}{27\bar{q}_2}$ and $z=z^*$ relative to the EFT string with charge $\bm{e}=(0,1)$ at $z=0$. Here we chose $|\bar q_2|>0$ and $1/432 > |\bar q_1|$. We also indicated the monodromies when encircling the three strings counter-clockwise with a common base point.}
    \label{fig:stringpositions}
\end{figure}

Let us now focus on the elementary charge $\bm{e} =(0,1)$ such that $M_\text{\tiny EFT}=M_{q_2=0}$. In this case we have two different kinds of regulator strings, as the two-cycle intersects two different components of $\Delta$. The first corresponds to the singularity $\Delta_2=0$ and its location is determined via  \eqref{ze2}. The second comes from the singularity $\Delta_1=0$, which in this case gives a single regulator string. To see this, note that for fixed $q_1$ the equation $\Delta_1=0$ is solved by 
\begin{align}
q_2^*=\frac{(1-432 q_1)^3}{27\cdot 432^3  q_1^3}\,,
\end{align} 
which is the only solution. Accordingly, the location of this second regulator string is given by 
\begin{align}
\frac{z^*}{z_0} =  \left(\frac{q_2^*}{\bar{q}_2} \right)\,.
\end{align}
We illustrated the position of the three regulator strings in figure~ \ref{fig:stringpositions}. 

Again, we are interested in the global tension of the system, which one should be able to read off from the asymptotic deficit angle of the monodromy $M_{z=\infty}$ at infinity. In the present case this monodromy is given by the monodromy $M_{q_2=\infty}$ around the divisor $\{q_2=\infty\}$. The inverse of this monodromy can be composed by the monodromies around the divisors $\{q_2=0\}$, $\{\Delta_1=0\}$ and $\{\Delta_2=0\}$. These three monodromies are illustrated in figure \ref{fig:stringpositions}. Their combined monodromy is given by 
\begin{align}\label{Mq2infty}
    (M_{q_2=\infty})^{-1}= M_{\Delta_2} M_{\Delta_1}M_{q_2=0}\,.
\end{align}
The monodromy around $\{\Delta_2=0\}$ can be calculated by noticing that at $\Delta_2=0$ the central charge of the zero section $E$ vanishes. In appendix~\ref{app:monodromies} we review how to obtain this monodromy and give an explicit representation for it through its action on type IIA D$(2p)$-branes.

In there we also give an explicit expression for $M_{q_2=\infty}$ which satisfies \cite{Aspinwall:2001zq}
\begin{align}
 (M_{q_2=\infty})^{18} = \text{Id} \,,
\end{align}
and is therefore of order 18. This reflects the $\mathbb{C}^5/\mathbb{Z}_{18}$ target space of the LG model. Accordingly, the asymptotic deficit angle at $z\rightarrow \infty$ is given by $2\pi/18$ and the tension of a string with general charges $\bm{e}=(0,1)$ is
\begin{align}\label{T0e2}
 \frac{T_{(0,e^2)}}{M_{\rm P}^2}=\frac{2\pi p_2}{18}\,.
\end{align}
As shown also in appendix~\ref{app:monodromies} the monodromies $M_{q_1=\infty}$ for (1,0) strings and $M_{q_2=\infty}$ for (0,1) are related via 
\begin{align}
    M_{q_1=\infty}=(M_{q_2=\infty})^3\,,
    \label{relqinf}
\end{align}
from where we deduce that $p_2 =1$. 

\subsubsection*{\texorpdfstring{EFT strings  $\bm{e} =(e^1,e^2)$}{EFT strings  e=(e1,e2)}}

Let us now come back to the more general setups with NS5-branes wrapped on $D$ as in \eqref{Dgeneralh2} with $e^1, e^2 > 0$, whose  backreaction is determined by \eqref{backh2}. In this case the necessary regulator strings are determined by the equation 
\begin{align}
    \Delta = \left[\left((1-432 \bar{q}_1 \left(\frac{z}{z_0}\right)^{e^1}\right)^3 -27 \cdot 432^3  \bar{q}_1^3\bar{q}_2\left(\frac{z}{z_0}\right)^{3 e^1+e^2}\right]\left[1+27 \bar{q}_2\left(\frac{z}{z_0}\right)^{e^2}\right]=0\,.
\end{align}
 This equation is a polynomial of degree 
\begin{align}
    d= 3e^1 + 2e^2\,,
\end{align} 
in $z$. Due to the factorised form of $\Delta$ we necessarily have a regulator string located at 
\begin{align}
    \frac{z^*}{z_0} = \left(-\frac{1}{27\bar{q}_2}\right)^{1/e^2}\,,
\end{align}
in agreement with \eqref{ze2}. The first factor in $\Delta$ is a polynomial of degree 
\begin{align}
    d' = 3e^1 +e^2\,.
\end{align}
This factor will therefore determine the position of $d'$ additional regulator strings. 

\begin{figure}[!t]
    \centering
    \includegraphics[width=0.8\textwidth]{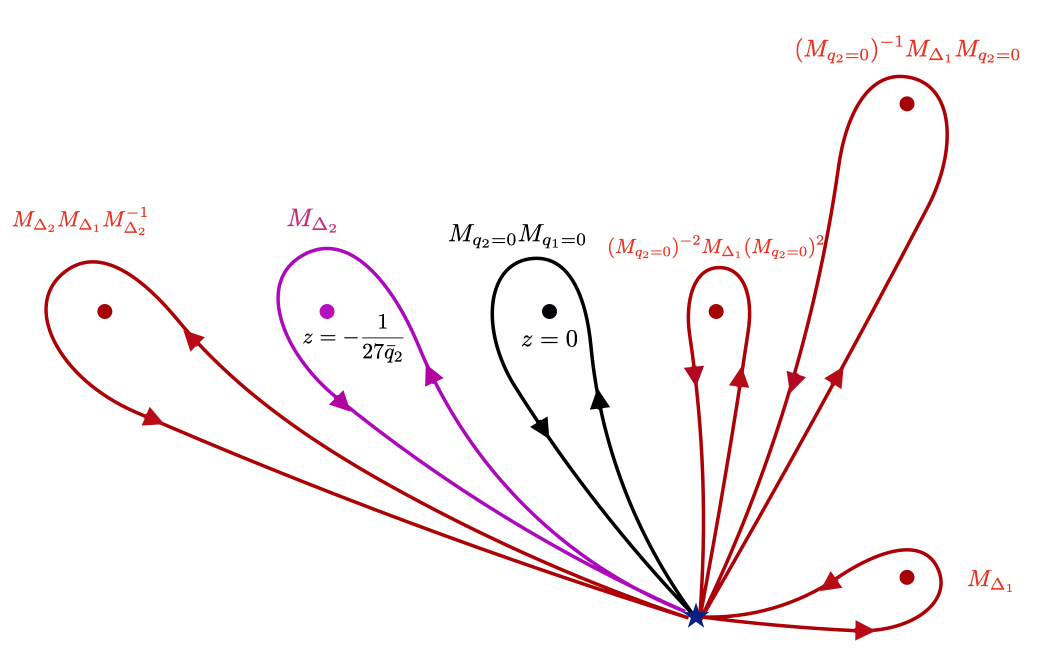}
    \caption{Positions of the regulator string for string solutions with $\bm{e}=(1,1)$ and $1\gg |\bar q_1|\gg |\bar q_2|>0$, together with the monodromies around them as measured along a loop encircling the strings counterclockwise.}
    \label{fig:stringpositions11}
\end{figure}

Let us consider the combined monodromy of the regulator strings and the NS5-brane strings. For simplicity let us focus on the case $\bm{e}=(1,1)$ and the region of parameter space $1\gg |\bar q_1|\gg |\bar q_2|>0$. In this case we have five regulator strings at positions illustrated in figure \ref{fig:stringpositions11}. From that figure we can read off the combined monodromy of the set of regulator strings and the NS5-string core at $z=0$ with monodromy $M_{\text{\tiny EFT}} = M_{q_2=0}M_{q_1=0}$: 
\begin{align}
    M_{\bm{e}=(1,1)} = [M_{\Delta_1}] M_{\Delta_2}  M_{q_2=0}M_{q_1=0}([M_{\Delta_1}])^3\,,
\end{align}
where, in order to avoid clutter, we did not distinguish the different representative of the conjugacy class $[M_{\Delta_1}]$, but refer to \eqref{M(1,0)} for the exact expression which yields $M_{q_1=0}([M_{\Delta_1}])^3=(M_{q_1=\infty})^{-1}$. We can further use \eqref{Mq2infty} to obtain
\begin{align}
M_{\bm{e}=(1,1)} = (M_{q_2=\infty} M_{q_1=\infty})^{-1} \,.
\end{align}
This monodromy is cancelled by the one of two finite-tension string cores corresponding to $\{q_1=\infty\}$ and $\{q_2=\infty\}$ with respective tensions $\delta_\infty = \frac{2\pi}{6}$ and $\delta_\infty = \frac{2\pi }{18}$. Recall that the monodromies $M_{q_1=\infty}$ and $M_{q_2=\infty}$ are related via \eqref{relqinf}, such that $M_{\bm{e}=(1,1)} = (M_{q_2=\infty})^{-4}=M_{z=\infty}^{-1}$. 

We can repeat this analysis in the case of general string charge $\bm{e}=(e^1,e^2)$. Since we know the tension of the elementary EFT string, by using BPSness we deduce that the tension of the full string solution is  given by 
\begin{align}
\frac{T_{(e^1,e^2)}}{2\pi M_{\rm P}^2} =\left(\frac{e^1}{6} + \frac{e^2}{18}\right)\,.
\end{align}
In geometric terms, the full solution corresponds to a holomorphic two-cycle $\Sigma_{(e^1,e^2)}$ whose class can be decomposed in terms of elementary classes 
\begin{align}
    [\Sigma_{(e^1,e^2)}] = e^1 [\Sigma_{(1,0)}] + e^2 [\Sigma_{(0,1)}]\,. 
\end{align}
The addition of tensions implies that the area as computed by \eqref{areatension} is  simply given by 
\begin{align}
    \text{Area}\left(\Sigma_{(e^1,e^2)}\right) = 2\pi\left(\frac{e^1}{6} + \frac{e^2}{18}\right)\,,
\end{align}
which is consistent with the fact that the two monodromies at infinity commute, as it follows from \eqref{relqinf}. The condition that the tension of this global string solution remains sub-Planckian constrains the cone of EFT string charges as
\begin{align}\label{constraintsubPlanckian}
     \frac{e^1}{6} + \frac{e^2}{18} < 1\,. 
\end{align}

Notice that to describe the backreaction of the EFT strings we do not need the full monodromy group $\Gamma$ but only a finite-index subgroup $\Gamma_\text{\tiny EFT}$. To see this, we note that in the backreaction none of the regulator strings associated to $\Delta_1$ or $\Delta_2$ appears on their own. Instead we either have groups of three regulator strings associated to $\Delta_1$, or one $\Delta_1$ regulator string together with a $\Delta_2$ regulator string. Accordingly, the subgroup $\Gamma_\text{\tiny EFT}$ is not generated by the elements in \eqref{generatorGammah2} but by the three elements 
\begin{align}
    \Gamma_\text{\tiny EFT} = \langle M_{q_1=0}, M_{q_2=0}, M_{q_2=\infty}\rangle\,.
\end{align}
The combined monodromy around the two sets of regulator strings appearing in the EFT string solutions can be related to these generators via 
\begin{eqn}
    M_{\rm reg} &=  M_{q_1=0}^{-1} M_{q_1=\infty}^{-1}=  M_{q_1=0}^{-1} M_{q_2=\infty}^{-3} 
    \,,\\
    M_{\Delta_2} M_{\Delta_1} &= M_{q_2=\infty}^{-1} M_{q_2=0}^{-1}\,,
\end{eqn}
where $M_{\rm reg}$ is given in \eqref{Mreg} and we used \eqref{relqinf}. This is in accordance with the general pattern outlined in section \ref{sss:general} that the fundamental string charges are associated to the EFT strings at infinite distance and to the finite-tension string at $q_i=\infty$. 

\subsubsection*{Non-EFT strings}

So far we restricted ourselves to solutions associated to NS5-branes wrapping Nef divisors, corresponding to \eqref{Dgeneralh2} with charges $e^1,e^2\geq 0$, i.e. to EFT strings. In the present example, the cone of EFT strings in the heterotic case is given by 
\begin{align}
    \mathcal{C}^{\text{\tiny EFT}}_{\rm S} = \mathbbm{Z}_{\geq 0} \oplus \text{Nef}^1(Y_3)_\mathbbm{Z} = \langle F1, D1, D2\rangle\,,
\end{align}
where $F1$ is the fundamental string. We may extend this to the cone of potential $\oh$BPS strings  \cite{Lanza:2021udy}
\begin{align}
    \mathcal{C}_{\rm S} = \mathbbm{Z}_{\geq 0} \oplus \text{Eff}^1(Y_3)_\mathbbm{Z} = \langle F1, E, D2\rangle\,,
\end{align}
where $E=D_1-3D_2$. An NS5-brane wrapped on an effective, but not Nef divisor corresponds to a string charge in $\mathcal{C}_{\rm S} - \mathcal{C}_{\rm S}^{\text{\tiny EFT}}$. As pointed out in \cite{Lanza:2021udy} this gives rise to a $\oh$BPS string that does not yield a weakly-coupled theory in the vicinity of its core. This difference becomes apparent if we use our previous analysis to describe the backreaction of such a string. Indeed, consider the non-EFT string obtained by wrapping an NS5-brane on the divisor $D'= e^1 D_1 + e^2 D_2$ with $e^1>0$ and $-3\leq e^2<0$ such that $\bm{e}\in \mathcal{C}_{\rm S}- \mathcal{C}_{\rm S}^{\text{\tiny EFT}}$. Let us assume that before taking into account quantum corrections the string core is located at  $z=0\in \C$. In this case, \eqref{backh2} yields the FI-parameters profile
\begin{align}\label{backnonEFT}
    q_1 = \left(\frac{z}{z_0}\right)^{e^1}\bar{q}_1 \,,\qquad q_2 = \left(\frac{z}{z_0}\right)^{-|e^2|} \bar{q}_2\, .
\end{align}
 As a consequence of the negative power in the second expression, we find that near $z=0$ 
\begin{align}
    (|q_1|, |q_2|)\,\stackrel{z\,\rightarrow\, 0}{\longrightarrow}\, (0, \infty)\,,
    \label{nonEFTz0}
\end{align}
whereas at asymptotic infinity we have 
\begin{align}
    (|q_1|, |q_2|)\,\stackrel{z\,\rightarrow\, \infty}{\longrightarrow}\, (\infty, 0)\,.
\end{align}

The motivation for this class of solutions comes from wrapping an NS5-brane on  effective divisors in the Calabi--Yau regime, so let us consider those string solutions that go through the CY phase in figure \ref{fig:modulispacelogq}. Then we can consider the counter-clockwise monodromy around a closed loop $C = \{|z| = r_C\}$ centred around the classical NS5-brane core at $z=0$, and such that it always remains in the Calabi--Yau phase. By large-volume physics such a monodromy must be 
\begin{align}
    M_C = M_{q_1=0}^{e^1} M_{q_2=0}^{-|e^2|}\,.
    \label{nonEFTLVmon}
\end{align}

Let us see if this is compatible with the profile \eqref{backnonEFT}, and in particular with the string monodromies contained in it. To do so we plug it in the equation $\Delta=0$, to obtain
\begin{align}
    \Delta = \left[\left((1-432 \bar{q}_1 \left(\frac{z}{z_0}\right)^{e^1}\right)^3 -27 \cdot 432^3  \bar{q}_1^3\bar{q}_2\left(\frac{z}{z_0}\right)^{3 e^1-|e^2|}\right]\left[1+27 \bar{q}_2\left(\frac{z}{z_0}\right)^{-|e^2|}\right]=0\,.
\end{align}
The first factor is a polynomial of degree $3e^1$ such that in general we will find $3e^1$ regulator strings corresponding to $\{\Delta_1=0\}$. Notice that this differs from the EFT string solution for which the number of $\Delta_1$ regulator strings depends on both string charges. The second factor corresponds to $|e^2|$ regulator strings coming from $\Delta_2$. From figure \ref{fig:modulispacelogq}, we expect the regulator strings $\Delta_2$ to appear as we approach the classical NS5-brane core at $z=0$, and the $\Delta_1$ regulator strings as we go in the opposite direction towards spatial infinity. 

To make this picture more precise let us  specify to the non-EFT string charge $\bm{e}=(1,-1)$. In this case we will have three strings with monodromy in the conjugacy class of $M_{\Delta_1}$  and an additional regulator string associated to $\Delta_2$ located at $z=-27\bar{q}_2$. Let us consider the regime $1\gg\bar{q}_2\gg \bar{q}_1 >1 $. In this case the regulator string associated to $\Delta_2$ is indeed much closer to the origin than the regulator strings associated to $\Delta_1$, and the Calabi--Yau phase is in between whenever $r_C>27q_2$. We first look at the  strings outside of the region bounded by $C$, namely within $r_C <  |z| <\infty$. In order to have a finite-energy solution, the combined monodromy of these strings must equal the inverse of the counter-clockwise monodromy around $C$. That is
\begin{align}
    M_C^{-1} = M_{q_1=\infty}. [M_{\Delta_1}]^3 . M_{q_2=0} \,,
    \label{nonEFTmoninf}
\end{align}
where the representatives of the conjugacy class $M_{\Delta_1}$ are as in \eqref{M(1,0)}. Using \eqref{M(1,0)} we find that the rhs of \eqref{nonEFTmoninf} indeed coincides with the inverse of \eqref{nonEFTLVmon} for $\bm{e}=(1,-1)$. 

Let us now compare $M_C$ with the combined effect of the monodromies in the interior of $C$, using again the profile \eqref{backnonEFT}. Note that this corresponds to follow the local non-EFT string flow towards the classical string core at $z=0$. As expected, when approaching the vicinity of $z=0$ the solution is not in the weakly-coupled CY phase but flows towards the hybrid orbifold phase in the lower right corner in figure \ref{fig:modulispacelogq}. In doing so and given the restriction on the string charges, the above flow only crosses the singular divisor $\Delta_2$, which signals the appearance of a regulator string as the flow enters the hybrid orbifold phase. Therefore, what from the Calabi--Yau phase perspective seems like an elementary object is in fact a bound state of several strings, of which the regulator string at $\Delta_2$ is one component. Of course, if the 4d string solution is coarse-grained below a certain accuracy, one is not able to resolve the components of the bound state as independent objects. The string profile intersects further divisors at $z=0$, cf. \eqref{nonEFTz0}. More precisely, it meets the intersection of two divisors at the boundary of the FI-parameter space: $\{q_1=0\}\cap \{q_2=\infty\}$. We can interpret this as the fact that $z=0$ hosts a bound state of strings, although for the profile \eqref{backnonEFT} this bound state cannot be resolved at any scale.

With this in mind one sees that the total monodromy generated by the string cores in the interior of $C$ is given by
\begin{align}
    M_C'=  M_{q_1=0}M_{q_2=\infty} M_{\Delta_2} \stackrel{\eqref{Mq2infty}}{=} M_{q_1=0} M_{q_2=0}^{-1}M_{\Delta_1}^{-1}\,. 
\end{align}
We clearly see that $M_C'\neq M_C$, due to an extra factor of $M_{\Delta_1}$. This tells us that whereas asymptotically the non-EFT string solution is well-regulated due to the three regulator strings associated to $\Delta_1$, the regulation close to $z=0$ fails. Thus, the naive solution \eqref{backnonEFT} needs to be corrected close to $z=0$ and, in particular, the string solution should intersect the divisor $\Delta_1=0$ once more in order for the monodromies $M_C$ and $M_C'$ to match. 

To fix this mismatch let us consider an alternative string profile, which differs from \eqref{backnonEFT} near $z=0$ but reduces to it in the CY phase $|z| = r_C$ and beyond. This is achieved by a multi-string solution with a string of charge $(1,0)$ located at $z=0$ and a second string with charge $(0,-1)$ located at $z=\epsilon\neq0$. In this case the backreaction profile reads 
\begin{align}
      q_1 = \frac{z}{z_0}\bar{q}_1 \,,\qquad q_2 = \left(\frac{z-\epsilon}{z_0}\right)^{-1} \bar{q}_2\,. 
\end{align}
Notice that now the equation $\Delta_1=0$ has an additional solution and thus there is an additional regulator string associated to $\Delta_1=0$ in the vicinity of $z=\epsilon$, as required for $M_C'=M_C$. 

We therefore see that in order to reproduce the large-volume monodromy of a non-EFT NS5-string we need to describe its classical core as a bound state of several strings, and in particular to modify the naive profile \eqref{backnonEFT}. That such a profile must be modified near the core should not come as a surprise, since the source equation \eqref{Bianchi} that led to it was derived in the large-volume regime, and for non-EFT strings the FI-term sources belong to a different phase. 

What is less clear is the microscopic meaning of the regulator strings that correspond to the divisor $\{q_2=\infty\}$.  By our previous discussion we know that they host a non-vanishing local deficit angle $|\delta_{q_2=\infty}| = \frac{\pi}{9}$ which, in fact, should correspond to a localised negative tension. To see this, let us consider the tension of a non-EFT string in the CY regime. There the modulus-dependent string tension of EFT and non-EFT strings  $T_{\bm{e}}(\bar{\bm{t}})$ can be computed via \eqref{thetension}, where $\ell_1$, $\ell_2$ are as in \cite[eq.(4.29)]{Lanza:2021udy}. Let us in particular consider the charge $\bm{e} = (1,-3)$, which corresponds to an NS5-brane wrapping the effective divisor $E$. In this case we have that
\begin{equation}
   M_{\rm P}^{-2} T_{(1,-3)}(\bar{\bm{t}}) =  \left( \ell_1 -3 \ell_3\right) =  \frac{(s^2)^2}{6 (s^1)^3+6(s^1)^2s^2+2s^1(s^2)^2} < \frac{1}{2s^1} \, ,
\end{equation}
where $s^i (z)= \im t^i(z)$ is given by \eqref{imt}, and $z$ is near the curve $C$. It is then easy to see that for large enough $s_0^1$, $ T_{(1,-3)}(\bar{\bm{t}})$ can be made parametrically small. As we flow towards $z \to 0$, this tension should become smaller and smaller, even if we take into account all quantum corrections. As a result, this system should be incompatible with any positive localised tension near $z=\epsilon$. 

As pointed out in section \ref{sec:toy}, it is not clear how to interpret a string core with localised tension from the viewpoint of a BPS string solution, and in particular a 4d string that hosts a negative tension that is not placed at spatial infinity. One possibility would be to exploit the analogy with the toy example of section \ref{sec:toy} and treat such objects as the Q7-branes of \cite{Bergshoeff:2007aa}. Whether this is or not a sensible approach in the present setup is beyond the scope of this work, and in this sense the full microscopic description of non-EFT strings still remains unclear.  
\subsection{\texorpdfstring{Calabi--Yau with $h^{1,1}=3$}{Calabi--Yau with h11=3}}
\label{sec:h113}

Let us now consider a Calabi--Yau with $h^{1,1}=3$. Although many properties of this example are similar to the one just  discussed, there are some new features that we would like  to highlight. First, due to the higher dimension of the moduli space, also the lattice of string charges is larger and hence the set of regulator strings becomes more involved as well. Most importantly, however, in this example we can have an NS5-string that is dual to a critical string, in contrast to all previous NS5-branes which are not obviously dual to one. The example that we are about to discuss thus illustrates how to describe the backreaction of critical strings in our language. 

To construct our example we consider a Calabi--Yau $Y_3$ that is an elliptic fibration over the Hirzebruch surface $\mathbbm{F}_2$, with associated GLSM charges (see e.g. \cite{Mayr:2000as} for more details)
\begin{align}\label{Qh113}
\underline{Q} = \left(\begin{matrix} -6&2&3&1&0&0&0&0 \\  0&0&0&-2&1&1&0&0 \\
0&0&0&0&0&-2&1&1 \end{matrix}\right)\,. 
\end{align}
Each row of the matrix corresponds to the charges of the GLSM matter fields under one of the three $U(1)$ factors. Again, we introduce FI-terms for each of the $U(1)$ factors with exponentiated FI-parameters given by $(q_1, q_2, q_3)$. In the large-volume regime these parameters translate respectively to the volume of the $T^2$-fibre, the fibral $\mathbbm{P}^1$ of $\mathbbm{F}_2$ and the base $\mathbbm{P}^1$ of $\mathbbm{F}_2$. Accordingly, we also have three generators $(D_1,D_2,D_3)$ of the K\"ahler cone on which we can wrap NS5-branes to obtain EFT strings in 4d. 

In order to analyse the set of regulator strings, let us consider the discriminant locus of the GLSM associated to $Y_3$. Here, the discriminant splits into three factors 
\begin{align}
\Delta = \Delta_1 \cdot \Delta_2 \cdot \Delta_3 \,,
\end{align}
with the principal component given by 
\begin{eqn}
\Delta_1\equiv \left[(1 - 432 q_1)^2 - 432^2\cdot 2^2 q_1^2  q_2)\right]^2 -
 432^4 \cdot 2^6 q_1^4 q_2^2 q_3 =0 \,.
\end{eqn}
Notice that $\Delta_1$ is a polynomial of multi-degree $(4,2,1)$ in $(q_1,q_2,q_3)$. The two additional singular divisors are given by 
\begin{eqn}
\Delta_2 &= (1-4q_2)^2- 64 q_2^2q_3=0\,,\\
\Delta_3 &=(1-4q_3)=0\,.
\end{eqn}
Notice that $\Delta_2$ is independent of $q_1$ and $\Delta_3$ independent of both $q_1$ and $q_2$, thereby reflecting the fibration structure of the Calabi--Yau. In general we expect seven regulator strings associated to $\Delta_1=0$, three regulator string associated to $\Delta_2=0$ and a single regulator string associated to $\Delta_3=0$. Finally, in this case the monodromy group $\Gamma$ has four generators
\begin{align}\label{Gammah3}
       \Gamma =\langle M_{q_3=0}, M_{\Delta_1}, M_{\Delta_2}, M_{\Delta_3} \rangle\,,
\end{align}
for which we give explicit expressions in appendix \ref{app:monodromies}. Again, the monodromies $M_{q_1=0}$ and $M_{q_2=0}$ are not independent generators of the monodromy group but can be expressed through the set of generators in \eqref{Gammah3}. 

Let us now consider NS5-branes wrapped on the divisors
\begin{align}
D_{\bm e} = e^1 D_1 + e^2 D_2 + e^3 D_3 \,,\qquad \bm{e} \in \mathbbm{Z}_{\geq 0}^3\,,
\end{align}
which constitute the cone of EFT strings build from NS5-branes, as in \cite{Lanza:2021udy}. In case we consider type IIA compactified on $Y_3$, an NS5-brane wrapped on $D_3$, i.e. a string with charge vector $\bm{e}=(0,0,1)$, is dual to a critical heterotic string compactified on $K3\times T^2$ with $K3$ instanton embedding $(14,10)$ \cite{Harvey:1995rn,Kachru:1995wm}. Since in the previous examples none of the NS5-branes were dual to a critical string, we are especially interested in the string with charge $\bm{e}=(0,0,1)$. However, due to the fibration structure of $Y_3$ in order to describe backreaction of this string, it is convenient to first discuss the backreaction of the strings with charge $\bm{e}=(1,0,0)$ and $\bm{e}=(0,0,1)$. As before, we consider each of the generators of the cone separately:

\subsubsection*{\texorpdfstring{EFT strings  $\bm{e} =(1,0,0)$}{EFT strings  e=(1,0,0)}}

In this case we have $M_\text{\tiny EFT}=M_{q_1=0}$. This kind of EFT string yields a profile for the FI-parameters given by 
\begin{align}
\bm{q} =\left( \bar{q}_1\frac{z}{z_0}, \, \bar{q}_2,\, \bar{q}_3 \right)\,.
\end{align}
Since $\Delta_2$ and $\Delta_3$ do not depend on $q_1$ we do not expect any regulator strings arising from $\Delta_2=0$ and $\Delta_3=0$ for this charge vector. To analyse the regulator strings coming from $\Delta_1=0$, let us start with the extreme cases and set $\bar{q}_2=\bar{q}_3=0$. In this case the equation $\Delta_1=0$ reduces to 
\begin{align}
 \Delta_1|_{\bar{q}_2=\bar{q}_3=0} = \left(1-432 \bar{q}_1 \frac{z}{z_0}\right)^4 =0 \,.
\end{align}
This indicates that the divisor $\{\Delta_1=0\}$ has a point of tangency of order four with $\{q_2=0\}\cap \{q_3=0\}$ corresponding a bound state of four regulator strings at
\begin{align}\label{zstarh113}
z^* = \frac{z_0}{432 \bar{q}_1}\,.
\end{align}
As in the previous example, these four strings have monodromies in the same conjugacy class $[M_{\Delta_1}]$, but in order to find a sensible string solution the monodromies around the four strings need to correspond to different representatives. In order to see which representatives correspond to the four regulator strings, let us understand how they split as we turn on $\bar q_2, \bar q_3\neq 0$. 

If we allow for $\bar{q}_3\neq0$ while keeping $\bar q_2=0$, the singularity $\Delta_1=0$ does not get deformed 
\begin{align}
    \Delta_1|_{\bar{q}_2=0} = \left(1-432 \bar{q}_1 \frac{z}{z_0}\right)^4\,,
\end{align}
indicating that the four regulator strings do not split once we turn on $\bar{q}_3$ only. This should not come as a surprise given that the first Chern class of the base $\mathbbm{F}_2$ is simply given by $c_1(\mathbbm{F}_2) = 2 D'_2$, 
where $D_2=\pi^* D_2'$ with $\pi$ the projection $Y_3\rightarrow \mathbbm{F}_2$. Accordingly, the fibration of the torus over $\mathbbm{F}_2$ is insensitive to the modulus dual to $D_3$. 

The situation is different if we choose $\bar{q}_2\neq 0$ while maintaining $\bar{q}_3=0$. In this case the polynomial $\Delta_1$ is given by
\begin{align}
 \Delta_1|_{\bar{q}_3=0} = \left[\left(1-432 \bar{q}_1\frac{z}{z_0}\right)^2 -4\cdot 432^2 \left(\bar{q}_1\frac{z}{z_0}\right)^2 \bar q_2 \right]^2=0\,.
\end{align}
In this case the four string cores at $z^*$ split into two sets of two string cores located at 
\begin{align}
 z^\pm = \frac{z_0\left(1 \pm 2\sqrt{\bar{q}_2}\right)}{432\bar{q}_1(1-4\bar{q}_2)}\,. 
\end{align}
Notice that none of the string cores remains at $z^*$ but both sets move away from $z^*$. Accordingly upon applying the map $\mathfrak{M}$ from the FI-parameter space to the K\"ahler moduli space the points corresponding to $z^\pm$ cannot be cusps and are thus not at infinite distance. Since in the limit $\bar q_2 \rightarrow 0$ both string cores coincide and give rise to a cusp in the moduli space, the monodromies $M_{\Delta_1^\pm}$ around the two sets have to be different representatives of the same conjugacy class. More precisely we have 
\begin{align}
    M_{\Delta_1^{+}} = (M_{q_2=0})^{-1} M_{\Delta_1^-} M_{q_2=0}\,,
\end{align}
which reflects the splitting of the four string cores once we turn on $\bar{q}_2$.

If we now additionally allow for $\bar q_3\neq 0$, both string cores at $z^\pm$ split further. Again, the monodromy around the two components in which e.g. the string core at $z^-$ splits need to be different representatives of the conjugacy class $[M_{\Delta_1}]$. To reflect the splitting of the two string cores at $z^-$ once we turn on $\bar{q}_3$ we find
\begin{align}
    M_{\Delta_1^-} = (M_{q_3=0})^{-1} M_{\Delta_1} M_{q_3=0} M_{\Delta_1}\,. 
\end{align}
Accordingly, the combined monodromy around all four regulator strings for the $\bm{e}=(1,0,0)$ solution is given by 
\begin{align}
    M_{\rm reg} = (M_{q_3=0} M_{q_2=0})^{-1} M_{\Delta_1} M_{q_3=0}  M_{\Delta_1} M_{q_2=0} (M_{q_3=0})^{-1} M_{\Delta_1} M_{q_3=0} M_{\Delta_1}\,. 
\end{align}
The combined monodromy around EFT and regulator strings equals the inverse monodromy around the divisor $\{q_1=0\}$
\begin{align}
    (M_{q_1=\infty})^{-1} = M_{q_1=0} M_{\rm reg} \,.
\end{align}
Using the representations for the monodromies in appendix~\ref{app:monodromies} one can show that the monodromy at spatial infinity $M_{z=\infty}=M_{q_1=\infty}$ is of order six. At $z=\infty$ we thus have an order-six orbifold singularity, such that the total tension of the string solution is similar to \eqref{Te1} with $p_1=1$: 
\begin{align}
\frac{T_{(1,0,0)}}{M_{\rm P}^2} =\frac{2\pi}{6}\, .
\end{align}

\subsubsection*{\texorpdfstring{EFT strings  $\bm{e} =(0,1,0)$}{EFT strings  e =(0,1,0)}}

We now consider the second elementary string with charge $\bm{e}=(0,1,0)$ corresponding to $M_\text{\tiny EFT}=M_{q_2=0}$ that leads to a backreaction of the form 
\begin{align}
\bm{q} =\left( \bar{q}_1, \, \bar{q}_2\frac{z}{z_0}, \, \bar{q}_3 \right)\,.
\end{align}
Let us consider $\bar{q}_3=0$. In this case, we have two regulator strings coming from $\Delta_1=0$ and $\Delta_2=0$ each since both are polynomials of degree two in $q_2$. Notice that since $\bar{q}_3=0$ the two regulator strings coming from $\Delta_1=0$ coincide and similar for the regulator strings from $\Delta_2=0$. As in the previous case, the singularity only fixes the conjugacy class of the monodromy around the strings but the actual representative might differ, e.g., for the two regulator strings associated to $\Delta_1=0$. From the previous section, we infer that in the present case the monodromy around the first $\Delta_1$ string needs to be conjugated by $M_{q_3=0}$ with respect to the monodromy around the second $\Delta_1$ string. A similar conclusion holds for the two $\Delta_2$ strings, which can be seen by noticing that if we tune $\bar{q}_3=0$ the two regulator strings corresponding to $\Delta_2=0$ coincide. The total monodromy equals the inverse of the monodromy around the divisor $\{q_2=\infty\}$ which is thus given by 
\begin{align}\label{010}
    M_{q_2=\infty}^{-1}=M_{q_2=0} (M_{q_3=0})^{-1} M_{\Delta_2} M_{q_3=0} M_{\Delta_2} (M_{q_3=0})^{-1} M_{\Delta_1} M_{q_3=0} M_{\Delta_1} \,. 
\end{align}
Using the explicit representations for the monodromies in appendix~\ref{app:monodromies}, one can calculate 
\begin{align}
    \left[M_{q_2=0}(M_{q_3=0})^{-1} M_{\Delta_2} M_{q_3=0} M_{\Delta_2}  \right] = M_{q_1=0}\,,
\end{align}
such that  $(M_{q_2=\infty})^2=M_{q_1=\infty}$ and hence the monodromy at spatial infinity $M_{z=\infty}=M_{q_2=\infty}$ is of order 12. Due to this relation between the monodromies at infinity we can infer 
\begin{align}
\frac{T_{(0,1,0)}}{M_{\rm P}^2} =\frac{1}{2}\frac{T_{(1,0,0)}}{M_{\rm P}^2} = \frac{2\pi}{12}\,.
\end{align}

\subsubsection*{\texorpdfstring{EFT strings  $\bm{e} =(0,0,1)$}{EFT strings  e =(0,0,1)}}

We can now turn to the last, and most interesting, elementary string with charge $\bm{e}=(0,0,1)$. In this case we have $M_\text{\tiny EFT}=M_{q_3=0}$ and the embedding of the solution in FI-parameter space reads 
\begin{align}
\bm{q} =\left( \bar{q}_1, \, \bar{q}_2, \ \bar{q}_3\frac{z}{z_0}  \right)\,.
\end{align}
The corresponding two-cycle yields three regulator strings, one for each of the factors of $\Delta$ such that the total monodromy is given by 
\begin{align}
    M_{q_3=\infty}^{-1} = M_{q_3=0} M_{\Delta_3} M_{\Delta_2} M_{\Delta_1}\,,
\end{align}
where we used that the total monodromy equals the inverse of the monodromy around $\{q_3=0\}$. Compared to the other string solutions discussed previously, the $\bm{e}=(0,0,1)$ string is special since the monodromy around the divisor $\{\Delta_3=0\}$ as given in \eqref{MDelta3} is of order two and therefore not of infinite order as the monodromies around the other components of the discriminant. Following the general logic in section \ref{sss:general} this yields a local deficit angle
\begin{align}
    |\delta_{\Delta_3=0}| = \pi\,. 
\end{align}
To understand this, we notice that at $\Delta_3=0$ an exceptional divisor shrinks to a curve instead of shrinking to a point as is the case for $\Delta_2=0$. In fact, along $\{\Delta_3=0\}$ the base $\mathbb{P}^1$ of the Hirzebruch surface $\mathbb{F}_2$ shrinks to zero size. To see the $\mathbb{Z}_2$ symmetry induced by $M_{\Delta_3}$ we notice that the mirror map $\mathfrak{M}$ maps $q_3$ to \cite{Aspinwall:1993xz} 
\begin{align}
   t^3= b^3+is^3 = \frac{1}{2\pi i} \log\left(\frac{1-2q_3-2\sqrt{1-4q_3}}{2q_3}\right)\,.
\end{align}
From this expression it is evident that for $q_3=1/4$ the K\"ahler modulus $t^3$ vanishes and furthermore transforms as $t^3\rightarrow -t^3$ when encircling the point $q_3=1/4$. 

The simplest global string solution for charge $\bm{e}=(0,0,1)$ thus necessarily contains finite tension strings. This is reminiscent of the toy example in section \ref{sec:toy} where the basic building blocks $1A$ and $1B$ also contain strings of finite tension that are not sent to infinity. As in the toy example, we can construct global solutions only involving regulator strings with unipotent monodromy by patching together the building blocks. In the present example, we can patch together a solution associated to the $\bm{e}=(0,0,1)$ EFT string with the solution associated its $M_{\Delta_3}$ conjugate. The monodromy around the former string is then still given by $M_{q_3=0}$ whereas the monodromy around the latter is given by y $M_{\Delta_3} M_{q_3=0} (M_{\Delta_3})^{-1}$.
Since we have now two strings with monodromy in the conjugacy class of $M_{q_3=0}$ we also need two regulator strings for each of the conjugacy classes of $M_{\Delta_1}$ and $M_{\Delta_2}$. The total monodromy is then given by 
\begin{align*}
    M_{\rm tot} =\left((M_{\Delta_3})^{-1} M_{q_3=0} M_{\Delta_3} M_{q_3=0}\right) (M_{q_3=0})^{-1} M_{\Delta_2} M_{q_3=0} M_{\Delta_2} (M_{q_3=0})^{-1} M_{\Delta_1} M_{q_3=0} M_{\Delta_1}\,. 
\end{align*}
Using \eqref{Mq2Mdelta3} we recognise this as the monodromy $M_{q_2=\infty}^{-1}$ in \eqref{010}. Thus the tension of a solution corresponding to a $q_3=0$ EFT string and its $M_{\Delta_3}$ conjugate has tension given by $T_{(0,1,0)}$, i.e. 
\begin{align}
    \frac{T_{(0,0,1)}}{2\pi} = \frac{1}{12} M_{\rm P}^2\,.
\end{align}
Notice, however, that similar to the $2AB$ solution of the toy-example the two-cycle corresponding to a $\bm{e}=(0,0,1)$ EFT string described here is a double-cover of the two-cycle in moduli space corresponding to constant $(\bar q_1, \bar{q}_2)$.

We observe that among the few cases we studied explicitly, only the toy model of section \ref{sec:toy} and the $\bm{e}=(0,0,1)$ string have finite tension regulator strings that are not sent to infinity. In the case of the toy-model we can identify \eqref{MSL} as the moduli space for the type IIB axio-dilaton, and the $S$-transformation that yields a finite tension regulator string in the $1B$ solution as the S-duality transformation of the type IIB F1 string. On the other hand, as mentioned before the $\bm{e}=(0,0,1)$ string discussed in this section is dual to the heterotic F1 string compactified on $K3\times T^2$ instanton embedding $(14,10)$. Under this duality the modulus $t^3$ gets mapped to the heterotic axio-dilaton such that the limit $t^3\rightarrow \ii \infty$ corresponds to weak coupling. The monodromy induced by the $\Delta_3=0$ singularity corresponds to the $\mathbb{Z}_2$ duality of the heterotic string with instanton embedding $(14,10)$ which was first encountered in \cite{Aldazabal:1996fm}. From that perspective, the string with monodromy $(M_{\Delta_3})^{-1} M_{q_3=0} M_{\Delta_3}$ is just the image of the critical heterotic F1 string under the non-trivial generator of this $\mathbb{Z}_2$ duality group. We thus observe that in the few examples studied in this paper, the finite tension regulator strings are associated to strong-weak coupling dualities for \textit{critical} strings. It would be interesting to see if this turns out to be true more generally. 

As in the previous example, we can consider the subgroup $\Gamma_\text{\tiny EFT}$ of the monodromy group $\Gamma$ that corresponds to physical string charges. Again, the unipotent monodromies corresponding to $\Delta_1=0$ and $\Delta_2=0$ always appear in certain combinations in the string solutions, such that only these combinations are part of the finite-index subgroup $\Gamma_\text{\tiny EFT}$, generated by 
\begin{align}
    \Gamma_\text{\tiny EFT} = \langle M_{q_1=0}, M_{q_3=0}, M_{\Delta_3}, M_{q_2=\infty} \rangle\,. 
\end{align}
Notice that again all four generators correspond either to EFT string charges or to finite tension string cores. Unlike in the previous example the monodromy $M_{q_2=0}$ is not a generator of the group $\Gamma_\text{\tiny EFT}$ since it can be expressed through $M_{q_3=0}$ and its $M_{\Delta_3}$ conjugate, cf. \eqref{Mq2Mdelta3}.

\section{Conclusions}
\label{s:conclu}

In this paper we have studied the backreaction of $\oh$BPS strings in 4d $\cN=1$ and $\cN=2$ EFTs coupled to gravity. The main goal of our analysis is to find global extensions of  local solutions valid around the core of fundamental axionic strings as analysed in \cite{Lanza:2020qmt,Lanza:2021udy}, dubbed EFT strings. The main obstacle to extending these local, weakly-coupled solutions arises from the fact that far away from the string core the backreaction necessarily flows towards strong coupling. In these regions the weak-coupling description breaks down, and for instance the naive tension of the string solution diverges.  
 However, strong coupling effects can in principle resolve this problem and regularise the string solutions in such a way to yield solutions with finite energy. Based on the results obtained in this paper, we conjecture that in any 4d $\cN=1$ or $\cN=2$ EFT it is always possible to find a unique extension of all local EFT string solutions to a finite-tension BPS string solution over $\mathbb{C}$. Moreover, the smallest or elementary EFT string charges should lead to BPS solutions with sub-Planckian tension.

Of course, gathering evidence for this conjecture requires information about the strong coupling behaviour of the underlying EFT, which in general is hard to obtain. The main evidence for our conjecture comes from considering EFT string in Calabi--Yau compactifications of type IIA or heterotic strings. In this case EFT strings arise from NS5-branes wrapping Nef divisors in the internal geometry. For these strings, the weak coupling regime corresponds to the large-volume limit for the underlying CY manifold. Away from the string core the solution flows towards the interior of the moduli space where non-perturbative world-sheet instanton corrections become important. These corrections can in principle be calculated via e.g. mirror symmetry. This is typically only feasible locally around special loci in the moduli space, although there are instances where expressions that are valid globally can be found \cite{Alvarez-Garcia:2021mzv}. In this paper we circumvent this problem by describing the backreaction of the NS5-branes in terms of the GLSM associated to the Calabi--Yau compactification. To that end, we interpret the NS5-branes as gauge instantons for the GLSM. These induce a gauge anomaly that, using the results of \cite{Quigley:2011pv,Blaszczyk:2011ib}, can be canceled through logarithmic FI parameters. In the large volume regime, the logarithmic profile for the FI parameters can then be identified with the logarithmic profile for the complex scalar fields obtained from the EFT-string backreaction. Away from the large volume limit, however, treating the backreaction of the NS5-branes in terms of the FI parameters constitutes a significant simplification to our problem. This allows us to explicitly study global string solutions without needing to know the details of the mirror map relating the FI parameters to the Calabi--Yau quantum K\"ahler moduli space.

Through the logarithmic profile for the FI parameters, the backreaction of an EFT string covers a two-cycle $\Xi$ in the FI-parameter space $\mathcal{M}_\tau$. This two-cycle, and thus the backreaction of the string beyond weak coupling, can be characterised by its intersection with special divisors of the closure of the FI-parameter space, such as the discriminant divisor $\Delta$ of the GLSM. Since the intersections with such divisors correspond to codimension-one defects in the string solution, we interpret them as additional string cores -- dubbed regulator strings -- that regulate the backreaction of the  EFT string solution. Besides the divisor $\{\Delta=0\}$, far away from the string core the string solution intersects further divisors at spatial infinity. Similar to the analysis of \cite{Bergshoeff:2006jj}, 
the BPS equations relate the order of the monodromy at infinity to the global tension of the string solution. A finite-order monodromy at infinity is a necessary condition for global BPS solution of finite energy, while we demand that all the regulator strings host no localised tension. The energy stored in the backreaction can then be identified with the area of the image of the two-cycle $\Xi$ under the mirror map. Conjecture~\ref{conj:SESC} then states that moduli spaces for which some elementary holomorphic curves of genus zero, i.e. curves corresponding to elementary strings, have area larger than $2\pi M_{\rm P}^{-2}$ should be in the Swampland. It would be interesting to test the validity of this criterium beyond the examples studied here. 

An important lesson drawn from the construction of global string solutions is the relevance of the non-Abelian structure of the monodromy group $\Gamma$. At the level of the local solutions/weak coupling regime, EFT strings form a cone of commuting charges ${\cal C}^{\text{\tiny EFT}}_{\rm S}$ that correspond to a free Abelian subgroup $\Gamma_{z=0} \coloneqq \langle M_{q_i=0} \rangle$ of $\Gamma$. As we extend the solution towards strong coupling regulator strings appear, that correspond to new elements of the monodromy group. A necessary condition to regulate the solution is that these new charges do not commute with those of ${\cal C}^{\text{\tiny EFT}}_{\rm S}$. This is easy to see because by forming a bound state with the EFT string core they convert a unipotent monodromy (the local EFT string monodromy $M_{q_i=0}$) to a finite-order monodromy (the global EFT string monodromy at spatial infinity $M_{q_i=\infty}$). Our results suggest that in general these finite-order monodromies also commute among each other and form a finite Abelian subgroup $\Gamma_{z=\infty} \coloneqq \langle M_{q_i=\infty} \rangle$. Therefore in both extreme regimes (weak coupling and strong coupling at infinity) we get a set of commuting EFT string charges, albeit of a very different nature since one set is torsional and the other one is not. It is tempting to speculate that the effect of an EFT string charge becoming torsional at strong coupling is a manifestation of the large tension that it has in this regime, in the sense that taking a few copies of the string already overcloses the transverse space. 

An amusing fact that we obtain from our analysis is that the monodromies involved in EFT string solutions do not necessarily generate $\Gamma$, but instead a finite-index subgroup $\Gamma_\text{\tiny EFT}$. This subgroup contains the full information of the EFT string charges through the entire moduli space of the theory. In our examples of section \ref{s:examples}, we have checked that  $\Gamma_\text{\tiny EFT}$ is generated by the subgroups $\Gamma_{z=0}$ and $\Gamma_{z=\infty}$, in addition to some finite-order elements of $\Gamma$. In other words, the unipotent monodromies around $\Delta=0$ are not necessary to generate the subgroup $\Gamma_\text{\tiny EFT}$ describing the EFT string charges. We have interpreted this physically as the fact the regulator string associated to unipotent monodromies of $\Delta=0$  correspond to conifold-like singularities, that should be resolved by integrating in a few  degrees of freedom into the EFT. Therefore, they should not correspond to fundamental string charges of the EFT. If this intuition is correct and the above statement is general, it implies a sharp statement  about the abelianised duality/monodromy group of an EFT, which we have summarised in Conjecture \ref{conj:abelian}.

While we have centred our analysis on global solutions for EFT strings, in a compactification there are other $\half$BPS strings that could in principle admit a global backreacted solution. In particular one can have a 4d $\half$BPS string which, when treated as a probe, looks perfectly normal, but that is not an EFT string because upon backreaction the flow generated towards its core drives the fields towards strong coupling. In our setup, this corresponds to an NS5-brane wrapping an effective, but non-Nef divisor of a Calabi--Yau. We have applied out techniques to try to find a sensible global solution that describes such a non-EFT string. We have found that this is possible at the expense at treating the non-EFT string core as a bound state of several strings, one of which hosts a finite, negative tension. Due to this last ingredient, it is not obvious to what extent one can understand this configuration as a valid microscopic BPS solution. However, it could be that in practice this negative tension is never perceived by a 4d EFT with cut-off $\Lambda$, as we need to coarse-grain the solution to an accuracy $1/\Lambda$. It would be interesting to see if this kind of setup could be related to a new kind of self-protection mechanism of EFTs. 

One way of interpreting Conjecture \ref{conj:SESC} is that we can describe global string solutions, i.e. objects that probe strong coupling regions of moduli space, entirely through their weak coupling data. This non-trivial observation is quite reminiscent of  the moduli space holography proposed in \cite{Grimm:2020cda} which states that the bulk of the moduli space can be reconstructed from its boundary data. In this context, \cite{Cecotti:2020rjq,Grimm:2020cda,Grimm:2021ikg} have related the variation of Hodge structures to the dynamics of auxiliary $\sigma$-models. In particular, \cite{Grimm:2021ikg} identifies the equations of motion of such an auxiliary $\sigma$-model with one-parameter variations of Hodge structures. In our setup something similar happens: we obtain one-parameter variations of the K\"ahler structure from anomaly cancellation in the GLSM. This can be interpreted as solving the space-time Bianchi equation, which by supersymmetry is equivalent to solving the equation of motions in the presence of EFT strings. It would be interesting to understand better the relation of our analysis capturing the backreaction of physical strings in 4d with the dynamics of the auxiliary $\sigma$-model considered in \cite{Grimm:2021ikg}. 

The examples considered in this work are certainly just a first step to understanding global string solutions in $\cN=1$ and $\cN=2$ EFTs. In fact, most of our intuition comes from the K\"ahler sector of type IIA CY compactifications or heterotic CY compactifications with standard embedding since in these cases the quantum moduli spaces are well-understood and its global structure is nicely encoded in the non-Abelian monodromy group $\Gamma$. A further simplification arose since we could relate the two-cycles in moduli space arising from string solutions in the examples discussed in section~\ref{s:examples} to the fundamental domain of $SL(2,\mathbb{R})$. This allowed us to confirm that the tension of the string solution is given by the order of the monodromy asymptotic infinity since the area of the fundamental domain is known explicitly. One obvious extension of the examples considered here would be to consider setups where the moduli space corresponds to modular curves for different congruence subgroups of $SL(2,\mathbb{Z})$. Such cases have been recently studied in the context of the refined Swampland Distance Conjecture in \cite{Klawer:2021ltm}.

On the other hand, it would be desirable to extend these results e.g. to string solutions in the hypermultiplet moduli space of type II CY compactifications. In general, quantum corrections to the hypermultiplet moduli space are less under control compared to the cases considered in this paper. Still, it would be interesting to see how local string solutions are regulated and extended to global solutions for this type of moduli space. Our analysis showed that in the vector multiplet moduli space of type IIA CY compactifications all points in the quantum K\"ahler moduli space can be reached for some choice of asymptotic string charges and background values for the saxionic fields. It would be interesting to see if this is also the case for the hypermultiplet moduli space. If so, the string solutions should also capture the obstructions to reach certain points in the classical moduli space that were observed and analysed in \cite{Marchesano:2019ifh,Baume:2019sry,Alvarez-Garcia:2021pxo} and play an important role in the realisation of the emergent string conjecture \cite{Lee:2019oct} for this case.

One could also consider EFT strings and their backreaction in F-theory compactifications on elliptically fibred Calabi--Yau four-folds. In these cases, EFT strings arise from wrapping D3-branes on curves on the base of the four-fold. It would be interesting to study global BPS string solutions in these setups as well, even though little is known about the non-perturbative K\"ahler moduli space of F-theory compactifications to 4d. Our analysis revealed that the presence of regulator strings signals that the backreaction leaves the perturbative regime, and that the EFT string should be replaced by a bound state that includes the regulator string charge. In the context of F-theory compactifications it was shown in \cite{Klaewer:2020lfg} that certain classical weak-coupling limits for D3-brane strings, that are dual to critical heterotic strings as first studied in \cite{Lee:2019jan}, are obstructed by perturbative quantum corrections. When interpreting such a D3-brane as an EFT string this obstruction, if it persists at the non-perturbative level, can also be related to the presence of a regulator string signalling a transition to a strong coupling region, as will be discussed in more detail in \cite{Wiesner:2022}. 

Finally, the examples considered here correspond to cases where the EFT field space is an actual moduli space, in the sense that there is no potential generated for the scalar fields. In more general  4d $\cN=1$ EFTs (e.g. heterotic CY compactifications without standard embedding) this does not need to be the case, and in general there will be a non-trivial superpotential \cite{Palti:2020qlc}. In the presence of a potential our 4d string solutions will be modified, and this modification will be significant as soon as the Hubble or mass scale of the potential is comparable to the accuracy of the solution. In fact, this is already true for local EFT string solutions in the presence of potentials generated by fluxes. In those cases the 4d string typically develops an anomaly, that is cured by a set of membranes ending on it, see e.g. \cite{Berasaluce-Gonzalez:2012awn,Herraez:2018vae} for the case of 4d strings obtained from NS5-branes. This configuration has a well-understood counterpart in terms of $p$-form gaugings in the 4d Lagrangian \cite{Dvali:2005an,Kaloper:2008fb,Marchesano:2014mla}, but only in weak coupling regimes. It would be interesting to see if our techniques could allow us to transcend this perturbative picture, and provide a more general meaning to a set of couplings that have recently played a major role in the physics of 4d string compactifications.

\bigskip

\bigskip

\centerline{\bf  Acknowledgments}

\vspace*{.5cm}

We thank I\~naki Garc\'ia-Etxebarria, Luca Martucci, Miguel Montero and Raffaele Savelli for discussions.  The work of F.M. is supported through the grants CEX2020-001007-S and PGC2018-095976-B-C21, funded by MCIN/AEI/10.13039/501100011033 and by ERDF A way of making Europe.


\appendix


\section{\texorpdfstring{Strings in $\cN=2$ } {Strings in N=2 } }
\label{ap:N=2}

The discussion of the chiral multiplet moduli space in $\cN=1$ EFTs has an analogue in the vector multiplet moduli space of $\cN=2$ supergravity theories. Due to the $\cN=2$ supersymmetry, the moduli space factorises as 
\begin{align}
    \cM_{\cN=2} = \cM_{\rm VM}\times \cM_{\rm HM}\,,
\end{align}
with the two factors corresponding to the vector- and the hypermultiplet moduli spaces. This factorisation is reflected in the general $\cN=2$ supergravity action 
\begin{eqn}
    S_{\cN=2} = M_P^2\int &\left(\frac{1}{2} R*1 + \frac{1}{2}\text{Im}\, \cN_{ij} F^i \wedge \star F^j + \frac{1}{2} \text{Re}\, \cN_{ij} F^i \wedge \star F^j\right.\\
    &\left. -\frac{1}{2} K_{i \bar j}dt^i \wedge * d\bar t^{\bar j} - h_{uv} dq^u \wedge *dq^v \right)\,,
\end{eqn}
where $K_{i\bar j} = \partial_i \bar \partial_{\bar j} K$, and  $K(t^i, \bar t^{\bar j})$ is the K\"ahler potential of the vector multiplet moduli space. On the other hand, $h_{uv}$ is the quaternionic K\"ahler metric on the hypermultiplet moduli space. 
Finally, $\cN_{ij}$ is the gauge kinetic function of the gauge fields in the vector multiplets. 

The scalar sector of both, the hyper- and the vector multiplets, contain a number of axions that are classically associated to a shift-symmetry. Dualising these axions, we can consider the strings charged under the corresponding two-form gauge fields. Due to the factorisation of the total moduli space, we can now consider strings that are charged either under the two-forms in the vector multiplets or those in the hypermultiplets. These two kinds of strings of $\cN=2$ supergravity have been analysed in detail in \cite{Bergshoeff:2007ij}. 

Let us focus on the vector sector since in this case the moduli space is K\"ahler as is the chiral multiplet moduli space in $\cN=1$ supergravity. In this case, we can drop the $h_{uv} dq^u\wedge * dq^v$ term in the action such that it reduces to 
\begin{eqn}
    S_{\rm VM}= M_P^2\int &\left(\frac{1}{2} R*1 + \frac{1}{2}\text{Im} \cN_{ij} F^i \wedge \star F^j + \frac{1}{2} \text{Re}\cN_{ij} F^i \wedge \star F^j -\frac{1}{2} K_{i \bar j}dt^i \wedge * d\bar t^{\bar j} \right)\,.
\end{eqn}
We are interested in solutions that are independent of two spatial directions and which take the general form~\cite{Bergshoeff:2007ij}
\begin{align}\label{solgeneral}
    ds^2 = -dv(du + Hdu) + 2e^{-2K(z,\bar z)} dzd\bar{z} \,,
\end{align}
where $H$ is the scalar dual to the gauge invariant 2-form field strength 
\begin{align}
    H = dB + \star \hat J\,.
\end{align}
Here $\hat J$ is the anomalous current whose divergence corresponds to
\begin{align}
   - \partial_\mu\left(\sqrt{g} \hat J^\mu\right)= \delta \mathcal{L}- \delta_F \mathcal{L}\,,
\end{align}
with $\delta$ corresponding to the total variation, given by
\begin{eqn}
    \delta t^i = \alpha^A k_A^i(t)\,,\\
    \delta F = \alpha^A T_A \,,
\end{eqn}
and $\delta_F$ is just the variation w.r.t. to the field strength. In the above expression, $\alpha^A$ are the infinitesimal variation, $k_A^i$ are holomorphic Killing vectors on the moduli space and $T_A\in \mathfrak{sp}(2n_V +2 , \mathbbm{R})$ generators of the Lie algebra of the isometry group of $\cM_{\rm VM}$. 

The solution \eqref{solgeneral} corresponds to the superposition of cosmic strings and electromagnetic and gravitational waves propagating in the $u-v$-plane. In order to obtain pure string solutions, we need to set $H=0$ and turn off the electro-magnetic fields, such that the action reduces to 
\begin{align}
    S_{\rm VM}= M_P^2\int &\left(\frac{1}{2} R*1  -\frac{1}{2} K_{i \bar j}dt^i \wedge * d\bar t^{\bar j} \right)\,,
\end{align}
which we identify as \eqref{effaction}. And indeed, the string solutions now have the form \cite{Bergshoeff:2007ij} 
\begin{align}
    ds^2= -dt^2 +dx^2 + e^{-K(z,\bar z)}|f(z)|^2 dz d\bar{z}\,,
\end{align}
for $f$ a holomorphic function of $z$ which agrees with the Ansatz \eqref{metric}. 

\section{\texorpdfstring{$SL(2,\mathbb{Z})$ conjugacy classes and tension}{SL(2,Z) conjugacy classes and tension}}
\label{ap:conju}

Let us consider the group $SL(2,\mathbb{Z})$ which in our convention is generated by the two elements 
\begin{align}
    T = \left(\begin{matrix} 1&1\\0&1  \end{matrix}\right)\,,\qquad S=\left(\begin{matrix} 0&-1\\1&0  \end{matrix}\right)\,.
\end{align}
A general matrix in $SL(2, \mathbb{Z})$ acts on $\tau$ as 
\begin{align}
    \tau \mapsto \frac{a\tau + b}{c\tau + b}\,,\qquad \left(\begin{matrix} a&b\\c&d  \end{matrix}\right) \in SL(2,\mathbb{Z})\,,
\end{align}
such that $T$ keeps $\tau=i\infty$ and $S$ keeps $\tau=i$ fixed. In addition there is an elliptic point at $\tau=(1+\ii \sqrt{3})/2\equiv \rho$ that is invariant under
\begin{align}
    TS=\left(\begin{matrix} 1&-1\\1&0 \end{matrix}\right)\,.
\end{align}
Let us consider the conjugacy classes of $SL(2,\mathbb{Z})$ with absolute value of the trace smaller than or equal to 2. First there are the two conjugacy classes with trace zero
\begin{align}
    S=\left(\begin{matrix} 0&-1\\1&0  \end{matrix}\right)\,,\qquad -S=\left(\begin{matrix} 0&1\\-1&0  \end{matrix}\right)\,.
\end{align}
For trace $\pm 1$ there are four conjugacy classes for which we choose the representatives
\begin{eqn}
    TS&=\left(\begin{matrix} 1&-1\\1&0 \end{matrix}\right)\,,\quad  \;ST^{-1}=\left(\begin{matrix} 0&-1\\1&1 \end{matrix}\right) \\
    -TS&=\left(\begin{matrix} -1&1\\-1&0 \end{matrix}\right)\,,\quad -ST^{-1}=\left(\begin{matrix} 0&1\\-1&-1 \end{matrix}\right)\,,
\end{eqn}
and finally for trace $2$ there is an infinite family of conjugacy classes 
\begin{align}
    T^n=\left(\begin{matrix} 1&n\\0&1 \end{matrix}\right)\,,
\end{align}
and similar for trace $-2$. Let us use these conjugacy classes to find string solutions covering the fundamental domain exactly once. In total we thus look at configurations with three strings, one for each special point. For $\tau=i\infty$ we consider the conjugacy class of $T$, corresponding to the minimal EFT string charge. We then have eight different possibilities to assign monodromy classes to the points $\tau=i,\rho$. Recall that for the string solution to be embedded into $\P^1$, we need the total monodromy of the three strings to cancel. Let us consider the ordering for which the strings at $\tau=i\infty, i, \rho$ are located at $z=0,1,\infty$, respectively. The combined monodromy of the strings is then given by 
\begin{align}
    M_{\rm tot} = M_{i\infty} M_{i} M_{\rho}\,.
\end{align}
For the string at $\tau=\infty$ we pick the representative $M_{i \infty}=T$. If we then choose $M_{i}=-S$, the only way to trivialise the total monodromy is by choosing $M_\rho=ST^{-1}$. 
Similarly, if we choose $M'_{i}=S$ we need $M'_\rho=-ST^{-1}$ for a trivial monodromy.  It turns out that we only have these two distinct choices to assign conjugacy classes to the points $\tau=i,\rho$ if we require the total monodromy to be trivial. Choosing other representatives for the given conjugacy classes either lead to string solutions that are related to the ones given here via an overall conjugation or to string solution that cover the moduli space \eqref{MSL} more than once. For our string solution the trivial monodromy implies that the asymptotic deficit angle vanishes as in \eqref{tadpole}. The absolute value of the deficit angle for the elements in the conjugacy classes with vanishing trace is $|\delta_{i}| = \frac{2\pi}{4}$ since these are elements of order four in $SL(2,\mathbb{Z})$. On the other hand, for the conjugacy classes with trace $\pm 1$ we have $|\delta_{\rho, +}| = \frac{2\pi}{6}$ and $|\delta_{\rho,-}|=\frac{2\pi}{3}$. The string core with monodromy $T$ only contributes to the deficit angle as $\frac{1}{2}A_{SL(2)}$, with $A_{SL(2)}$ the area of the fundamental domain, via its backreaction (cf. \eqref{tadpole}). Imposing that this area is positive allows us to fix the sign of the deficit angles at $\tau=i,\rho$. In the first case, we have 
\begin{align}
    \frac{1}{2}A_{SL(2)} + (-1)^k \frac{2\pi}{4} + (-1)^l \frac{2\pi}{3} =0 \,,\qquad k,l=0,1\,. 
\end{align}
The condition $A_{SL(2)}\geq0$ requires $(k,l)=(0,1)$ or $(k,l)=(1,1)$ yielding $A_{SL(2)}=2\pi/6$ or $A_{SL(2)}=14\pi/6$, respectively. In the second case, we find 
\begin{align}
    \frac{1}{2}A_{SL(2)}+(-1)^{k'} \frac{2\pi}{4} + (-1)^{l'} \frac{2\pi}{6} =0\,,\qquad k',l'=0,1\,. 
\end{align}
which for $A_{SL(2)}\geq 0$ requires $(k',l')=(1,0)$ or again $(k',l')=(1,1)$ leading to $A_{SL(2)}=2\pi/6$ or $A_{SL(2)}=10\pi/6$, respectively. Since for both cases the area of the fundamental domain has to be the same, this selects the $(k,l)=(1,0)$ and $(k',l')=(0,1)$ solutions such that $A_{SL(2)}=2\pi/6$. 

In the first case we thus assign a negative deficit angle to the string with $\tau=\rho$. We can interpret this as a negative tension at $z=\infty$ and identify the corresponding string solution as the solution 1B. In the second case we have a negative tension for the string at $\tau=i$. However, given our ordering of strings this string is not at spatial infinity. To get a physical solution, we need to exchange the string at $z=1$ and $z=\infty$. By doing that we need to conjugate the monodromy around the string with $\tau=\rho$ by $S$ such that we obtain 
\begin{align}
    M'_{\rho}\rightarrow -S^{-1} S T^{-1}S = - T^{-1}S\,. 
\end{align}
We can identify the string solution for this choice of conjugacy classes as the building block 1A. 

\section{Basics of GLSMs}
\label{ap:GLSM}

In this appendix, we want to give some background on the Gauged Linear Sigma Models (GLSMs) that are used in the main text to describe the backreaction of  NS5-branes. In this work we exclusively consider Abelian GLSMs, i.e. two-dimensional supersymmetric gauge theories with gauge group $U(1)^r$ with either $(2,2)$ or $(0,2)$ supersymmetry. GLSMs were introduced in \cite{Witten:1993yc}, see e.g. \cite{Melnikov:2019tpl} for a review. 

Let us work in $(0,2)$ superspace spanned by the coordinate $(z,\bar z, \theta^+ , \bar \theta^+)$. To each $U(1)$ factor we have a pair of gauge superfields $V_+^i$, $V_-^i$, $i=1,\dots, h$. The associated field strength is contained in a Fermi multiplet $\Upsilon^i$ given by 
\begin{eqn}
\Upsilon^i= [\bar{\mathcal{D}}_+^i, \nabla_-^i]\,,
\end{eqn}
where $\mathcal{D}_+^i$ and $\nabla_-^i$ are gauge covariant (super-)derivatives defined as 
\begin{eqn}
\mathcal{D}_+^i&= \partial_{\theta^+} -i\bar \theta^+\left(\partial_{+} +i Q V_+^i\right) \,,\qquad \bar{\mathcal{D}}_+^i=- \partial_{\bar \theta^+} + \theta^+\left(\partial_{+} +i Q V_+^i\right) \,,\\
 \nabla_-^i&= \partial_-+iQ V_-^i\,. 
\end{eqn}
In addition, we have neutral chiral bosonic multiplets $S_i$, $i=1\,,\dots ,h$  and charged chiral matter fields $\Phi^\alpha$, $\alpha=1, \dots, n$ with charges $Q_\alpha^i$ under the gauge fields $V^i$. The chirality constraint reads
\begin{align}
    \bar{D}_+ \Phi^\alpha = (-\partial_{\bar \theta^+} + i \theta^+ \partial_+) \Phi^\alpha=0\, .
\end{align}
Finally, we have Fermi fields $\Gamma^a$, $a=1,\dots, d$, with charges $Q_a^i$ under the $U(1)$ gauge factors satisfying 
\begin{align}\label{ap:Gamma}
    \bar D_+ \Gamma^a = \sqrt{2}E^a(\Phi^\alpha, S_i)\,. 
\end{align}
The couplings $E^a$ are polynomials in the chiral fields. The kinetic term in the Lagrangian for these fields reads 
\begin{align}
    \mathcal{L}_{\rm kin} =  \int d^2 \theta^+ \left[-\frac{1}{8e^2} \bar \Upsilon_i \Upsilon_i -\frac{i}{2e ^2} \bar{S}_i \partial_- S- -\frac{i}{2}\bar \Phi^\alpha \nabla_-^i \Phi^\alpha -\frac{1}{2} \bar \Gamma^a \Gamma^a\right]\,.
\end{align}
In addition to that for the $U(1)$ gauge factors we can introduce a Fayet-Iliopoulos term 
\begin{align}
    \mathcal{L}_\text{F-I} = \frac{1}{4} \int d\theta^+ \tau^i \Upsilon^i|_{\bar \theta^+=0} + {\rm h.c.} = -\varrho^i D_i + \theta^i F_{01,i}\,,
\end{align}
where we expanded $\Upsilon_i$ in its components in Wess--Zumino gauge. On top of that we can have an additional coupling 
\begin{align}
    \mathcal{L}_J = \int d\theta^+ \Gamma^a J_a + {\rm h.c.}\,,
\end{align}
where the $J_a$ are functions of the chiral field $\Phi^\alpha$ and supersymmetry requires $\sum_{a = 1}^d E^a J_a=0$. Integrating out the auxiliary fields one derives the scalar potential 
\begin{align}\label{app:Ubos}
    U_{\rm bos} = \sum_a |E^a|^2 + \sum_{a} |J_a|^2 +\frac{e^2}{2} \sum_{i=1}^h D_i^2 \,,
\end{align}
with the D-term given by
\begin{align}
    D_i = \varrho^i - \sum Q_\alpha^i |\phi^\alpha|^2\,,
\end{align}
where $\phi^\alpha$ are the leading components of $\Phi^\alpha$. For general choices of the charge matrices $Q_\alpha^i$ and $Q_a^i$, the $U(1)$ gauge theories might suffer from anomalies. Under a gauge transformation parameterised by $\Lambda_i$, the action is in general not invariant but suffers from a gauge anomaly 
\begin{align}\label{ap:anomaly}
    \delta S = \frac{\mathcal{A}^{ij}}{16 \pi}\left(\int d^2z d\theta^+ \Lambda_i \Upsilon_j + {\rm h.c.}\right)\,,
\end{align}
where the anomaly matrix only depends on the charges 
\begin{align}
 \mathcal{A}^{ij} = \sum_{\alpha=1}^n Q_\alpha^i Q_\alpha^j - \sum_{a=1}^d Q_a^i Q_a^j \,. 
\end{align}
On the $(2,2)$ locus, this anomaly is automatically cancelled since here the bosonic and fermionic multiplets $(\Phi, \Gamma)$ combine into a single $(2,2)$ chiral multiplet. Accordingly, we need to have the same number of $\Phi$-fields as $\Gamma$-fields with the same charge and $\mathcal{A}^{ij}$ vanishes trivially. 

\section{Quintic fibration}
\label{apsec:quintic}

In this appendix we show that for the example of the quintic considered in section \ref{ss:quintic} we can fix $p=1$ in equation \eqref{AreaSigma5}. This implies that  the generator of the EFT string charge lattice yields a global solution with sub-Planckian tension, as predicted by Conjecture \ref{conj:SESC}.  As discussed in detail in section \ref{ss:EFTGLSM} the presence of the NS5-branes induces a profile for the FI-parameter $q$ as a function of $z$. According to \cite{Green:1993zr}, the energy stored in the backreaction can be identified with the degree of $q(z)$ 
\begin{align}\label{backdeg}
    \mathcal{E}_{\rm back} = 2\pi \,\text{deg}(q(z))\,. 
\end{align}
 To compute the degree of $q(z)$ we note that the variation of the FI-parameter $q$ over the transverse space can be equivalently described as a variation of the complex structure parameter $\psi$ of the mirror quintic. The mirror quintic is given by the equation 
\begin{align}
    P= x_1^2 + x_2^2+x_3^2 +x_4^3 + x_5^5 - \psi x_1 x_2x_3x_4 x_5 =0\,,
\end{align}
in the projective space $\mathbb{P}^5$ spanned by the coordinates $x_i$, $i=1,\dots 5$. To calculate the degree of the map $q(z)$ we thus have to find the degree of the line bundle $\mathcal{L}$ describing the variation of the holomorphic $(3,0)$-form of the mirror quintic as encoded in $\psi$ over $\mathbb{P}^1$. Notice that, via the mirror map, $\psi$ and $q$ are related via 
\begin{align}\label{qpsi}
    q = \frac{1}{\psi^5}\,. 
\end{align}
To find the degree of the line bundle $\mathcal{L}$, we follow \cite{Greene:1989ya} and construct a section of it. Therefore consider the hypersurface in $\mathbb{P}^1\times \mathbb{P}^5$ given by the equation
\begin{align}
     y_1(x_1^2 + x_2^2+x_3^2 +x_4^3 + x_5^5) - y_2( x_1 x_2x_3x_4 x_5) =0\,,
\end{align}
where $(y_1,y_2)$ are coordinates on $\mathbb{P}^1$. A section of $\mathcal{L}$ in the patch where $y_1\neq 0$ is given by
\begin{align}
   s_{(1)}= \frac{dx_3\wedge dx_4 \wedge dx_5}{5x_2^4-z x_3x_4x_5}\,,
\end{align}
where $z=y_2/y_1$ and we set $x_1=1$. On the other hand in the patch $y_2\neq 0$ we have the section
\begin{align}
     s_{(2)} = \frac{dx_3\wedge dx_4 \wedge dx_5}{5z' x_2^4- x_3x_4x_5}\,,
\end{align}
where $z'=y_1/y_2=z^{-1}$. Since $s_{(2)}=z s_{(1)}$, we find $c_1(\mathcal{L})=1$. Thus the fibration corresponds to $\psi$ being a polynomial of degree one in $z$. On the other hand, by \eqref{backdeg} this implies that the backreaction energy is given by
\begin{align}\label{app:ebackfibration}
    \mathcal{E}_{\rm back} = 2\pi\,.
\end{align}
Since $\psi$ is a degree-1 polynomial in $z$ via \eqref{qpsi} the backreaction of NS5-branes is hence described by a degree-5 polynomial in $q$. From the discussion in section \ref{ss:quintic} we know that a degree-5 polynomial corresponds to the backreaction of five NS5-brane strings. Using \eqref{app:ebackfibration} we hence can fix the integer $p$ in \eqref{AreaSigma5} to $p=1$.

\section{Representations of monodromy matrices}\label{app:monodromies}

In this appendix we collect explicit representation for the monodromy matrices of the $h^{1,1}=2$ example discussed in section \ref{sec:h112} and the $h^{1,1}=3$ example of section \ref{sec:h113}.

\subsection{\texorpdfstring{Example with $h^{1,1}=2$}{Example with h11=2}}

 Consider the action of the monodromies on B-branes given by the basis of sheaves 
\begin{align}\label{basissheafs}
    (\mathcal{O}_{\rm pt}, \mathcal{C}^i, \mathcal{O}_{D_i}, \mathcal{O}_{Y_3})\,,
\end{align}
where the $\mathcal{O}_{Y_3}$ is the structure sheaf of the CY $Y_3$, $D_i$ are the K\"ahler cone generators, $\mathcal{O}_{\rm pt}$ the skyscraper sheaf and the Mori cone generators $C^i$ are related to the push-forwards
\begin{align}\label{Cisheaf}
    \mathcal{C}^i={\iota_!}\mathcal{O}_{C^i}\left(K_{C^i}^{1/2}\right)\,.
\end{align}
In the $h^{1,1}=2$ example discussed in section \ref{sec:h112} the Mori cone generators are given by 
\begin{align}
    C^1=D_2.D_2\,,\qquad C^2=(D_1-3D_2).D_2\,.
\end{align}
In the basis \eqref{basissheafs}, the large volume monodromies $M_{q_i=0}$ are given by 
\begin{align}\label{largevolMono}
    M_{q_i=0} = \left(\begin{matrix} 1&-\delta_{ia} &0&0\\ 0 &\delta_{ab} & -\kappa_{iab}& \frac{\kappa_{iia}+\kappa_{aai}}{2} \\ 0&0&\delta_{ab} &-\delta_{ia}\\ 0&0&0&1\end{matrix}\right)^{-1}\,,
\end{align}
with $\kappa_{ijk}$ the triple self-intersection numbers. Using that in the $h^{1,1}=2$ example discussed in section \ref{sec:h112} the non-vanishing intersection numbers are 
\begin{align}
    \kappa_{111}=9\,,\quad \kappa_{112}=3\,,\quad \kappa_{122}=1\,,
\end{align}
we find the large volume monodromies to be given by 
\begin{align}
    M_{q_1=0} = \left(
\begin{array}{cccccc}
 1 & 1 & 0 & 3 & 9 & 0 \\
 0 & 1 & 0 & 3 & 9 & 0 \\
 0 & 0 & 1 & 1 & 3 & 1 \\
 0 & 0 & 0 & 1 & 0 & 0 \\
 0 & 0 & 0 & 0 & 1 & 1 \\
 0 & 0 & 0 & 0 & 0 & 1 \\
\end{array}
\right)\,, \quad M_{q_2=0} = \left(
\begin{array}{cccccc}
 1 & 0 & 1 & 0 & 1 & 0 \\
 0 & 1 & 0 & 1 & 3 & -1 \\
 0 & 0 & 1 & 0 & 1 & 0 \\
 0 & 0 & 0 & 1 & 0 & 1 \\
 0 & 0 & 0 & 0 & 1 & 0 \\
 0 & 0 & 0 & 0 & 0 & 1 \\
\end{array}
\right)\,. 
\end{align}
In addition we have the monodromies around the two components of the discriminant. First, the monodromy around $\Delta_1=0$ is triggered by the D6-brane wrapped in $Y_3$ becoming massless. The action on the basis \eqref{basissheafs} is then  (cf. \cite{Aspinwall:2001zq})
\begin{align}\label{Delta1mono}
    \text{ch}(\mathcal{F}) \rightarrow \text{ch}(\mathcal{F}) - \langle\mathcal{F},\mathcal{O}_{Y_3}\rangle \text{ch}\left(\mathcal{O}_{Y_3}\right)\,,
\end{align}
where the inner product is 
\begin{align}
    \langle \mathcal{E}, \mathcal{F}\rangle = \int_{Y_3} {\rm Td}(Y_3) \text{ch}(\mathcal{E})^\vee \text{ch}(\mathcal{F})\,,
\end{align}
with the Todd class for three-folds  given by $\text{Td}(Y_3) = 1+\frac{1}{12} c_2(Y_3)$. 
Furthermore, the operator $\vee:\oplus_i H^{i,i}(Y_3) \rightarrow\oplus_i H^{i,i}(Y_3)$ acts as $\delta^\vee =(-1)^i \delta$ for $\delta\in H^{i,i}(Y_3)$. For the three-fold in question the second Chern class is  
\begin{align}
    \int_{Y_3} c_2(Y_3) . D_1 = 102\,,\quad \int_{Y_3} c_2(Y_3) . D_2 = 36\,.
\end{align}
Using $\text{ch}(\mathcal{O}_{Y_3})=1$ and 
\begin{align}
    \text{ch}(D_i) = 1-e^{-D_i}\,,\qquad \text{ch}\left(\mathcal{C}^i\right)=C^i\, ,
\end{align}
we can find $M_{\Delta_1}$ and thus  $M_{q_1=\infty}$ via \eqref{M(1,0)} to be
\begin{align}
    M_{\Delta_1}= \left(
\begin{array}{cccccc}
 1 & 0 & 0 & 0 & 0 & 0 \\
 -10 & 1 & 0 & 0 & 0 & 0 \\
 -3 & 0 & 1 & 0 & 0 & 0 \\
 0 & 0 & 0 & 1 & 0 & 0 \\
 0 & 0 & 0 & 0 & 1 & 0 \\
 -1 & 0 & 0 & 0 & 0 & 1 \\
\end{array}
\right)\,,\quad 
    M_{q_1=\infty} = \left(
\begin{array}{cccccc}
 1 & -1 & 3 & -3 & -3 & 0 \\
 1 & 0 & 0 & -3 & 0 & 0 \\
 0 & 0 & 1 & -1 & 0 & -1 \\
 0 & 0 & 1 & 0 & -1 & 0 \\
 0 & 0 & 0 & 0 & 1 & -1 \\
 0 & 0 & 0 & 0 & 1 & 0 \\
\end{array}
\right)\,.
\end{align}
which indeed satisfies $M_{q_1=\infty}^6={\rm Id}$. The monodromy around $\Delta_2=0$ can now be calculated along the same lines by noticing that along $\Delta_2=0$ the central charge of the zero section $E=D_1-3D_2$ vanishes. Accordingly, the action of the monodromy $M_{\Delta_2}$ on the basis \eqref{basissheafs} is now given by 
\begin{align}\label{delta2mono}
    \text{ch}(\mathcal{F}) \rightarrow \text{ch}(\mathcal{F}) - \langle\mathcal{F},\mathcal{O}_{E}\rangle\, \text{ch}(\mathcal{O}_{E})\,.
\end{align}
To find the explicit expression for $M_{\Delta_2}$ let us first trade $\mathcal{O}_{D_i}$ for $\mathcal{O}_{E}$ in the basis \eqref{basissheafs}. The basis change can be achieved through the matrix
\begin{align}
    B= \left(
\begin{array}{cccccc}
 1 & 0 & 0 & 0 & 0 & 0 \\
 0 & 1 & 3 & -3 & -3 & 0 \\
 0 & 0 & 1 & 0 & 0 & 0 \\
 0 & 0 & 0 & 1 & 0 & 0 \\
 0 & 0 & 0 & 0 & 1 & 0 \\
 0 & 0 & 0 & 0 & 0 & 1 \\
\end{array}
\right)\,. 
\end{align}
In this basis it is straight-forward to find 
\begin{align}
    M_{\Delta_2}'=\left(
\begin{array}{cccccc}
 1 & 1 & 0 & 0 & 0 & 0 \\
 0 & 1 & 0 & 0 & 0 & 0 \\
 0 & -2 & 1 & 0 & 0 & 0 \\
 0 & -3 & 0 & 1 & 0 & 0 \\
 0 & 1 & 0 & 0 & 1 & 0 \\
 0 & 0 & 0 & 0 & 0 & 1 \\
\end{array}
\right)\,,
\end{align}
such that $M_{\Delta_2}= B M_{\Delta_2}' B^{-1}$. A direct computation reveals $(M_{\Delta_2}M_{q_2=0})^3=M_{q_1=0}$. We can further calculate the monodromy $M_{q_2=\infty}$ in \eqref{Mq2infty} 
\begin{align}
    M_{q_2=\infty} = (M_{\Delta_2} M_{\Delta_1} M_{q_2=0})^{-1} = \left(
\begin{array}{cccccc}
 -2 & 0 & -1 & 0 & 0 & 0 \\
 12 & -11 & 36 & -37 & -39 & 2 \\
 3 & 0 & 1 & 0 & -1 & 0 \\
 -1 & 4 & -12 & 13 & 12 & -1 \\
 0 & -1 & 3 & -3 & -2 & 0 \\
 1 & -1 & 3 & -3 & -3 & 1 \\
\end{array}
\right)\,. 
\end{align}
From this expression we find $(M_{q_2=\infty})^3=M_{q_1=\infty}$,  so $(M_{q_2=\infty})^{18}=\text{Id}$ and  $[M_{q_1=\infty}, M_{q_2=\infty}]=0$.

\subsection{\texorpdfstring{Example with $h^{1,1}=3$}{Example with h11=3}}
For the second example with $h^{1,1}=3$ we can proceed in the same way. In this case the intersection polynomial is given by 
\begin{align}
    I(Y_3)=8D_1^3 +4D_1^2D_2 + 2 D_1^2D_3 + 2 D_1 D_2^2 + D_1D_2D_3\,,
\end{align}
and the second Chern class gives 
\begin{align}
        \int_{Y_3} c_2(Y_3) . D_1 = 92\,,\quad \int_{Y_3} c_2(Y_3) . D_2 = 48\,,\quad \int_{Y_3} c_2(Y_3) . D_3= 24\,.
\end{align}
The Mori cone generators can be expressed as 
\begin{align}
    C^1 = D_2.D_3\,,\qquad C^2 = (D_1-2D_2).D_3\,,\qquad C^3 = (D_1-2D_2).(D_2-3D_3)\,. 
\end{align}
From \eqref{largevolMono} large volume monodromies are 
\begin{eqn}
    M_{q_1=0}&= \left(
\begin{array}{cccccccc}
 1 & 1 & 0 & 0 & 2 & 4 & 8 & 0 \\
 0 & 1 & 0 & 0 & 2 & 4 & 8 & 0 \\
 0 & 0 & 1 & 0 & 1 & 2 & 4 & 1 \\
 0 & 0 & 0 & 1 & 0 & 1 & 2 & 1 \\
 0 & 0 & 0 & 0 & 1 & 0 & 0 & 0 \\
 0 & 0 & 0 & 0 & 0 & 1 & 0 & 0 \\
 0 & 0 & 0 & 0 & 0 & 0 & 1 & 1 \\
 0 & 0 & 0 & 0 & 0 & 0 & 0 & 1 \\
\end{array}
\right) \,,\quad M_{q_2=0} = \left(
\begin{array}{cccccccc}
 1 & 0 & 1 & 0 & 0 & 0 & 2 & 0 \\
 0 & 1 & 0 & 0 & 1 & 2 & 4 & -1 \\
 0 & 0 & 1 & 0 & 0 & 0 & 2 & 0 \\
 0 & 0 & 0 & 1 & 0 & 0 & 1 & 0 \\
 0 & 0 & 0 & 0 & 1 & 0 & 0 & 0 \\
 0 & 0 & 0 & 0 & 0 & 1 & 0 & 1 \\
 0 & 0 & 0 & 0 & 0 & 0 & 1 & 0 \\
 0 & 0 & 0 & 0 & 0 & 0 & 0 & 1 \\
\end{array}
\right)\,,\vspace{1cm}
\end{eqn}
and 
\begin{eqn}
M_{q_3=0} = \left(
\begin{array}{cccccccc}
 1 & 0 & 0 & 1 & 0 & 0 & 0 & 0 \\
 0 & 1 & 0 & 0 & 0 & 1 & 2 & -1 \\
 0 & 0 & 1 & 0 & 0 & 0 & 1 & 0 \\
 0 & 0 & 0 & 1 & 0 & 0 & 0 & 0 \\
 0 & 0 & 0 & 0 & 1 & 0 & 0 & 1 \\
 0 & 0 & 0 & 0 & 0 & 1 & 0 & 0 \\
 0 & 0 & 0 & 0 & 0 & 0 & 1 & 0 \\
 0 & 0 & 0 & 0 & 0 & 0 & 0 & 1 \\
\end{array}
\right)\,.
\end{eqn}
As before, the monodromies around the singular divisors $\{\Delta_1=0\}$ and $\{\Delta_2=0\}$ can be calculated using \eqref{Delta1mono} and \eqref{delta2mono} to find 
\begin{align*}
    M_{\Delta_1} = \left(
\begin{array}{cccccccc}
 1 & 0 & 0 & 0 & 0 & 0 & 0 & 0 \\
 -9 & 1 & 0 & 0 & 0 & 0 & 0 & 0 \\
 -4 & 0 & 1 & 0 & 0 & 0 & 0 & 0 \\
 -2 & 0 & 0 & 1 & 0 & 0 & 0 & 0 \\
 0 & 0 & 0 & 0 & 1 & 0 & 0 & 0 \\
 0 & 0 & 0 & 0 & 0 & 1 & 0 & 0 \\
 0 & 0 & 0 & 0 & 0 & 0 & 1 & 0 \\
 -1 & 0 & 0 & 0 & 0 & 0 & 0 & 1 \\
\end{array}
\right),\quad M_{\Delta_2}=\left(
\begin{array}{cccccccc}
 1 & 1 & -2 & 0 & 2 & 4 & 2 & 0 \\
 0 & 1 & 0 & 0 & 0 & 0 & 0 & 0 \\
 0 & -3 & 7 & 0 & -6 & -12 & -6 & 0 \\
 0 & -1 & 2 & 1 & -2 & -4 & -2 & 0 \\
 0 & 0 & 0 & 0 & 1 & 0 & 0 & 0 \\
 0 & -2 & 4 & 0 & -4 & -7 & -4 & 0 \\
 0 & 1 & -2 & 0 & 2 & 4 & 3 & 0 \\
 0 & 0 & 0 & 0 & 0 & 0 & 0 & 1 \\
\end{array}
\right). 
\end{align*}
The monodromy around $\{q_1=\infty\}$ in this case is given by 
\begin{eqn}
    M_{q_1=\infty} &= \big[M_{q_1} \left(M_{q_2=0} M_{q_3=0} \right)^{-1} M_{\Delta_1} M_{q_3} M_{\Delta_1} M_{q_2} (M_{q_3})^{-1} M_{\Delta_1}M_{q_3}M_{\Delta_1}\big]^{-1}\\
    &=\left(
\begin{array}{cccccccc}
 1 & -1 & 2 & 0 & -2 & -4 & -2 & 0 \\
 1 & 0 & 0 & 0 & -2 & -4 & 0 & 0 \\
 0 & 0 & 1 & 0 & -1 & -2 & 0 & -1 \\
 0 & 0 & 0 & 1 & 0 & -1 & 0 & -1 \\
 0 & 0 & 1 & -2 & 0 & 0 & 1 & 0 \\
 0 & 0 & 0 & 1 & 0 & 0 & -1 & 0 \\
 0 & 0 & 0 & 0 & 0 & 0 & 1 & -1 \\
 0 & 0 & 0 & 0 & 0 & 0 & 1 & 0 \\
\end{array}
\right)\,,
\end{eqn}
which satisfies $(M_{q_1=\infty})^6 =\text{Id}$. On the other hand, we have 
\begin{align}
    \left[M_{q_2} (M_{q_3})^{-1} M_{\Delta_2} M_{q_3} M_{\Delta_2}\right]^2 = M_{q_1}\, .
\end{align}
With this input, we can calculate the monodromy $M_{q_2=\infty}$ for the $h^{1,1}=3$ example as given in \eqref{010}
\begin{align}
  M_{q_2=\infty}=\left(
\begin{array}{cccccccc}
 -1 & 1 & -1 & -2 & 1 & 4 & 4 & -1 \\
 1 & 0 & 1 & -8 & -2 & 4 & 2 & 4 \\
 0 & 1 & -1 & -4 & 1 & 4 & 4 & -1 \\
 0 & 0 & 0 & -1 & 0 & 1 & 1 & 0 \\
 -1 & 1 & -1 & -1 & 2 & 3 & 3 & -1 \\
 0 & 0 & 0 & 0 & 0 & -1 & 0 & -1 \\
 0 & 0 & 0 & 0 & 0 & 1 & 1 & 0 \\
 0 & 0 & 0 & -1 & 0 & 1 & 1 & 1 \\
\end{array}
\right)\,.  
\end{align}
Direct computation then reveals that  $M_{q_2=\infty}^2=M_{q_1=\infty}$. 

To calculate the last generator of the monodromy group $M_{\Delta_3}$ we cannot simply use \eqref{delta2mono} since at $\Delta_3=0$ no divisor shrinks to a point. To describe the associated monodromy we can use \cite[eq.(28)]{Aspinwall:2001zq} that gives the action of the monodromy around a point in moduli space where a divisor $E'$ shrinks to a curve $Z$ of genus $g$ as 
\begin{align}
    \text{ch}(\mathcal{E}) \rightarrow \text{ch}(\mathcal{E}) -\langle \mathcal{O}_{E'} +(1-g) \mathcal{O}_\Gamma, \mathcal{E}\rangle \,\text{ch}(\mathcal{O}_\Gamma) + \langle \mathcal{O}_\Gamma , \mathcal{E}\rangle \text{ch}(\mathcal{O}_{E'})\,. 
\end{align}
In the present case the curve $\Gamma\equiv C^3$ and the divisor $E'=D_2-2D_3$. Notice that the sheaf $\mathcal{O}_{C^3}$ differs from the sheaf $\mathcal{C}^3$ as in \eqref{Cisheaf} by the twist with the canonical bundle $K_{C^3}^{1/2}$. Taking this subtlety for the choice of basis into account, we find that in the basis \eqref{basissheafs} the monodromy $M_{\Delta_3}$ is given by 
\begin{align}\label{MDelta3}
    M_{\Delta_3} =\left(
\begin{array}{cccccccc}
 1 & 0 & 0 & 0 & 0 & 0 & 0 & 0 \\
 0 & 1 & 0 & 0 & -1 & 0 & 0 & -1 \\
 0 & 0 & 1 & 0 & 0 & 0 & 0 & 0 \\
 0 & 0 & 1 & -1 & 0 & 0 & 1 & 0 \\
 0 & 0 & 0 & 0 & -1 & 0 & 0 & -2 \\
 0 & 0 & 0 & 0 & 1 & 1 & 0 & 1 \\
 0 & 0 & 0 & 0 & 0 & 0 & 1 & 0 \\
 0 & 0 & 0 & 0 & 0 & 0 & 0 & 1 \\
\end{array}
\right)\,. 
\end{align}
Notice that $M_{\Delta_3}$ is of finite order, i.e. $(M_{\Delta_3})^2=\text{Id}$. On the other hand we have the relation 
\begin{align}\label{Mq2Mdelta3}
    M_{q_2=0} =\left(M_{\Delta_3}\right)^{-1} M_{q_3=0} M_{\Delta_3}M_{q_3=0} \,. 
\end{align}


\bibliographystyle{JHEP2015}
\bibliography{papers}

\end{document}